\documentclass[11pt,a4paper]{article}
\usepackage{boondox-cal}
\usepackage{float}
\usepackage{diagbox}
\usepackage{jheppub}
\usepackage{graphicx}
\usepackage{amsmath}
\usepackage{amssymb}
\usepackage{amsthm}
\usepackage{longtable}
\usepackage{mathtools}
\usepackage{multirow}
\usepackage{mathrsfs}
\input epsf
\usepackage{epsfig}
\usepackage{tikz}
\usepackage{subfigure,tikz}
\usetikzlibrary{backgrounds,automata}
\usepackage{graphics}
\usepackage{subfigure}
\usepackage[active]{srcltx}
\usepackage[T1]{fontenc} 
\usepackage{graphicx}
\usepackage{natbib}
\usepackage{hyperref}

\usepackage[title]{appendix}
\allowdisplaybreaks

\newcommand{\bea}{\begin{eqnarray}}
	\newcommand{\eea}{\end{eqnarray}}
\newcommand{\bean}{\begin{eqnarray*}}
	\newcommand{\eean}{\end{eqnarray*}}

\def\a{{\alpha}}

\allowdisplaybreaks

\def\Label#1{\label{#1}%
	\smash{\hbox to0pt{\raise1ex\hbox{\tiny[#1]}\hss}}}

\newcommand{\parall}[2]{{#1}\ /\kern -0.8em / \  {#2}}

\newcommand{\red}[1]{\textcolor{red}{{}#1}}

\title{\boldmath Direct Expression for One-Loop Tensor Reduction with Lorentz Indices via Generating Function}

\author[a,b,c]{Chang Hu*,}

\affiliation[a]{College of Physics Science and Technology, Hebei University, Baoding 071002, China}

\affiliation[b]{Hebei Key Laboratory of High-precision Computation and Application of Quantum Field Theory, Baoding,
071002, China
}
\affiliation[c]{Hebei Research Center of the Basic Discipline for Computational Physics, Baoding, 071002, China
}

\author[d,e,f]{Yifan Hu,}
\author[d,e,f]{Jiyuan Shen.}

\affiliation[d]{School of Fundamental Physics and Mathematical Sciences, Hangzhou Institute for Advanced Study, UCAS, Hangzhou 310024, China}

\affiliation[e]{University of Chinese Academy of Sciences, Beijing 100049, China}

\affiliation[f]{Institute of Theoretical Physics, Chinese Academy of Sciences, Beijing 100190, China}

\emailAdd{isiahalbert@126.com} 
\emailAdd{huyifan23@mails.ucas.ac.cn}
\emailAdd{isaiahyshen@gmail.com}

\abstract{ In \cite{Hu:2023mgc}, we derived a direct expression for one-loop tensor reduction using generating functions and Feynman parametrization in projective space, avoiding recursive relations. However, for practical applications, this expression still presents two challenges: (1) While the final reduction coefficients are expressed in terms of the dimension 
D and Mandelstam variables, the given expression explicitly contains irrational functions; (2) The expression involves an auxiliary vector 
R, which can be eliminated via differentiation $\frac{\partial}{\partial R}$, but the presence of irrational terms making differentiation cumbersome. (3) Most practical applications require the tensor form with Lorentz indices.

In this paper, we provide a rational form of the reduction coefficients with Lorentz indices, free from recursion. Additionally, We provide a pure Wolfram Mathematica implementation of the code. Our practical tests demonstrate that this direct expression achieves significantly higher computational efficiency compared to the traditional Passarino-Veltman (PV) reduction or other recursion-based methods.

}

\renewcommand{\thefootnote}{\fnsymbol{footnote}}
\footnotetext{*Corresponding author}

\usepackage{amsfonts}
\begin{document}
\maketitle

\renewcommand{\thefootnote}{\arabic{footnote}}
\setcounter{footnote}{0} 
	\flushbottom
	
\section{Motivation and Introduction}\label{sec:intro}

With the increasing precision of experiments at the Large Hadron Collider(LHC), precise calculations have become increasingly crucial in the search for new physics. A significant challenge in precise calculations is the computation of loop diagrams, or Feynman integrals. The primary method for addressing Feynman integrals is through loop integral reduction, where any loop integral is expressed as a linear combination of a set of basic integrals (commonly referred to as master integrals). The coefficients of this reduction, known as reduction coefficients, are rational functions of external momenta, polarization vectors, masses, and spacetime dimensions. Master integrals are typically determined solely by the structure of the Feynman integrals themselves and are often independent of the specific physical theory. Consequently, the computation of Feynman integrals involves two steps: first, determining the reduction coefficients (including identifying the master integrals), and second, computing the master integrals. 

The study of reduction methods primarily follows two approaches. The first is based on analytic properties, focusing on the mathematical structure of Feynman integrals and proposing new methodologies from a more general perspective. The second approach stems from practical computational needs, emphasizing the development of efficient algorithms and tools to enhance computational efficiency. These two approaches are complementary: analytic research provides theoretical guidance for methodology, while practical tools can, in turn, feedback into and inspire further theoretical studies. Reduction methods can be broadly classified into two categories: integrand-level reduction and integral-level reduction. The former fundamentally involves algebraic decomposition, with representative techniques including the OPP method \cite{Ossola:2006us} and computational algebraic geometry methods \cite{Mastrolia:2011pr,Badger:2012dp,Zhang:2012ce,Larsen:2015ped,Zhang:2016kfo}. The latter includes well-known approaches such as PV reduction \cite{Passarino:1978jh}, the IBP method \cite{Chetyrkin:1981qh,Tkachov:1981wb,Laporta:2000dsw,vonManteuffel:2012np,Maierhofer:2017gsa,Smirnov:2019qkx,vonManteuffel:2014ixa}, unitarity cuts \cite{Bern:1994cg,Bern:1994zx,Britto:2004nc,Britto:2005ha,Britto:2006sj,Anastasiou:2006gt,Anastasiou:2006jv,Britto:2006fc,Britto:2007tt,Britto:2010um,Giele:2008ve}, and the intersection theory method \cite{Mastrolia:2018uzb,Frellesvig:2019uqt,Mizera:2019vvs,Frellesvig:2020qot,Caron-Huot:2021iev,Caron-Huot:2021xqj}. Correspondingly, several software packages have been developed for these methods. For example, integrand-level reduction is implemented in \texttt{CutTools} \cite{Ossola:2007ax}, while integral-level reduction benefits from tools like \texttt{Kira}\cite{Maierhofer:2017gsa,Maierhofer:2018gpa,Klappert:2020nbg}, \texttt{FIRE}\cite{Smirnov:2019qkx,Smirnov:2008iw,Smirnov:2013dia,Smirnov:2014hma}, \texttt{FIESTA}\cite{Smirnov:2008iw,Smirnov:2008py,Smirnov:2009pb,Smirnov:2013eza,Smirnov:2015mct,Smirnov:2021rhf}, \texttt{FireFly}\cite{Klappert:2019emp,Klappert:2020aqs}, \texttt{Reduze}\cite{Studerus:2009ye,vonManteuffel:2012np}, \texttt{LiteRed}\cite{Lee:2012cn,Smirnov:2013dia} (based on IBP and the Laporta algorithm), \texttt{NeatIBP} \cite{Wu:2023upw}(which integrates the advantages of computational algebraic geometry), \texttt{Blade}\cite{Guan:2019bcx,Guan:2024byi}, which specializes in numerical reduction, \texttt{OPEITeR}\cite{Goode:2024mci,Goode:2024cfy}, which focus on tensor integral reduction, and \texttt{AmpRed}\cite{Chen:2024xwt}, which leverages the specific features of parameterizations. Other reduction methods and software packages are not detailed here\cite{Peraro:2019svx,Belitsky:2023qho,Smirnov:2020quc,Usovitsch:2020jrk,Anastasiou:2004vj,Abreu:2018zmy}.

Although the reduction process involves both calculating reduction coefficients and identifying master integrals, by around the year 2000, the master integrals for one-loop integrals were already well-established\cite{Ellis:2007qk,Carrazza:2016gav}. This paper focuses solely on the computation of reduction coefficients for one-loop tensor integrals and does not address the identification of master integrals. Typically, the reduction of tensor integrals is carried out using the PV method \cite{Passarino:1978jh,Denner:2002ii,Denner:2016kdg}, which calculates the coefficients through the analysis of possible tensor structures with Lorentz indices and some algebraic transformations. Another approach involves decomposing the loop momentum into components parallel and orthogonal to the subspace formed by external momenta \cite{Mastrolia:2016dhn}. In \cite{Feng:2021enk,Hu:2021nia,Feng:2022uqp,Feng:2022iuc,Feng:2022rwj,Feng:2022rfz,Li:2022cbx,Zhang:2023jzv}, we also explored an improved PV reduction method by introducing an auxiliary vector $R$, effectively transforming tensor integrals into scalar integrals for further analysis.

Regardless of the reduction method employed, a key characteristic shared by most is the presence of an iterative framework. For instance, through IBP relations, a higher-rank tensor integral can be expressed as a linear combination of lower-rank tensor integrals. Effectively leveraging these iterative structures while avoiding redundant calculations plays a crucial role in improving the efficiency of the reduction process.

In \cite{Feng:2022hyg}, the author systematically introduced the concept of generating functions into the computation of loop reduction coefficients.\footnote{The potential of generating functions in handling Feynman integrals has previously begun to emerge in \cite{Ablinger:2014yaa,Kosower:2018obg}. Subsequent follow-up works include \cite{Guan:2023avw,Hu:2023mgc,Feng:2024qsa,Li:2024rvo,Chen:2024bpf,Brandhuber:2024qdn}.} Specifically, the reduction coefficients are treated as the Taylor expansion coefficients of a generating function with respect to the parameter $t$. This idea is widely used in both mathematics and physics, particularly in algorithms with recursive structures. Generating functions not only provide deeper insights into the analytic structure of mathematical forms but also significantly improve computational efficiency. For example, the closed-form formula for Fibonacci numbers demonstrates how integers can be expressed using irrational numbers, showcasing an innovative use of analytic structures. Similarly, the Hermite polynomials $H_n(x)$, which are otherwise complex to represent directly, become remarkably simple and elegant when expressed through a generating function, highlighting the power of this approach in achieving concise representations.

\begin{equation}
    e^{2tx-t^2}=\sum_{n=0}^{\infty}H_n(x)\frac{t^n}{n!}.
\end{equation}
Therefore, rather than directly tackling the complexity of reduction coefficients themselves, we can focus on studying their generating functions

\begin{equation}
    \sum_{r=0}^{\infty}\int d^Dl\frac{(l\cdot R)^r\cdot t^r}{D_1D_2\cdots D_n}=\int\frac{d^Dl}{1-t(l\cdot R)}\cdot \frac{1}{D_1D_2\cdots D_n}.
\end{equation}

In \cite{Li:2022cbx}, the author introduced a method that places Feynman parametrization in projective space, deriving a recursion relation for tensor integrals of rank $r$. Building on this recursion relation, we presented in \cite{Hu:2023mgc} a generating function for reducing $n$-gon tensor integrals to $(n-k)$-gon scalar integrals in arbitrary dimensions $D$. Furthermore, we derived a closed-form formula for the reduction coefficients of tensor integrals with arbitrary tensor rank $r$, which does not rely on recursive calculations. 
However, in practical applications, the following two issues remain.
\begin{itemize}
    \item In our formulas, there are irrational terms of the following form:
    \begin{equation}
        \sqrt{(\overline{LL})R^2+(\overline{VL})^2-(\overline{LL})(\overline{VV})}.
    \end{equation}

The meaning of the symbols $(\overline{LL})$, $(\overline{VL})$, and $(\overline{VV})$ will be explained in detail in Section \ref{sec:n&r}. Although these irrational terms vanish after simplification, the final reduction coefficients return to a rational form.

\item In our expression, an auxiliary vector $R$ is introduced. However, in practical applications, the tensor form with Lorentz indices is typically required
\begin{equation}
    \int d^Dl\frac{l^{\mu_1}l^{\mu_2}\cdots l^{\mu_r}}{D_1D_2\cdots D_n}.
\end{equation}
Although the auxiliary vector $R$ can be eliminated using the differential operator $\frac{\partial}{\partial R}$
\begin{equation}
    \frac{1}{r!}\frac{\partial}{\partial R^{\mu_1}}\frac{\partial}{\partial R^{\mu_2}}\cdots \frac{\partial}{\partial R^{\mu_r}}\left(l\cdot R\right)^r=l^{\mu_1}l^{\mu_2}\cdots l^{\mu_r},
\end{equation}

the presence of irrational terms, as mentioned in the first point, makes differentiation inconvenient.

\end{itemize}
Therefore, in this paper, we present a method to directly provide the general expression for the coefficients of one-loop tensor integrals with Lorentz indices reduced to their master integrals. In fact, for terms that consist only of polynomials in the form of $(R^2)^{a_0}(R\cdot q_1)^{a_1}\cdots (R\cdot q_n)^{a_n}$, it is straightforward to translate them into forms with Lorentz indices. For example:
    \begin{align}
        \frac{\partial}{\partial R^{\mu_1}}\frac{\partial}{\partial R^{\mu_2}}\cdots \frac{\partial}{\partial R^{\mu_5}}(R^2)(R\cdot q_1)^2(R\cdot q_2)\to &g^{\mu_1\mu_2}q_1^{\mu_3}q_1^{\mu_4}q_2^{\mu_5}+\sigma(\mu_1,\mu_2,\cdots,\mu_5),\notag\\
        \frac{\partial}{\partial R^{\mu_1}}\frac{\partial}{\partial R^{\mu_2}}\cdots \frac{\partial}{\partial R^{\mu_8}}(R^2)^2(R\cdot q_1)(R\cdot q_2)^3\to &g^{\mu_1\mu_2}g^{\mu_3\mu_4}q_1^{\mu_5}q_2^{\mu_6}q_2^{\mu_7}q_2^{\mu_8}+\sigma(\mu_1,\mu_2,\cdots,\mu_8),
        \label{eq:1-01}
    \end{align}
where $\sigma(\mu_1,\mu_2,\cdots,\mu_r)$ represents a full permutation of the Lorentz indices $(\mu_1,\mu_2,\cdots,\mu_r)$. So far, we have identified that the primary issue lies in the presence of irrational terms in the expressions. As discussed in \cite{Hu:2023mgc}, expressing rational terms using irrational ones is a common practice in mathematics, as exemplified by the well-known closed-form formula for the Fibonacci sequence

\begin{equation}
\begin{aligned}
    F_n=\frac{z_+^n-z_-^n}{z_+-z_-},\ \ \ z_{\pm}=\frac{1\pm \sqrt{5}}{2}.
\end{aligned}
\end{equation}
However, if we perform the variable substitution $A=\frac{z_++z_-}{2}$ and $B=(z_+-z_-)^2$, we obtain:
\begin{equation}
    F_n=\sum_{i=0}^{\lfloor\frac{n-1}{2}\rfloor}\mathbf{C}^{2i+1}_n A^{n-1}B^i,
    \label{eq:1-2}
\end{equation}
where $\mathbf{C}^i_j=\frac{j!}{i!(j-i)!}$ is the binomial coefficient. Equation \eqref{eq:1-2} turns out to consist entirely of rational terms. This implies that by choosing appropriate variables, it is possible to obtain the desired fully rational expression. Once the expression is entirely in a rational form, it becomes straightforward to translate it into a form with Lorentz indices. Therefore, we need to carefully analyze the mathematical structure of our expression and attempt to construct a form composed of a linear combination of terms like \eqref{eq:1-01}. This forms the main idea of this paper, with the detailed approach provided in the main text.

The structure of this paper is as follows: In Section \ref{sec:n&r}, we define the notations required for this work and present the general expression for the reduction coefficients given in \cite{Hu:2023mgc}. In Section \ref{sec:coef with index}, we describe how to efficiently transform these coefficients into a general expression with Lorentz indices based on their analytic structure, along with some simple illustrative examples. Section \ref{sec:compare} provides a comparison between the proposed method and traditional PV reduction, including a discussion of the computational efficiency of the \textbf{Wolfram Mathematica} code based on this work compared to other packages capable of reducing tensors with Lorentz indices. Section \ref{sec:outlook} offers some outlook and further discussion. In the Appendix, we provide numerical verifications to confirm the correctness of the expressions, usage instructions for the supporting code, and some formulas required in the derivations of this paper.

\section{Notations and Review}\label{sec:n&r}

In this section, we will define some notations and provide a comprehensive review of the general expression for the reduction coefficients from our previous work \cite{Hu:2023mgc}. This review is designed to be self-contained, allowing readers to fully understand the necessary background without needing to refer to the previous paper. Readers interested in the derivation of this general formula are encouraged to refer to our previous work.

\subsection{Notations}\label{subsec：notations}

Let us begin by introducing the notations that will be used throughout this paper:

\begin{itemize}
    \item \textbf{Some $n$-dimensional vectors:}
    \[
    \begin{aligned}
        &\boldsymbol{L}=\{1,1,1,\cdots ,1,1\}, \\
        &\boldsymbol{V}=\{R \cdot q_1, R\cdot q_2, R\cdot q_3,\cdots ,R\cdot q_n\}, \\
        &\boldsymbol{H}_b = \{0, \dots, 0, 1, 0, \dots, 0\},
    \end{aligned}
    \]
    where the value $1$ appears in the $b$-th position of $\boldsymbol{H}_b$.

    \item \textbf{The $Q$-matrix:}
    The $Q$-matrix is an $n \times n$ symmetric matrix defined as:
    \[
    Q_{ij} = \frac{M_i^2 + M_j^2 - (q_i - q_j)^2}{2}.
    \]

    \item \textbf{Scalar products with subscripted labels:}
    For two vectors $A$ and $B$ and a label list $\textbf{b}=\{b_1,b_2,...,b_k\}$, we define the following scalar product notations:
    \[
    \begin{aligned}
        (\overline{AB}) &= A \cdot Q^{-1} \cdot B, \\
        (\overline{AB})_{\textbf{b}} &= A_{\widehat{\textbf{b}}} \cdot \left(Q_{\widehat{\textbf{b}} \widehat{\textbf{b}}}\right)^{-1} \cdot B_{\widehat{\textbf{b}}},
    \end{aligned}
    \]
    where $A_{\widehat{\textbf{b}}}$ and $B_{\widehat{\textbf{b}}}$ are derived by removing all components of $A$ and $B$ corresponding to the indices in the label set $\textbf{b} = \{b_1, b_2, \dots, b_k\}$. Similarly, $Q_{\widehat{\textbf{b}} \widehat{\textbf{b}}}$ is obtained by deleting the $b_i$-th rows and columns from $Q$. For instance, with $n = 4$ and $\textbf{b} = \{2, 3\}$, we have:
    \[
    (\overline{VV})_{2,3} = \begin{pmatrix}
        R \cdot q_1 & R \cdot q_4
    \end{pmatrix}
    \begin{pmatrix}
        M_1^2 & \frac{M_1^2 + M_4^2 + (q_1 - q_4)^2}{2} \\
        \frac{M_4^2 + M_1^2 + (q_4 - q_1)^2}{2} & M_4^2
    \end{pmatrix}^{-1}
    \begin{pmatrix}
        R\cdot q_1 \\ R\cdot q_4
    \end{pmatrix}.
    \]

    \item \textbf{Appended labels for analytic expressions:}
    If $\Omega$ is an analytic expression involving terms such as $(\overline{AB})$ or $(\overline{AB})_{\textbf{b}}$, the notation $[\Omega]_{\textbf{a}}$ indicates the expression where each term of the form $(\overline{AB})$ or $(\overline{AB})_{\textbf{b}}$ is appended with subscript $\textbf{a}$. For example:
    \[
    \begin{aligned}
        [(\overline{VL})]_{1,2} = (\overline{VL})_{1,2}, \ \ [(\overline{LL})_2]_{3} = (\overline{LL})_{2,3}.
    \end{aligned}
    \]
    For a more complex expression, such as:
    \begin{equation}
        P = \frac{(\overline{H_1L})(\overline{VV})_3 + (\overline{LL})_2 R^2}{(\overline{VL})_2 (H_3V)_1},\ \ [P]_{4,5} = \frac{(\overline{H_1L})_{4,5} (\overline{VV})_{3,4,5} + (\overline{LL})_{2,4,5} R^2}{(\overline{VL})_{2,4,5} (H_3V)_{1,4,5}}.
    \end{equation}

    \item \textbf{Some frequently used building blocks:}
    \begin{equation}
    \begin{aligned}
        x_{ \pm}=&\frac{2\left((\overline{V L}) \pm \sqrt{(\overline{L L}) R^2+(\overline{V L})^2-(\overline{L L})(\overline{V V})}\right)}{(\overline{L L})},\\
 X^{(b)}=&\left(\left(\overline{H_b L}\right)(\overline{V L})_b-\left(\overline{H_b V}\right)(\overline{L L})_b\right) /(\overline{L L}), \\
Y^{(b)}=&\left(\left(\overline{H_b L}\right) R^2+\left(\overline{H_b V}\right)(\overline{V L})_b-\left(\overline{H_b L}\right)(\overline{V V})_b\right) /(\overline{L L}) .
\end{aligned}
\end{equation}

\item \textbf{Pochhammer symbol}:
\begin{equation}
(x)_n \equiv \frac{\Gamma(x+n)}{\Gamma(x)}=\prod_{i=1}^n(x+(i-1)) .
\end{equation}

\end{itemize}

\subsection{The expression for the reduction coefficients}\label{subsec:coef}
In this section, we directly use the notations introduced above to present the general formula for the reduction coefficients from our previous paper \cite{Hu:2023mgc}. 

\subsubsection{$n$-gon to $n$-gon}
The one-loop tensor Feynman integral that we aim to reduce is given by:

\begin{equation}
I_n^{(r)}\equiv\int d^D l\frac{( R \cdot l)^r}{\prod_{j=1}^n\left(l-q_j\right)^2-M_j^2}.
\end{equation}
At this point, the chosen master scalar integral is
\begin{equation}
I_n\equiv\int d^D l\frac{1}{\prod_{j=1}^n\left(l-q_j\right)^2-M_j^2}.
\end{equation}
The reduction coefficients are\footnote{The factor $\frac{1}{2^r}$ arises because, in \cite{Hu:2023mgc}, the numerator of the reduced integral was $(2R\cdot l)^r$. Besides, note that for some special binomial coefficients, we define:
\begin{equation}
\begin{aligned}
       \mathbf{C}_n^0=1,\ \ \mathbf{C}^{-1}_{-1}=1,\ \ \mathbf{C}^{-1}_m=0, \text{with m}\neq -1.
\end{aligned}
\end{equation}
Under this definition, \eqref{eq:red n to n} in this paper and Eq (3.14) in \cite{Hu:2023mgc} yield the same result.
}:
\begin{equation}
C_{n \rightarrow n}^{(r)}=\frac{1}{2^r}\sum_{j=0}^r \sum_{i=0}^j\left(\frac{\left(\frac{D-n}{2}\right)_i}{(D-n)_i}\left(x_{-}-x_{+}\right)^i x_{+}^{r-i} \mathbf{C}_{j-1}^{i-1}\right) .
\label{eq:red n to n}
\end{equation}

\subsubsection{$n$-gon to $(n-k)$-gon}
Now we chose the master scalar integral as 
\begin{equation}
I_{n, \widehat{\mathbf{I}_k}}=\int d^D l \frac{1}{\prod_{j=1, j \notin \mathbf{I}_k}^n\left(l-q_j\right)^2-M_j^2},
\end{equation}
where the propagators within label set $\mathbf{I}_k=\{ a_1, a_2, a_3, \cdots, a_k \}$ are removed. The reduction coefficients are shown:
\begin{align}
  C^{(r)}_{n\to n;\widehat{\mathbf{I}_k}}=&\frac{1}{2^r}\sum_{m_1,...,m_k=0}^{\sum_{i=1}^km_i+k\leq r}\sum_{l_1+l_2+l_3+\sum_{i=1}^km_i+k=r}\Bigg\{\mathbf{M}_1^{(l_1)}\Bigg(\textbf{M}_2^{(l_2)}(m_1,...,m_k)\notag\\
    \times &\sum_{\{a'_1,...,a'_k\}\in \sigma(\textbf{I}_k),}\sum_{b_1,b_2,...,b_{k-1}=0}^1\textbf{C}_{\{b_1,...,b_{k-1}\}}^{(a'_1,...,a'_k)}(n;m_1,m_2,\cdots,m_k)\notag\\
   \times&\Big([\mathbf{C}_{a'_1}^{(1)}(n-k+1;m_1)]_{a'_2,...,a'_k} \cdot \mathbf{M}^{(l_3)}_{3;\{b_1,b_2,...,b_{k-1}\}}(m_1,...,m_k)\notag\\
   &\ +[\mathbf{C}_{a'_1}^{(2)}(n-k+1;m_1)]_{a'_2,...,a'_k}\cdot \mathbf{M}^{(l_3)}_{4;\{b_1,b_2,...,b_{k-1}\}}(m_1,...,m_k)\Big)\Bigg)\Bigg\}.
   \label{eq:red n to n-k}
\end{align}

Next, we will explain each term in the above expression one by one.
\begin{itemize}
    \item $\mathbf{M}^{(l_1)}_1$ simply depends on $l_1$\footnote{This expression differs slightly from Eq(6.18) of \cite{Hu:2023mgc}, but it can be obtained using the following Taylor series:
    \begin{equation}
\left(1+\mathbf{A} x+\mathbf{B} x^2\right)^c=\sum_{m=0}^{\infty}\left\{\sum_{i=0}^{\left\lfloor\frac{m}{2}\right\rfloor} \frac{(c-m+i+1)_{(m-i)}}{(m-2 i)!\cdot i!} \mathbf{A}^{m-2 i} \mathbf{B}^i\right\} x^m.
\end{equation}
}:
    \begin{equation}
        \mathbf{M}^{(l_1)}_1\equiv \sum_{i=0}^{\lfloor\frac{l_1}{2}\rfloor}\frac{\left(\frac{D-n-2}{2}-l_1+i+1\right)_{(l_1-i)}}{(l_1-2i)!i!}\left(-\left(x_++x_-\right)\right)^{l_1-2i}\left(x_+\cdot x_-\right)^i.
        \label{eq: M1}
    \end{equation}

\item $\mathbf{M}_2^{\left(l_2\right)}\left(m_1, \ldots, m_k\right)$ depends on $l_2$ and parameters $m_1,m_2,\cdots, m_k$:
\begin{equation}
\mathbf{M}_2^{\left(l_2\right)}\left(m_1, \ldots, m_k\right)\equiv \frac{\left(\left[x_{+}\right]_{\mathbf{I}_k}\right)^{l_2}\left(\sum_{i=1}^k m_i+D-n+k-1\right)_{l_2}}{\left(\sum_{i=1}^k m_i+D-n+k-1\right) \cdot l_{2}!}.
\label{eq: M2}
\end{equation}

\item  In addition to relying on  $l_2$ and parameters $m_1,m_2,\cdots, m_k$, the $\mathbf{M}_{3 ;\left\{b_1, b_2, \ldots, b_{k-1}\right\}}^{\left(l_3\right)}\left(m_1, \ldots, m_k\right)$ and $\mathbf{M}_{4 ;\left\{b_1, b_2, \ldots, b_{k-1}\right\}}^{\left(l_3\right)}\left(m_1, \ldots, m_k\right)$ also depend on $\left\{b_1, b_2, \ldots, b_{k-1}\right\}$ which are the first $k-1$ elements of a binary array of length $k$, consisting of 0s and 1s.
    \begin{align}
        &\mathbf{M}_{3 ;\left\{b_1, b_2, \ldots, b_{k-1}\right\}}^{\left(l_3\right)}\left(m_1, \ldots, m_k\right)\notag\\
        \equiv&\sum_{i=0}^{l_3} \frac{\mathbf{C}_{l_3-1}^{i-1}\left(\frac{D-n+k}{2}\right)_i\left(\left[x_{-}-x_{+}\right]_{\mathbf{I}_k}\right)^i\left(\left[x_{+}\right]_{\mathbf{I}_k}\right)^{l_3-i}}{(D-n+k)_i} \cdot \mathcal{S}_i\left(\left\{b_1, \ldots, b_{k-1}\right\}\right),\\
        &\mathbf{M}_{4 ;\left\{b_1, b_2, \ldots, b_{k-1}\right\}}^{\left(l_3\right)}\left(m_1, \ldots, m_k\right)\notag\\
        \equiv&\sum_{i=0}^{l_3} \frac{\mathbf{C}_{l_3-1}^{i-1}\left(\frac{D-n+k+2}{2}\right)_i\left(\left[x_{-}-x_{+}\right]_{\mathbf{I}_k}\right)^i\left(\left[x_{+}\right]_{\mathbf{I}_k}\right)^{l_3-i}(i+1)}{(D-n+k+1)_i} \cdot \mathcal{S}_i\left(\left\{b_1, \ldots, b_{k-1}\right\}\right),
    \end{align}
    \label{eq:M3 M4}
where 
\begin{equation}
\begin{aligned}
\mathcal{S}_i\left(\left\{b_1, \ldots, b_{k-1}\right\}\right) \equiv \begin{cases}
    1,&k=1\\
    \frac{\left(\prod_{\alpha=1}^{k-1}\left(\sum_{\beta=1}^\alpha m_\beta+D-n+k-b_\alpha\right)_i\right)\left(\sum_{\beta=1}^k m_\beta+D-n+k-1\right)_i}{\left(\prod_{\alpha=1}^{k-1}\left(\sum_{\beta=1}^\alpha m_\beta+D-n+k\right)_i\right)\left(\sum_{\beta=1}^k m_\beta+D-n+k\right)_i},&k\geq 2.
\end{cases}
\end{aligned}
\end{equation}
\end{itemize}

The two summation symbols in the first line of Expression \eqref{eq:red n to n-k} are straightforward to understand. The two summation symbols in the second line are explained as follows: 
\begin{itemize}
    \item For the first summation, the summation runs over all permutations of $\mathbf{I}_k$. For example for $k = 3$, $\mathbf{I}_3=\{a_1,a_2,a_3\}$ this summation should runs over
\begin{equation}
\left\{a_1^{\prime}, a_2^{\prime}, a_3^{\prime}\right\}=\Big\{\left\{a_1, a_2, a_3\right\},\left\{a_1, a_3, a_2\right\},\left\{a_2, a_1, a_3\right\},\left\{a_2, a_3, a_1\right\},\left\{a_3, a_1, a_2\right\},\left\{a_3, a_2, a_1\right\}\Big\}.
\end{equation}
Incidentally, it is easy to see from \eqref{eq:red n to n-k} that the entire expression is symmetric with respect to this permutation. Therefore, the summation symbol can be moved to the very beginning of the expression.

\item The second summation symbol in the second line of \eqref{eq:red n to n-k} runs over all possible configurations of the first $k-1$ elements of a binary array of length $k$, consisting of 0s and 1s. For example, for $k=4$, it runs over the following 8 terms:
\begin{equation} 
\left\{b_1, b_2, b_3\right\}=\Big\{\{0,0,0\},\{0,0,1\},\{0,1,0\},\{0,1,1\},\{1,0,0\},\{1,0,1\},\{1,1,0\},\{1,1,1\}\Big\} .
\end{equation}
Besides, for the case $k=1$, this summation no longer exists.
\end{itemize}
Are you ready, readers? Next, we will explain the most complex coefficients, 
$\mathbf{C}^{(a'_1,a'_2,\cdots,a'_k)}_{\{b_1,\cdots,b_{k-1}\}}$ and $[\mathbf{C}^{(1)/(2)}_{a'_1}(n-k+1;m_1)]_{a'_2,\cdots,a'_k}$. During this explanation, we will define a significant number of intermediate coefficients. Fully understanding these coefficients is not an easy task, but we will provide a package at the end of the paper to facilitate their use. 
\begin{itemize}
    \item The coefficient $ \mathbf{C}^{(a'_1,a'_2,\cdots,a'_k)}_{\{b_1,b_2,\cdots,b_{k-1}\}}(n;m_1,m_2,\cdots,m_k) $ is related to three factors: first, a permutation $(a'_1,a'_2,\cdots,a'_k)$ of the label list $\mathbf{I}_k$ (note that this permutation is ordered); second, a binary array $\{b_1,b_2,\cdots,b_{k-1}\}$ of length $k-1$ consisting of 0s and 1s; thrid, parameter $(n;m_1,m_2,\cdots,m_k)$. Before providing the detailed definition of this coefficient, we first define two operators, referred to as the $\mathbf{0}$-operator and the $\mathbf{1}$-operator as follows:
\begin{align}
& (\mathbf{1})_{m^{\prime}}\left(n^{\prime} ; k^{\prime} ; \mathbf{a} ; b ; sum\right)=\frac{\left(D-n^{\prime}-1\right) \cdot\left(\mathbf{P}_1(\mathbf{a} ; b) \cdot \mathbf{N}_{m^{\prime}-1}^{(1)}(\mathbf{a} ; b;n')+\mathbf{P}_2(\mathbf{a} ; b) \cdot \mathbf{N}_{m^{\prime}-2}^{(1)}(\mathbf{a} ; b;n')\right)}{2\left(D-n^{\prime}+k^{\prime}+sum\right)}, \\
& (\mathbf{0})_{m^{\prime}}\left(n^{\prime} ; k^{\prime} ; \mathbf{a} ; b \right)=\mathbf{Q}_0(\mathbf{a} ; b) \cdot \mathbf{N}_{m^{\prime}}^{(0)}(\mathbf{a} ; b;n')+\mathbf{Q}_1(\mathbf{a} ; b) \cdot \mathbf{N}_{m^{\prime}-1}^{(0)}(\mathbf{a} ; b;n') .
\end{align}
\label{eq:OP 01}
The \textbf{0}-operator and \textbf{1}-operator are each determined by six parameters: The integers $m',n',k',sum$, a single index $b$, and a set of indices $\mathbf{a}$. The corresponding coefficients also depend on these corresponding parameters and are defined as follows:
\begin{align}
			\textbf{N}^{(1)}_{m'}(\mathbf{a};b;n')=&\sum_{l=0}^{m'}\sum_{i=0}^{\lfloor\frac{l}{2}\rfloor}\sum_{j=0}^{\lfloor\frac{m'-l}{2}\rfloor}\frac{(-\frac{D-n'}{2}-l+i+1)_{(l-i)}(\frac{D-n'-3}{2}-m'+l+j+1)_{(m'-l-j)}}{(l-2i)!\cdot i!\cdot (m'-l-2j)!\cdot j!}\notag\\
   &\times \textbf{A}_1(\mathbf{a};b)^{l-2i}\textbf{A}_2(\mathbf{a};b)^{i}\textbf{B}_1(\mathbf{a};b)^{m'-l-2j}\textbf{B}_2(\mathbf{a};b)^{j},\\
			\textbf{N}^{(0)}_{m'}(\mathbf{a};b;n')=&\sum_{l=0}^{m'}\sum_{i=0}^{\lfloor\frac{l}{2}\rfloor}\sum_{j=0}^{\lfloor\frac{m'-l}{2}\rfloor}\frac{(-\frac{D-n'}{2}-l+i+1)_{(l-i)}(\frac{D-n'-1}{2}-m'+l+j+1)_{(m'-l-j)}}{(l-2i)!\cdot i!\cdot(m'-l-2j)!\cdot j!}\notag\\
   &\times \textbf{A}_1(\mathbf{a};b)^{l-2i}\textbf{A}_2(\mathbf{a};b)^{i}\textbf{B}_1(\mathbf{a};b)^{m'-l-2j}\textbf{B}_2(\mathbf{a};b)^{j}\label{eq:define N},
\end{align}
and 
\begin{equation}
		\begin{aligned}
			&\textbf{P}_1(\mathbf{a};b)=-\Big([x_{+}+x_{-}]_{b}X^{(b)}+4Y^{(b)}\Big),	\\	
			&\textbf{P}_2(\mathbf{a};b)=2Y^{(b)}([x_{+}+x_{-}]_{b}-2[x_{+}]_{\mathbf{a},b})+X^{(b)}(2[x_{+}\cdot x_{-}]_{b}-[x_{+}+x_{-}]_{b}[x_{+}]_{\mathbf{a},b}),\\
                &\textbf{Q}_0(\mathbf{a};b)=X^{(b)},\\                &\textbf{Q}_1(\mathbf{a};b)=2Y^{(b)}+[x_{+}]_{\mathbf{a},b}\cdot X^{(b)},\\
		\end{aligned}
		\label{eq:define PQ}
	\end{equation}
with
\begin{equation}
		\begin{aligned}
			&\textbf{A}_1(\mathbf{a};b)=-(x_{+}+x_{-})+2[x_{+}]_{\mathbf{a},b},\\
			&\textbf{A}_2(\mathbf{a};b)=[x^2_{+}]_{\mathbf{a},b}+x_{+}\cdot x_{-}-(x_{+}+x_{-})[x_{+}]_{\mathbf{a},b},\\
			&\textbf{B}_1(\mathbf{a};b)=-[x_{+}+x_{-}]_{b}+2[x_{+}]_{\mathbf{a},b},\\
			&\textbf{B}_2(\mathbf{a};b)=[x^2_{+}]_{\mathbf{a},b}+[x_{+}\cdot x_{-}]_{b}-[x_{+}+x_{-}]_{b}\cdot[x_{+}]_{\mathbf{a},b}.
		\end{aligned}
		\label{eq:define AB}
	\end{equation}
After defining these intermediate coefficients, we can finally provide the construction of $ \mathbf{C}^{(a'_1,a'_2,\cdots,a'_k)}_{\{b_1,b_2,\cdots,b_{k-1}\}}(n;m_1,m_2,\cdots,m_k) $ as follows:
		\begin{align}
			&\textbf{C}^{(a'_1,...,a'_k)}_{\{b_1,...,b_{k-1}\}}(n;m_1,m_2,\cdots,m_k)\notag\\
			=&[(\textbf{b}_1)_{m_2}(n-k+2;1;a'_1;a'_2;m_1)]_{a'_3,...,a'_k}\notag\\
			\cdot &[(\textbf{b}_2)_{m_3}(n-k+3;2;\{a'_1,a'_2\};a'_3;m_1+m_2)]_{a'_4,...,a'_k}\notag\\
			&\cdots\notag\\
			\cdot & [(\textbf{b}_i)_{m_{i+1}}(n-k+i+1;i;\{a'_1,...,a'_i\};a'_{i+1};m_1+\cdots+m_i)]_{a'_{i+2},...,a'_k}\notag\\
			&\cdots\notag\\
			\cdot &(\textbf{b}_{k-1})_{m_k}(n;k-1;\{a'_1,...,a'_{k-1}\};a'_k;m_1+...+m_{k-1}).
            \label{eq:define CCC}
		\end{align}
        
     Let us provide an example to illustrate this expression \eqref{eq:define CCC} more concretely. For $k=4$, $(a'_1,a'_2,a'_3,a'_4)=(1,2,3,4)$, $\{b_1,b_2,b_3\}=\{0,1,0\}$ as an example, the $\textbf{C}^{(1,2,3,4)}_{\{0,1,0\}}(n;m_1,m_2,m_3,m_4)$ is\footnote{Remember notation $[\Delta]_{\textbf{b}}$ means we add a subscript $\textbf{b}$ on each term in $\Delta$ with form $(\overline{AB})$ or $(\overline{AB})_{\textbf{a}}$. As we explained in Section \ref{subsec：notations}.}
		\begin{align}
			&\textbf{C}^{(1,2,3,4)}_{\{0,1,0\}}(n;m_1,m_2,\cdots,m_4)\notag\\
            =&[(\textbf{0})_{m_2}(n-2;1;\{1\};2;m_1)]_{3,4}\cdot [(\textbf{1})_{m_3}(n-1;2;\{1,2\};3;m_1+m_2)]_{4}\notag\\
			\cdot &(\textbf{0})_{m_4}(n;3;\{1,2,3\};4;m_1+m_2+m_3).
		\end{align}
    \textbf{An additional point to note} is that when $k=1$, coefficient $\textbf{C}^{(a'_1,...,a'_k)}_{\{b_1,...,b_{k-1}\}}(n;m_1,m_2,\cdots,m_k)$ is directly set to 1.

\item The other two coefficients, $\mathbf{C}_{a'_1}^{(1)/(2)}(n-k+1;m_1)$, depend on the first element $a'_1$ of the permutation $(a'_1,a'_2,\cdots,a'_k)$ and two integers $n-k+1$, $m_1$, which are defined as follows:
		\begin{align}
			\textbf{C}_{a'_1}^{(1)}(n-k+1;m_1)=&\textbf{P}_{a'_1}^{(0)}(n-k+1)\cdot \sum_{i=0}^{\lfloor\frac{m_1}{2}\rfloor}\frac{(\frac{n-k+1-D}{2}-m_1+i+1)_{(m_1-i)}}{(m_1-2i)!\cdot i!}\textbf{A}_{a'_1}^{m_1-2i}\textbf{B}_{a'_1}^i\notag\\
			+&\textbf{P}_{a'_1}^{(1)}\cdot \sum_{i=0}^{\lfloor\frac{m_1-1}{2}\rfloor}\frac{(\frac{n-k+1-D}{2}-m_1+i+2)_{(m_1-1-i)}}{(m_1-1-2i)!\cdot i!}\textbf{A}_{a'_1}^{m_1-1-2i}\textbf{B}_{a'_1}^i,\\
			\textbf{C}_{a'_1}^{(2)}(n-k+1;m_1)=
			&\textbf{Q}_{a'_1}^{(1)}\cdot \sum_{i=0}^{\lfloor\frac{m_1-1}{2}\rfloor}\frac{(\frac{n-k+1-D}{2}-m_1+i+2)_{(m_1-1-i)}}{(m_1-1-2i)!\cdot i!}\textbf{A}_{a'_1}^{m_1-1-2i}\textbf{B}_{a'_1}^i\notag\\
			+&\textbf{Q}_{a'_1}^{(2)}\cdot \sum_{i=0}^{\lfloor\frac{m_1-2}{2}\rfloor}\frac{(\frac{n-k+1-D}{2}-m_1+i+3)_{(m_1-2-i)}}{(m_1-2-2i)!\cdot i!}\textbf{A}_{a'_1}^{m_1-2-2i}\textbf{B}_{a'_1}^i,
             \label{eq:c1 and c2}
		\end{align}
    where
    \begin{equation}
		\begin{aligned}
			&\textbf{A}_{a'_1}=-(x_{+}+x_{-})+2[x_{+}]_{a'_1},\\
			&\textbf{B}_{a'_1}=x_{+}\cdot x_{-}-(x_{+}+x_{-})\cdot [x_{+}]_{a'_1}+[x_{+}^2]_{a'_1},\\
   &\textbf{P}^{(0)}_{a'_1}(n-k+1)=(D-(n-k+1))\cdot X^{(a'_1)},\\&\textbf{P}^{(1)}_{a'_1}=2Y^{(a'_1)}+[x_{+}]_{a'_1}\cdot X^{(a'_1)},\\
			&\textbf{Q}^{(1)}_{a'_1}= -[x_{+}-x_{-}]_{a'_1}\cdot X^{(a'_1)}/2,\\
            &\textbf{Q}^{(2)}_{a'_1}=-\frac{1}{2}\cdot (2Y^{(a'_1)}+[x_{+}]_{a'_1}\cdot X^{(a'_1)})\cdot [x_{+}-x_{-}]_{a'_1}.
		\end{aligned}
  \label{eq: define abpq}
	\end{equation}
    It should be noted that in \eqref{eq:red n to n-k}, these two terms also need to be assigned additional subscripts $a'_2,\cdots, a'_k$.
\end{itemize}
At this point, we have provided the definitions of all the notations and coefficients in the general formula for the reduction coefficients \eqref{eq:red n to n-k}. Although our previous paper focused on the generating functions for the reduction coefficients instead of reduction coefficients themselves, \eqref{eq:red n to n} and \eqref{eq:red n to n-k} were fully derived in that work. In this paper, we will directly study the mathematical analytic structure of the general formulas for these two reduction coefficients, rather than the generating functions explored in the previous work. Nevertheless, we must not overlook the critical role that the idea of generating functions plays in this context.

\section{The general formula for the reduction coefficients with Lorentz indices}\label{sec:coef with index}

In this section, we will present the general formula for the reduction coefficients with Lorentz indices. Compared to Feynman integrals involving auxiliary vector $R$, the target of reduction in this section is:
\begin{equation}
    I^{\mu_1\mu_2\cdots\mu_r}_n=\int d^Dl \frac{l^{\mu_1}l^{\mu_2}l^{\mu_3}\cdots l^{\mu_r}}{\prod_{j=1}^n\left(l-q_j\right)^2-M_j^2}
    \label{eq:GF fey}.
\end{equation}
As mentioned in Section 1, expressions in the form of $(R^2)^{a_0}(R\cdot q_1)^{a_1}(R\cdot q_2)^{a_2}\cdots (R\cdot q_n)^{a_n}$ can be easily translated into forms with Lorentz indices. By examining the definitions in Section \ref{subsec：notations}, it is not difficult to observe that:
\begin{itemize}
    \item $\left(\overline{LL}\right)$ contains no auxiliary vector $R$, Therefore, there is no change during the conversion into the form with Lorentz indices.
    \item  $\left(\overline{VL}\right)$ is linear with respect to the auxiliary vector $R$. Therefore, during the conversion into the form with Lorentz indices, this term will carry a Lorentz index. If we define an $n$-dimensional vector with a Lorentz index as 
    \begin{equation}
        \boldsymbol{V^{\mu}}=\{q_1^{\mu},q_2^{\mu},q_3^{\mu},\cdots,q_n^{\mu}\}.
    \end{equation}
    It is easy to see that:
    \begin{equation}
      \frac{\partial}{\partial R^{\mu}}\left(\overline{VL}\right)\rightarrow \boldsymbol{V}^{\mu}\cdot Q^{-1}\cdot \boldsymbol{L}\equiv \left(\overline{VL}\right)^{\mu}.
    \end{equation}
    \item $\left(\overline{VV}\right)$is quadratic with respect to vector $R$. Then by same means we have 
    \begin{equation}
       \frac{1}{2!}\frac{\partial}{\partial R^{\mu}}\frac{\partial}{\partial R^{\nu}}\left(\overline{VV}\right)\rightarrow  \frac{1}{2!}\cdot \left( \boldsymbol{V}^{\mu}\cdot Q^{-1}\cdot \boldsymbol{V}^{\nu}+\boldsymbol{V}^{\nu}\cdot Q^{-1}\cdot \boldsymbol{V}^{\mu}\right)=\boldsymbol{V}^{\mu}\cdot Q^{-1}\cdot \boldsymbol{V}^{\nu}\equiv \left(\overline{VV}\right)^{\mu\nu}.
    \end{equation}
    \item For $R^2$, it is easy to see
     \begin{equation}
        \frac{1}{2!}\frac{\partial}{\partial R^{\mu}}\frac{\partial}{\partial R^{\nu}}R^2\rightarrow   \frac{1}{2!}(g^{\mu\nu}+g^{\nu\mu})=g^{\mu\nu},
    \end{equation}
    where $g^{\mu\nu}$ is the metric of the $D$-dimensional Minkowski spacetime.
    \item For those with appended labels, we have the similar "translation" rules, such as
    \begin{equation}
      \frac{\partial}{\partial R^{\mu}}[\left(\overline{VL}\right)]_{\mathbf{a}}\rightarrow \boldsymbol{V}^{\mu}_{\widehat{\mathbf{a}}}\cdot (Q_{\widehat{\mathbf{a}}\widehat{\mathbf{a}}})^{-1}\cdot \boldsymbol{L}_{\widehat{\mathbf{a}}}\equiv [\left(\overline{VL}\right)]_{\mathbf{a}}^{\mu}.
    \end{equation}
\end{itemize}
With the above basic "translation" rules, it becomes straightforward to similarly "translate" expressions that are homogeneous in the auxiliary vector $R$ (It is evident that the reduction coefficients must satisfy the condition of being homogeneous with respect to $R$) and generated by addition and multiplication of the above building blocks. For example\footnote{Note that Lorentz indices and exponents should not be confused. In this paper, all Lorentz indices are represented using Greek letters $\mu,\nu$ or $\mu_1,\mu_2,\cdots $.},
\begin{equation}
    \begin{aligned}
      &\frac{1}{3!}\frac{\partial}{\partial R^{\mu_1}}\frac{\partial}{\partial R^{\mu_2}}\frac{\partial}{\partial R^{\mu_3}}\left(\left(\overline{VL}\right)R^2+\left(\overline{VV}\right)\left(\overline{VL}\right)+\left(\overline{H_b V}\right)^3\right)\\
      \rightarrow& \frac{1}{3!}\bigg\{\left(\overline{VL}\right)^{\mu_1}g^{\mu_2\mu_3}+\left(\overline{VV}\right)^{\mu_1\mu_2}\left(\overline{VL}\right)^{\mu_3}+\left(\overline{H_b V}\right)^{\mu_1}\left(\overline{H_b V}\right)^{\mu_2}\left(\overline{H_b V}\right)^{\mu_3}+\sigma(\mu_1,\mu_2,\mu_3)\bigg\},
    \end{aligned}
\end{equation}
where $\sigma(\mu_1,\mu_2,\mu_3)$ is the permutation of Lorentz indices. It should be noted that the reduction coefficients \eqref{eq:red n to n} and \eqref{eq:red n to n-k} are always homogeneous of degree 
$r$ in $R$. Now, let the game begins.

\subsection{$n$-gon to $n$-gon}\label{subsec:n2n}
We begin with the reduction coefficients of $n$-gon to $n$-gon as a warm up. Let us repeat the expression 
\begin{equation}
C_{n \rightarrow n}^{(r)}=\frac{1}{2^r}\sum_{j=0}^r \sum_{i=0}^j\left(\frac{\left(\frac{D-n}{2}\right)_i}{(D-n)_i}\left(x_{-}-x_{+}\right)^i x_{+}^{r-i} \mathbf{C}_{j-1}^{i-1}\right),
\label{eq:red n to n 01}
\end{equation}
where 
\begin{equation}
x_{ \pm}=\frac{2\left((\overline{V L}) \pm \sqrt{(\overline{L L}) R^2+(\overline{V L})^2-(\overline{L L})(\overline{V V})}\right)}{(\overline{L L})},
\end{equation}
which is linear to $R$. Similar to the approach we used for the Fibonacci sequence in Section 1, if we perform the following variable substitution:
\begin{equation}
\begin{aligned}
     \mathbf{\Delta}=&x_+-x_-=\frac{4\sqrt{(\overline{L L}) R^2+(\overline{V L})^2-(\overline{L L})(\overline{V V})}}{(\overline{L L})},\\
     \mathbf{K}=&x_++x_-=\frac{4\left(\overline{VL}\right)}{\left(\overline{LL}\right)}.
\end{aligned}
\label{eq:defin K delta}
\end{equation}
Then, it is evident that we can transform \eqref{eq:red n to n 01} into a polynomial of the following form:
\begin{equation}
    C^{(r)}_{n\to n}=\sum_{i+j=r}\Omega(i,j) \mathbf{K}^{i} \mathbf{\Delta}^{j}.
\end{equation}
The coefficient $\Omega(i,j)$ is determined only by the parameters $i,j$, the spacetime $D$ and the number of propagators $n$, independent of any Mandelstam variables $q_{i'}\cdot q_{j'}$ or masses $M_{i'}$. Next, we can rewrite the above expression as two terms by separating it into odd and even powers of the variable $\mathbf{\Delta}$ as
\begin{equation}
    C^{(r)}_{n\to n}=\Big[\sum_{i+2j=r}\Omega(i,2j)\mathbf{K}^i\mathbf{\Delta}^{2j}\Big]+\Big[\sum_{i+2j=r-1}\Omega(i,2j+1)\mathbf{K}^i\mathbf{\Delta}^{2j}\Big]\cdot \mathbf{\Delta}.
\end{equation}
Here comes an interesting point: we know that the left-hand side of the above equation must be a rational function, and the terms inside the two square brackets on the right-hand side are also rational functions. However, $\mathbf{\Delta}$ is inherently an irrational term. This means nothing but that the terms inside the second square bracket on the right-hand side of the equation must vanish! In other word, although so far we do not know the exact expression of $\Omega(i,j)$, we are certain that\footnote{Readers can also rigorously verify this.}
\begin{equation}
    \Big[\sum_{i+2j=r-1}\Omega(i,2j+1)\mathbf{K}^i\mathbf{\Delta}^{2j}\Big]=0.
\end{equation}
Therefore, after after selecting a new variable $\mathbf{T}=\mathbf{\Delta}^2$ and performing the specific calculations, we obtain\footnote{The detailed derivation can be found in the Appendix \ref{sec:solve n2n}.}:
\begin{equation}
    C^{(r)}_{n\to n}=\sum_{i=0}^{\lfloor\frac{r}{2}\rfloor}\Omega(r-2i,i)\mathbf{K}^{r-2i}\mathbf{T}^i\equiv\sum_{i=0}^{\lfloor\frac{r}{2}\rfloor}\frac{(2i-1)!!\mathbf{C}^{2i}_r }{2^{2r+i}\left(\frac{D-n+1}{2}\right)_{i}}\mathbf{K}^{r-2i}\mathbf{T}^i,
    \label{eq:n to n 02}
\end{equation}
where $(2i-1)!!$ is the double factorial.
Next, we "translate" the above expression with the auxiliary vector $R$ into the form with Lorentz indices, which becomes straightforward. This is because,  for $\mathbf{K}$ and $\mathbf{T}$, one can apply the following "translation" rules:	
\begin{equation}
    \begin{aligned}
        \mathbf{K}\to &\mathbf{K}^{\mu}\equiv\frac{4\left(\overline{VL}\right)^{\mu}}{\left(\overline{LL}\right)},\\
        \mathbf{T}\to & \mathbf{T}^{\mu\nu}\equiv\frac{16\left(\left(\overline{LL}\right)g^{\mu\nu}+\left(\overline{VL}\right)^{\mu}\left(\overline{VL}\right)^{\nu}-\left(\overline{LL}\right)\left(\overline{VV}\right)^{\mu\nu}\right)}{\left(\overline{LL}\right)^2}.
    \end{aligned}
\label{eq:trans K T}
\end{equation}
The tensor structure defined in \eqref{eq:trans K T} forms the building blocks of the reduction coefficients $C^{\mu_1\mu_2\cdots \mu_r}_{n\to n}$ with Lorentz indices. The remaining task is simply to assign these $r$ Lorentz indices into the structure in order, and finally, sum over all permutations of these Lorentz indices:
\begin{equation}
\begin{aligned}
     &C^{\mu_1\mu_2\cdots \mu_r}_{n\to n}\\
     =&\frac{1}{r!}\Bigg\{\sum_{i=0}^{\lfloor\frac{r}{2}\rfloor}\frac{(2i-1)!!\mathbf{C}^{2i}_r }{2^{2r+i}\left(\frac{D-n+1}{2}\right)_{i}}\mathbf{K}^{\mu_1}\mathbf{K}^{\mu_2}\cdots \mathbf{K}^{\mu_{r-2i}}\mathbf{T}^{\mu_{r-2i+1},\mu_{r-2i+2}}\mathbf{T}^{\mu_{r-2i+3},\mu_{r-2i+4}}\cdots\mathbf{T}^{\mu_{r-1},\mu_{r}}\\
     +&\sigma(\mu_1,\mu_2,\cdots,\mu_r)\Bigg\}.
\end{aligned}
\label{eq:n to n with LI}
\end{equation}
It is worth noting that the tensor structure $\mathbf{T}^{\mu\nu}$ is symmetric with respect to its two Lorentz indices. During the permutation summation, many identical terms will inevitably appear. For example, when assigning four Lorentz indices to $\mathbf{T}^2$, in the permutation $(\mu_1, \mu_2, \mu_3, \mu_4)$, the term $\mathbf{T}^{\mu_1\mu_2} \mathbf{T}^{\mu_3\mu_4}$ will occur. However, this term will also appear in the permutation $(\mu_3, \mu_4, \mu_1, \mu_2)$. Therefore, although writing the expression as in \eqref{eq:n to n with LI} is more concise, in practical implementation, one can leverage these symmetries to combine identical terms in advance, thereby reducing the computational cost in the program.

\subsubsection{Analytic Examples}	
The expression \eqref{eq:n to n with LI} remains somewhat abstract. To demonstrate the computation process more concretely, let us consider a massless bubble as an example:
\begin{equation}
    \int d^Dl \frac{l^{\mu_1}l^{\mu_2}l^{\mu_3}\cdots l^{\mu_r}}{l^2(l-p)^2}.
\end{equation}
In this case, the $Q$-matrix is a $2\times 2$ symmetric matrix:
\begin{equation}
    Q=\begin{pmatrix}
0 & -\frac{p^2}{2} \\
-\frac{p^2}{2} & 0
\end{pmatrix},\  Q^{-1}=\begin{pmatrix}
0 & -\frac{2}{p^2} \\
-\frac{2}{p^2} & 0
\end{pmatrix}.
\end{equation}
Then we have
\begin{equation}
\begin{aligned}
    \left(\overline{LL}\right)=&\begin{pmatrix}
1 & 1
\end{pmatrix}\begin{pmatrix}
0 & -\frac{p^2}{2} \\
-\frac{p^2}{2} & 0
\end{pmatrix}\begin{pmatrix}
    1 \\ 1
\end{pmatrix}=-\frac{4}{p^2},\\
\left(\overline{VL}\right)^{\mu}=&\begin{pmatrix}
0 & p^{\mu}
\end{pmatrix}\begin{pmatrix}
0 & -\frac{p^2}{2} \\
-\frac{p^2}{2} & 0
\end{pmatrix}\begin{pmatrix}
    1 \\ 1
\end{pmatrix}=-\frac{2p^{\mu}}{p^2},\\
\left(\overline{VV}\right)^{\mu\nu}=&\begin{pmatrix}
0 & p^{\mu}
\end{pmatrix}\begin{pmatrix}
0 & -\frac{p^2}{2} \\
-\frac{p^2}{2} & 0
\end{pmatrix}\begin{pmatrix}
    0 \\ p^{\nu}
\end{pmatrix}=0.
\end{aligned}
\end{equation}
Then two tensor building blocks are
\begin{equation}
\begin{aligned}
    \mathbf{K}^{\mu}=2p^{\mu},\ \ \ \mathbf{T}^{\mu\nu}=4 p^{\mu}p^{\nu}-4p^2 g^{\mu\nu}.
\end{aligned}
\label{eq:KT bub}
\end{equation}
Then the results are:
\begin{itemize}

    \item For $r=1$, by \eqref{eq:n to n 02} we have  $C^{(1)}_{2\to 2}=\mathbf{K}/4$. Thus, $C^{\mu_1}_{2\to 2}=p^{\mu_1}/2$.
    
    \item For $r=2$, by \eqref{eq:n to n 02} we have    
    
        \begin{equation}
C^{2}_{2\to 2}=\frac{\mathbf{K}^2}{16}+\frac{\mathbf{T}}{16(D-1)},
\end{equation}
which carries two Lorentz indices. Assigning the two Lorentz indices into the corresponding tensor structure, with \eqref{eq:n to n with LI} and \eqref{eq:KT bub}, we obtain:
\begin{align}
    C^{\mu_1\mu_2}_{2\to 2}=&\frac{1}{2!}\Bigg\{\Bigg(\frac{\mathbf{K}^{\mu_1}\mathbf{K}^{\mu_2}}{16}+\frac{\mathbf{T}^{\mu_1\mu_2}}{16(D-1)}\Bigg)+\Bigg(\frac{\mathbf{K}^{\mu_2}\mathbf{K}^{\mu_1}}{16}+\frac{\mathbf{T}^{\mu_2\mu_1}}{16(D-1)}\Bigg)\Bigg\}\notag\\
    =&\frac{\mathbf{K}^{\mu_1}\mathbf{K}^{\mu_2}}{16}+\frac{\mathbf{T}^{\mu_1\mu_2}}{16(D-1)}\notag\\
    =&\frac{(2 p^{\mu_1})(2 p^{\mu_2})}{16}+\frac{4p^{\mu_1}p^{\mu_2}-4p^2 g^{\mu_1\mu_2}}{16(D-1)}\notag\\
    =&-\frac{1}{4(D-1)}p^2 g^{\mu_1\mu_2}+\frac{D}{4(D-1)}p^{\mu_1}p^{\mu_2}.
\end{align}

  \item For $r=3$, by \eqref{eq:n to n 02} we have
\begin{equation}
C^{3}_{2\to 2}=\frac{\mathbf{K}^3}{64}+\frac{3 \mathbf{K T}}{64(D-1)},
\end{equation}
which carries three Lorentz indices. Assigning the three Lorentz indices into the corresponding tensor structure, with \eqref{eq:n to n with LI} and \eqref{eq:KT bub}, we obtain:
\begin{equation}
\begin{aligned}
   C^{\mu_1\mu_2\mu_3}_{2\to 2}=&\frac{1}{3!}\Bigg\{\Bigg(\frac{\mathbf{K}^{\mu_1}\mathbf{K}^{\mu_2}\mathbf{K}^{\mu_3}}{64}+\frac{3 \mathbf{K}^{\mu_1}\mathbf{T}^{\mu_2\mu_3}}{64(D-1)}\Bigg)+\sigma(\mu_1,\mu_2,\mu_3)\Bigg\}\\
   =&\frac{1}{3!}\Bigg\{\Bigg(\frac{(2p^{\mu_1})(2p^{\mu_2})(2p^{\mu_3})}{64}+\frac{3 (2p^{\mu_1})(4p^{\mu_2}p^{\mu_3}-4p^2 g^{\mu_2\mu_3})}{64(D-1)}\Bigg)+\sigma(\mu_1,\mu_2,\mu_3)\Bigg\}\\
    =&-\frac{1}{8(D-1)}\left(p^2p^{\mu_1}g^{\mu_2\mu_3}+p^2p^{\mu_2}g^{\mu_1\mu_3}+p^2p^{\mu_3}g^{\mu_1\mu_2}\right)+\frac{D+2}{8(D-1)}p^{\mu_1}p^{\mu_2}p^{\mu_3}.
\end{aligned}
\end{equation}
\item For $r=4$, by \eqref{eq:n to n 02} we have
\begin{equation}
C^{4}_{2\to 2}=\frac{\mathbf{K}^4}{256}+\frac{3 \mathbf{K^2 T}}{128(D-1)}+\frac{3 \mathbf{T^2}}{256(D^2-1)},
\end{equation}
which carries four Lorentz indices. Assigning the four Lorentz indices into the corresponding tensor structure, with \eqref{eq:n to n with LI} and \eqref{eq:KT bub}, we obtain:
\begin{align}
   C^{\mu_1\mu_2\mu_3\mu_4}_{2\to 2}=&\frac{1}{4!}\Bigg\{\Bigg(\frac{\mathbf{K}^{\mu_1}\mathbf{K}^{\mu_2}\mathbf{K}^{\mu_3}\mathbf{K}^{\mu_4}}{256}+\frac{3 \mathbf{K}^{\mu_1}\mathbf{K}^{\mu_2}\mathbf{T}^{\mu_3\mu_4}}{128(D-1)}+\frac{3 \mathbf{T}^{\mu_1\mu_2}\mathbf{T}^{\mu_3\mu_4}}{256(D^2-1)}\Bigg)+\sigma(\mu_1,\mu_2,\mu_3,\mu_4)\Bigg\}\notag\\
    =&\frac{1}{4!}\Bigg\{\Bigg(\frac{(2p^{\mu_1})(2p^{\mu_2})(2p^{\mu_3})(2p^{\mu_4})}{256}+\frac{3 (2p^{\mu_1})(2p^{\mu_2})(4p^{\mu_3}p^{\mu_4}-4p^2 g^{\mu_3\mu_4})}{128(D-1)}\notag\\
   & \ \ \ \ \ \ \ \ +\frac{3 (4p^{\mu_1}p^{\mu_2}-4p^2 g^{\mu_1\mu_2})(4p^{\mu_3}p^{\mu_4}-4p^2 g^{\mu_3\mu_4})}{256(D^2-1)}\Bigg)+\sigma(\mu_1,\mu_2,\mu_3,\mu_4)\Bigg\}\notag\\
  =&\frac{1}{16(D^2-1)}\left((p^2)^2g^{\mu_1\mu_2}g^{\mu_3\mu_4}+(p^2)^2g^{\mu_1\mu_3}g^{\mu_2\mu_4}+(p^2)^2g^{\mu_1\mu_4}g^{\mu_2\mu_3}\right)\notag\\
  &-\frac{D+2}{16(D^2-1)}\big(p^2p^{\mu_1}p^{\mu_2}g^{\mu_3\mu_4}+p^2p^{\mu_1}p^{\mu_3}g^{\mu_2\mu_4}+p^2p^{\mu_1}p^{\mu_4}g^{\mu_2\mu_3}\notag\\
&\ \ \ \ \ \ \ \ \ \ \ \ \ \ \ \ \ \ \ \ +p^2p^{\mu_2}p^{\mu_3}g^{\mu_1\mu_4}+p^2p^{\mu_2}p^{\mu_4}g^{\mu_1\mu_3}+p^2p^{\mu_3}p^{\mu_4}g^{\mu_1\mu_2}\big)\notag\\
  &+\frac{D^2+6D+8}{16(D^2-1)}p^{\mu_1}p^{\mu_2}p^{\mu_3}p^{\mu_4}.
\end{align}

\end{itemize}
The above results have all been compared with package \texttt{AmpRed}\cite{Chen:2024xwt}, and the outcomes match perfectly.

\subsection{$n$-gon to $(n-k)$-gon}\label{subsec:n2n-k}
After this warm-up, we move to the more complex case of reducing the $n$-gon to an $(n-k)$-gon. Without loss of generality, we consider the master integral obtained by removing the first $k$ propagators, i.e, $\mathbf{I}_k=\{1,2,\cdots,k\}$:
\begin{equation}
    \int d^Dl \frac{1}{\prod_{j=k+1}^n (l-q_j)^2-M^2_j}.
\end{equation}
To handle reduction coefficient \eqref{eq:red n to n-k}, we also use a similar "variable substitution method." From the above section, we observe that two building blocks are ultimately obtained. Therefore, we now aim to determine which building blocks are required to construct the $n$-gon to $(n-k)$-gon reduction coefficient. This requires a careful analysis of the structure of the expression and the composition of the intermediate coefficients. First, Let us focus on the irrational terms. In fact, as mentioned in the discussion section of \cite{Hu:2023mgc}, the only irrational term that appears is:
\begin{equation}
   [\mathbf{\Delta}]_{\mathbf{I}_k}=[x_+-x_-]_{\mathbf{I}_k}=\Bigg[\frac{4\sqrt{\left(\overline{LL}\right)R^2+\left(\overline{VL}\right)^2-\left(\overline{LL}\right)\left(\overline{VV}\right)}}{\left(\overline{LL}\right)}\Bigg]_{\mathbf{I}_k}.
\end{equation}
Thus, similar to the discussion in the previous section, we can expand the reduction coefficient $C^{(r)}_{n\to n-k}$ according to whether  $[\mathbf{\Delta}]_{\mathbf{I}_k}$ appears with odd or even powers, yielding the following form:
\begin{equation}
    C^{(r)}_{n\to n-k}=\Bigg[ \sum\omega(2i)\left([\mathbf{\Delta}]_{\mathbf{I}_k}\right)^{2i}\Bigg]+\Bigg[ \sum\omega(2i+1)\left([\mathbf{\Delta}]_{\mathbf{I}_k}\right)^{2i}\Bigg][\mathbf{\Delta}]_{\mathbf{I}_k},
\end{equation}
where those $\omega(j)$ are purely rational. It can be observed that the reduction coefficient $C^{(r)}_{n\to n-k}$ must be rational, and the two terms inside the square brackets on the right-hand side of the equation are also rational. This means that even if we do not know the exact form of $\omega(j)$, we can still conclude that the terms inside the second square bracket in the above equation must vanish:
\begin{equation}
    \Bigg[ \sum\omega(2i+1)\left([\mathbf{\Delta}]_{\mathbf{I}_k}\right)^{2i}\Bigg]=0.
\end{equation}
Thus, once changing the variable as $[\mathbf{T}]_{\mathbf{I}_k}=\left([\mathbf{\Delta}]_{\mathbf{I}_k}\right)^{2}$, we have
\begin{equation}
    C^{(r)}_{n\to n-k}=\sum_i\Omega(i)[\mathbf{T}]_{\mathbf{I}_k}^{i},
\end{equation}
which is a purely rational form. Next, to determine $\Omega(i)$, we will analyze each term based on the definitions of the intermediate coefficients to identify the rest building blocks. First, from Equation \eqref{eq:red n to n-k}, we observe that the entire formula is symmetric with respect to permutations of $\mathbf{I}_k$. Therefore, without loss of generality, we can focus on one specific permutation. Let us take $\mathbf{I}_3 = \{1, 2, 3\}$ and $(a'_1, a'_2,a'_3) = (1, 2, 3)$ as the example objects of analysis.

\begin{itemize}
    \item $\mathbf{M}^{(l_1)}_1$, by \eqref{eq: M1}, it is easy to see it's built by two building blocks:
    \begin{equation}
        \begin{aligned}
            \mathbf{K}\equiv&x_++x_-=\frac{4\left(\overline{VL}\right)}{\left(\overline{LL}\right)},\\
            \mathbf{M}\equiv&x_+\cdot x_-=\frac{4\left((\overline{VV})-R^2\right)}{\left(\overline{LL}\right)}.
        \end{aligned}
    \end{equation}
    Here, we define the product of $x_+$ and $x_-$ as $\mathbf{M}$, and this notation will be used throughout the rest of the paper.

    \item $\mathbf{M}_2^{(l_2)}(m_1,m_2,m_3)$, $\mathbf{M}^{(l_3)}_{3;\{b_1,b_2\}}(m_1,m_2,m_3)$ and $\mathbf{M}^{(l_3)}_{4;\{b_1,b_2\}}(m_1,m_2,m_3)$. From \eqref{eq: M2} and \eqref{eq:M3 M4}, we can see they depends on $[x_+]_{123}$ and $[x_+-x_-]_{123}$. By
    \begin{equation}
        [x_+]_{123}=\frac{1}{2}[x_++x_-]_{123}+\frac{1}{2}[x_+-x_-]_{123}\equiv \frac{[\mathbf{K}]_{123}+[\mathbf{\Delta}]_{123}}{2},
    \end{equation}
     we see their building blocks are $[\mathbf{K}]_{123}$ and $[\mathbf{\Delta}]_{123}$.

     \item $\mathbf{C}^{(1,2,3)}_{\{b_1,b_2\}}(n;m_1,m_2,m_3)$. First, according to \eqref{eq:define CCC}, this is composed of the product of two $\mathbf{0}$ or $\mathbf{1}$ operators. From \eqref{eq:OP 01} and \eqref{eq:define N}, it can be seen that these operators are polynomials composed of the intermediate coefficients $\mathbf{A}$, $\mathbf{B}$, $\mathbf{P}$, and $\mathbf{Q}$. Furthermore, from \eqref{eq:define PQ}, \eqref{eq:define AB}, it follows that the building blocks of these intermediate coefficients are:
     \begin{equation}
         X^{(b)},\ Y^{(b)},\ [\mathbf{K}]_b,\ [\mathbf{M}]_b, \ [\mathbf{K}]_{\mathbf{a},b},\ [\mathbf{\Delta}]_{\mathbf{a},b}.
     \end{equation}
    Moreover, we can observe that whether it is the $\mathbf{0}$-operator or the $\mathbf{1}$-operator, they have the form $f_1 \cdot X^{(b)} + f_2 \cdot Y^{(b)}$, meaning they are linear with respect to $X^{(b)}$ and $Y^{(b)}$. In our example: 
     \begin{equation}
         \mathbf{C}^{(1,2,3)}_{\{b_1,b_2\}}(n;m_1,m_2,m_3)=[(\mathbf{b_1})_{m_2}(n-1;1;\{1\};2;m_1)]_3\cdot (\mathbf{b_2})_{m_3}(n;2;\{1,2\};3;m_1+m_2).
     \end{equation}
      Thus, all the building blocks included here are:
      \begin{equation}
          X^{(3)},\ [X^{(2)}]_3,\ Y^{(3)},\ [Y^{(2)}]_3,\  [\mathbf{K}]_3,\ [\mathbf{K}]_{23},\ [\mathbf{K}]_{123},\ [\mathbf{\Delta}]_{123},\ [\mathbf{M}]_3, \ 
          [\mathbf{M}]_{23}.
      \end{equation}

  \item $\mathbf{C}_1^{(1)/(2)}(n-2;m_1)$. According to \eqref{eq:c1 and c2} and \eqref{eq: define abpq}, through a similar analysis as before, we can derive the following two properties:  
  
(1) Similar to the 0 and 1 operators, $\mathbf{C}_1^{(1)/(2)}(n-2;m_1)$ has the form $f_1 \cdot X^{(a'_1)} + f_2 \cdot Y^{(a'_1)}$.  
(2) Note that this coefficient also needs to be assigned the subscripts $a'_2, \cdots, a'_k$. Therefore, the building blocks of them are:
\begin{equation}
    [X^{(1)}]_{23},\ [Y^{(1)}]_{23},\ [\mathbf{K}]_{23},\ [\mathbf{K}]_{123}, ,\ [\mathbf{\Delta}]_{123},\ [\mathbf{M}]_{23}.
\end{equation}   
\end{itemize}
In summary, the building blocks that appear in this example are:
\begin{equation}
  X^{(3)},\ [X^{(2)}]_3,\ [X^{(1)}]_{23},\ 
  Y^{(3)},\ [Y^{(2)}]_3,\ [Y^{(1)}]_{23},\ 
  \mathbf{K},\ [\mathbf{K}]_3,\ [\mathbf{K}]_{23},\ [\mathbf{K}]_{123},\ 
  \mathbf{M},\ [\mathbf{M}]_{3},\ [\mathbf{M}]_{23},\ [\mathbf{\Delta}]_{123}.
\end{equation}
Through a similar analysis, it can be easily seen that for the general case, when the permutation is $(a'_1, a'_2, \cdots, a'_k)$, and considering that the final expression for the reduction coefficients must be rational, all the \textbf{rational} building blocks are in five types:
\begin{equation}
    \begin{aligned}
        &X^{(a'_k)},\ [X^{(a'_{k-1})}]_{a'_k},\ [X^{(a'_{k-2})}]_{a'_{k-1}a'_{k}},\ [X^{(a'_{k-3})}]_{a'_{k-2},a'_{k-1}a'_k},\ \cdots, [X^{(a'_{1})}]_{a'_2a'_3\cdots a'_k},\\
        &Y^{(a'_k)},\ [Y^{(a'_{k-1})}]_{a'_k},\ [Y^{(a'_{k-2})}]_{a'_{k-1}a'_{k}},\ [Y^{(a'_{k-3})}]_{a'_{k-2},a'_{k-1}a'_k},\ \cdots, [Y^{(a'_{1})}]_{a'_2a'_3\cdots a'_k},\\
        &\mathbf{K},\ [\mathbf{K}]_{a'_k},\ [\mathbf{K}]_{a'_{k-1}a'_k},\ [\mathbf{K}]_{a'_{k-2}a'_{k-1}a'_k},\ \cdots,\ [\mathbf{K}]_{a'_1a'_2\cdots a'_k},\\
        &\mathbf{M},\ [\mathbf{M}]_{a'_k},\ [\mathbf{M}]_{a'_{k-1}a'_k},\ [\mathbf{M}]_{a'_{k-2}a'_{k-1}a'_k},\ \cdots,\ [\mathbf{M}]_{a'_2\cdots a'_k},\\
        &[\mathbf{T}]_{\mathbf{I}_k}.
    \end{aligned}
    \label{eq:building blocks}
\end{equation}
From the above discussion, it can be seen that the only irrational term is indeed $[\mathbf{\Delta}]_{\mathbf{I}_k}$. Next, regarding these building blocks, the following points need further clarification:
\begin{itemize}
    \item (1) From \eqref{eq:building blocks}, it can be observed that we have listed a total of five types, categorized by rows: the $X$-type (first row), $Y$-type (second row), $K$-type (third row), $M$-type (fourth row. Note that the $M$-type contains one fewer term $[\mathbf{M}]_{\mathbf{I}_k}$ compared to the $K$-type.), and the unique $T$-type. Among these, the $Y$-type, $M$-type, and $T$-type are quadratic in the auxiliary vector $R$, while the $K$-type and $X$-type are linear in $R$. This implies that when "translating" into forms with Lorentz indices, $X$ and $K$ carry only one Lorentz index, whereas $Y$, $M$, and $T$ carry two Lorentz indices. Furthermore, all Mandelstam variables $q_i\cdot q_j$ and masses $M_i$ in the reduction coefficients are encapsulated within these building blocks. The remaining parts depend only on the number of propagators $n$, the tensor rank $r$, the number of propagators in the master integral $n-k$, and the spacetime dimension $D$.

    \item (2) The reduction coefficients $C^{(r)}_{n\to n-k}$ are  homogeneous \textbf{polynomials} in these building blocks. In other words, these building blocks will not appear in the denominator. Thus, when taking the derivative with respect to $R$, $\frac{\partial}{\partial R}$, it suffices to assign the corresponding $r$ Lorentz indices to these building blocks in order, and finally perform a summation over all permutations of the Lorentz indices.

    \item (3) From the above discussion, it can be observed that the $\mathbf{0}$ and $\mathbf{1}$ operators, as well as $\mathbf{C}^{(1)/(2)}_{a'_1}(n-k_1;m_1)$, all have the form $(**)\cdot X + (**)\cdot Y$ (Here, $(**)$ represents some polynomials composed of $K$-, $M$-, and $T$-type building blocks, independent of $X$- and $Y$-type blocks.). Therefore, under the permutation $(a'_1, a'_2, \cdots, a'_k)$, the entire reduction coefficient $C^{(r)}_{n\to n-k}$ takes the form:
    \begin{align}
        &\sum_{(a'_1,\cdots,a'_k)\in \sigma(\mathbf{I}_k)}(**)\prod_{i=1}^k[(**)X^{(a'_i)}+(**)Y^{(a'_i)}]_{a'_{i+1}a'_{i+2}\cdots a'_{k}}\notag\\
        =&\sum_{(a'_1,\cdots,a'_k)\in \sigma(\mathbf{I}_k)}(**)\prod_{i=1}^k[(X\ \text{or}\ Y)^{(a'_i)}]_{a'_{i+1}a'_{i+2}\cdots a'_{k}}.
    \end{align}
    In other words, there are exactly $k$ $X$- or $Y$-type blocks, and for a specific $i$ in the above equation, either $[X^{(a'_i)}]_{a'_{i+1}a'_{i+2}\cdots a'_{k}}$  appears once or $[Y^{(a'_i)}]_{a'_{i+1}a'_{i+2}\cdots a'_{k}}$ appears once. Since $X$ occupies one Lorentz index and $Y$ occupies two, and there must be exactly $k$ $X$- and $Y$-type blocks in total, it can be observed that when the tensor rank $r$ is equal to or not much larger than $k$, the entire reduction coefficient becomes very concise. This is because the $X$- and $Y$-type blocks alone will consume most of the Lorentz indices, leaving very few, if any, for the $K$-, $M$-, and $T$-type blocks. As a result, the possible forms of $K$-, $M$-, and $T$-type blocks are highly constrained. For example, for the case $r=k=3$ and permutation $(a'_1,a'_2,a'_3)=(1,2,3)$, the only term is $X^{(3)}[X^{(2)}]_3[X^{(1)}]_{23}$.
\end{itemize}
Combining the above three points and noting that the entire reduction coefficient is completely symmetric with respect to permutations of the index set $\mathbf{I}_k$, the general formula for the reduction coefficients is\footnote{Note that $\alpha_j$, $\beta_j$, $\gamma$, and $\theta_j$ here denote exponents, not Lorentz indices.
}:
    \begin{align}
        &C^{(r)}_{n\to n-k}\notag\\ 
        =&\sum_{\alpha_0+\cdots+\alpha_k+2(\beta_1+\cdots+\beta_k)+2\gamma+2k-(\theta_1+\cdots+\theta_k)=r}\Omega^{{r}}_{n\to n-k}(\alpha_0,\alpha_1,\cdots,\alpha_k,\beta_1,\cdots,\beta_k,\gamma,\theta_1,\cdots,\theta_k)\notag\\
        \times &\sum_{(a'_1,a'_2,\cdots,a'_k)\in \sigma(\mathbf{I}_k)}\Bigg\{\mathbf{K}^{\alpha_k}[\mathbf{K}]_{a'_k}^{\alpha_{k-1}}[\mathbf{K}]_{a'_{k-1}a'_k}^{\alpha_{k-2}} [\mathbf{K}]_{a'_{k-2}a'_{k-1}a'_k}^{\alpha_{k-3}}\cdots [\mathbf{K}]_{a'_1a'_2\cdots a'_k}^{\alpha_0}\notag\\
        &\ \ \ \ \  \ \ \ \ \ \ \ \ \ \ \ \ \ \ \ \ \ \times \mathbf{M}^{\beta_k}[\mathbf{M}]_{a'_k}^{\beta_{k-1}}[\mathbf{M}]_{a'_{k-1}a'_k}^{\beta_{k-2}} [\mathbf{M}]_{a'_{k-2}a'_{k-1}a'_k}^{\beta_{k-3}}\cdots [\mathbf{M}]_{a'_2\cdots a'_k}^{\beta_1}[\mathbf{T}]_{\mathbf{I}_k}^{\gamma}\notag\\
       &\ \ \ \ \ \ \ \ \ \ \ \ \ \ \ \ \ \ \ \ \ \ \times (X^{(a'_k)})^{\theta_k}[X^{(a'_{k-1})}]_{a'_k}^{\theta_{k-1}}[X^{(a'_{k-2})}]_{a'_{k-1}a'_{k}}^{\theta_{k-2}}[X^{(a'_{k-3})}]_{a'_{k-2},a'_{k-1}a'_k}^{\theta_{k-3}}\cdots [X^{(a'_{1})}]_{a'_2a'_3\cdots a'_k}^{\theta_1}\notag\\
       &\ \ \ \ \ \ \ \ \ \ \ \ \ \ \ \ \ \ \ \ \ \ \times (Y^{(a'_k)})^{1-\theta_k}[Y^{(a'_{k-1})}]_{a'_k}^{1-\theta_{k-1}}[Y^{(a'_{k-2})}]_{a'_{k-1}a'_{k}}^{1-\theta_{k-2}}[Y^{(a'_{k-3})}]_{a'_{k-2},a'_{k-1}a'_k}^{1-\theta_{k-3}}\cdots [Y^{(a'_{1})}]_{a'_2a'_3\cdots a'_k}^{1-\theta_1}
        \Bigg\}.
         \label{eq:red n to n-k 01}
    \end{align}
where all parameters $\alpha_j,\beta_j,\gamma$ range must be non-negative, and all $\theta_j$ can only take values of $0$ or $1$. The first summation ensures that each term contains exactly $r$ Lorentz indices, while the summation over permutations is placed after the coefficient $\Omega$ because the reduction coefficients are symmetric with respect to permutations of set label $\mathbf{I}_k$. Next, we will translate the above expression into the reduction coefficients in the form with Lorentz indices. We only need to provide the "translation" rules for $M$-type, $X$-type, and $Y$-type blocks:
\begin{equation}
    \begin{aligned}
        &\mathbf{M}\to \mathbf{M}^{\mu\nu}\equiv\frac{4\left(\left(\overline{VV}\right)^{\mu\nu}-g^{\mu\nu}\right)}{\left(\overline{LL}\right)},\\
&X^{(b)}\to(X^{(b)})^{\mu}\equiv\frac{\left(\overline{H_bL}\right)\left(\overline{VL}\right)_b^{\mu}-\left(\overline{H_bV}\right)^{\mu}\left(\overline{LL}\right)_b}{\left(\overline{LL}\right)},\\
        &Y^{(b)}\to (Y^{(b)})^{\mu\nu}\equiv\frac{2\left(\overline{H_bL}\right)g_{\mu\nu}+\left(\overline{H_bV}\right)^{\mu}\left(\overline{VL}\right)_b^{\nu}+\left(\overline{H_bV}\right)^{\nu}\left(\overline{VL}\right)_b^{\mu}-2\left(\overline{H_bL}\right)\left(\overline{VV}\right)_b^{\mu\nu}}{2\left(\overline{LL}\right)}.
    \end{aligned}
    \label{eq:trans MXY}
\end{equation}
Next, following the approach of \eqref{eq:n to n with LI}, we assign the $r$ Lorentz indices sequentially to the corresponding tensor blocks, perform a summation over all permutations, and finally multiply by the factor $\frac{1}{r!}$ to obtain the reduction coefficients in the form with Lorentz indices $C^{\mu_1\mu_r\cdots \mu_r}_{n\to n-k}$. Since the formula is too lengthy, we will not explicitly write it here but instead illustrate the process with the following example:
\begin{equation}
  \mathbf{K}^1[\mathbf{K}]_2^1[\mathbf{K}]_{12}^2\mathbf{M}^1[\mathbf{M}]_{2}^2[\mathbf{T}]_{12}^1X^{(2)}[Y^{(1)}]_2.
\end{equation}
The corresponding form with $15$ Lorentz indices is written as:
\begin{equation}
    \begin{aligned}
        \frac{1}{15!}\Bigg\{&\mathbf{K}^{\mu_1}[\mathbf{K}]^{\mu_2}_2[\mathbf{K}]_{12}^{\mu_3}[\mathbf{K}]_{12}^{\mu_4}\mathbf{M}^{\mu_5\mu_6}[\mathbf{M}]_2^{\mu_7\mu_8}[\mathbf{M}]_2^{\mu_9\mu_{10}}[\mathbf{T}]_{12}^{\mu_{11}\mu_{12}}(X^{(2)})^{\mu_{13}}[Y^{1}]_{2}^{\mu_{14}\mu_{15}}\\
        +&\sigma(\mu_1,\mu_2,\mu_3,\cdots,\mu_{15})\Bigg\}.
    \end{aligned}
\end{equation}

\subsubsection{Coefficient $\Omega$ }

The derivation of the general expression for the coefficient $\Omega$ is a purely mathematical process and will therefore not be included in this paper. In practice, the coefficient can be directly accessed using \textbf{Mathematica}\footnote{See section \ref{subsec:mathematica}} and \eqref{eq:red n to n-k}. Although the coefficient may appear complicated, the simplified result is unexpectedly simple, demonstrating that the use of tensor building blocks can lead to significant simplifications. Below, we provide several extreme examples with high tensor rank $r=15$ and $k=3$ to illustrate that the coefficient $\Omega$ is much simpler than one might initially expect. For example, for $\boldsymbol{\alpha}\equiv\{\alpha_0,\alpha_1,\alpha_2,\alpha_3\}=\{0,2,0,1\}$, $\boldsymbol{\beta}\equiv\{\beta_1,\beta_2,\beta_3\}=\{0,2,1\}$, $\boldsymbol{\theta}\equiv\{\theta_1,\theta_2,\theta_3\}=\{1,1,0\}$ and $\gamma=1$, coefficient $\Omega^{15}_{n\to n-3}(\{0,2,0,1\},\{0,2,1\},1,\{1,1,0\})$ is
    \begin{equation}
        \begin{aligned}
            \frac{-2079(112+9 D-9 n)(26+D-n)}{524288(4+D-n)(5+D-n)(6+D-n)(11+D-n)(12+D-n)(13+D-n)(14+D-n)}.
        \end{aligned}
    \end{equation}
The other examples are shown in Table \ref{table:omega}.
\begin{longtable}{|cccc|}
\hline
\multicolumn{1}{|c|}{$\quad\quad\quad\quad\quad \boldsymbol{\alpha}\quad\quad\quad\quad\quad$} & \multicolumn{1}{c|}{$\quad\quad\quad\boldsymbol{\quad\beta}\quad\quad\quad\quad$} & \multicolumn{1}{c|}{$\quad\quad\quad\quad\boldsymbol{\gamma}\quad\quad\quad\quad$} & $\boldsymbol{\theta}$ \\ \hline
\multicolumn{1}{|c|}{0,2,0,1}  & \multicolumn{1}{c|}{0,2,1}          & \multicolumn{1}{c|}{1}        & 0,1,1    \\ \hline
\multicolumn{4}{|c|}{ $\Omega^{15}_{n\rightarrow n-3}(\{0,2,0,1\},\{0,2,1\},1,\{0,1,1\})$}\\ \hline
\multicolumn{4}{|c|}{\multirow{3}{*}{\Large $\frac{-2079(9D-9n+112)(D-n+26)}{524288(D-n-4)(D-n-5)(D-n-6)(D-n+11)(D-n+12)(D-n+13)(D-n+14)}$}}\\
\multicolumn{4}{|c|}{}\\
\multicolumn{4}{|c|}{}\\ \hline
\multicolumn{1}{|c|}{$\boldsymbol{\alpha}$} & \multicolumn{1}{c|}{$\boldsymbol{\beta}$} & \multicolumn{1}{c|}{$\boldsymbol{\gamma}$} & $\boldsymbol{\theta}$ \\ \hline
\multicolumn{1}{|c|}{2,2,1,1}  & \multicolumn{1}{c|}{0,0,1}          & \multicolumn{1}{c|}{2}        & 1,1,1    \\ \hline
\multicolumn{4}{|c|}{$\Omega_{n\rightarrow n-3}^{15}(\{2,2,1,1\},\{0,0,1\},2,\{1,1,1\})$}\\ \hline
\multicolumn{4}{|c|}{\multirow{3}{*}{\Large $\frac{45 \left(5 (D-n)^4+170 (D-n)^3+1378 (D-n)^2-15143 (D-n)-87171\right) (D-n+16) (D-n+18)}{134217728 (D-n+4) (D-n+6) (D-n+9) (D-n+10) (D-n+11) (D-n+12) (D-n+13) (D-n+14)}$}}\\
\multicolumn{4}{|c|}{}\\
\multicolumn{4}{|c|}{}\\ \hline
\multicolumn{1}{|c|}{$\boldsymbol{\alpha}$} & \multicolumn{1}{c|}{$\boldsymbol{\beta}$} & \multicolumn{1}{c|}{$\boldsymbol{\gamma}$} & $\boldsymbol{\theta}$ \\ \hline
\multicolumn{1}{|c|}{0,0,0,12} & \multicolumn{1}{c|}{0,0,0}          & \multicolumn{1}{c|}{0}        & 1,1,1    \\ \hline
\multicolumn{4}{|c|}{$\Omega_{n\rightarrow n-3}^{15}(\{0,0,0,12\},\{0,0,0\},0,\{1,1,1\})$}\\ \hline
\multicolumn{4}{|c|}{\multirow{3}{*}{\Large $\frac{(D-n+16) (D-n+18) (D-n+20) (D-n+22) (D-n+24) (D-n+26) (D-n+28)}{134217728 (D-n+3) (D-n+4) (D-n+5) (D-n+7) (D-n+9) (D-n+11) (D-n+13)}$}}\\
\multicolumn{4}{|c|}{}\\
\multicolumn{4}{|c|}{}\\ \hline
\multicolumn{1}{|c|}{$\boldsymbol{\alpha}$} & \multicolumn{1}{c|}{$\boldsymbol{\beta}$} & \multicolumn{1}{c|}{$\boldsymbol{\gamma}$} & $\boldsymbol{\theta}$ \\ \hline
\multicolumn{1}{|c|}{0,0,0,0}  & \multicolumn{1}{c|}{0,0,0}          & \multicolumn{1}{c|}{6}        & 1,1,1    \\ \hline
\multicolumn{4}{|c|}{$\Omega_{n\rightarrow n-3}^{15}(\{0,0,0,0\},\{0,0,0\},6,\{1,1,1\})$}\\ \hline
\multicolumn{4}{|c|}{\multirow{3}{*}{\Large $\frac{10395}{134217728 (D-n+4) (D-n+6) (D-n+8) (D-n+10) (D-n+12) (D-n+14)}$}}\\
\multicolumn{4}{|c|}{}\\
\multicolumn{4}{|c|}{}\\ \hline
\multicolumn{1}{|c|}{$\boldsymbol{\alpha}$} & \multicolumn{1}{c|}{$\boldsymbol{\beta}$} & \multicolumn{1}{c|}{$\boldsymbol{\gamma}$} & $\boldsymbol{\theta}$ \\ \hline
\multicolumn{1}{|c|}{0,0,0,0}  & \multicolumn{1}{c|}{0,0,5}          & \multicolumn{1}{c|}{0}        & 0,0,1    \\ \hline
\multicolumn{4}{|c|}{$\Omega_{n\rightarrow n-3}^{15}(\{0,0,0,0\},\{0,0,5\},0,\{0,0,1\})$}\\ \hline
\multicolumn{4}{|c|}{\multirow{3}{*}{\Large $-\frac{945}{32 (D-n+3) (D-n+4) (D-n+6) (D-n+8) (D-n+10) (D-n+12) (D-n+14)}$}}\\
\multicolumn{4}{|c|}{}\\
\multicolumn{4}{|c|}{}\\ \hline
\caption{Some $\Omega$ with $r=15$, $k=3$}
\label{table:omega}
\end{longtable}

In fact, it can be observed that these coefficients $\Omega$ are rational functions of the parameter $D-n$, having the form $\frac{Poly_1(D-n)}{Poly_2(D-n)}$, where $Poly_1(D-n)$ and $Poly_2(D-n)$ are polynomials in $D-n$. After simplification, the highest degree of the polynomials in both the numerator and the denominator with respect to $D-n$ typically does not exceed $r-k$. For example, in the case we discussed, the degree of the polynomials in both the numerator and denominator will not exceed $r-k = 15-3 = 12$ (and in most cases, it will be even lower). In other words, the number of terms in both the numerator and the denominator is at most 13. Therefore, we can conclude that the simplified form of the coefficient $\Omega$ is always quite simple.

\subsubsection{Analytic Result}

Now we present the reduction of an tensor triangle to tadpole with rank $r=2,3$ as examples. Without loss of generality, we choose the label list $\mathbf{I}_2=\{2,3\}$. Then by \eqref{eq:red n to n-k} the reduction coefficients are
\begin{align}
C_{3 \rightarrow 1, \widehat{2,3}}^{(2)} & =\frac{1}{4}X^{(2)} \cdot\left[X^{(3)}\right]_2+(2\leftrightarrow3), \\
C_{3 \rightarrow 1, \widehat{2,3}}^{(3)} & =\Bigg\{\frac{D}{16(D-1)}\left(X^{(2)} \cdot\left[X^{(3)}\right]_2 \cdot\left[\mathbf{K}\right]_2\right) +\frac{1}{16}\left(X^{(2)} \cdot\left[X^{(3)}\right]_2 \cdot\left[\mathbf{K}\right]_{2,3}\right)\notag\\
& +\frac{D+1}{16(D-1)}\left(X^{(2)} \cdot\left[X^{(3)}\right]_2 \cdot \mathbf{K}\right) +\frac{1}{2(D-1)}\left(Y^{(2)} \cdot\left[X^{(3)}\right]_2\right)\notag\\
&+\frac{1}{4(D-1)}\left(X^{(2)} \cdot\left[Y^{(3)}\right]_2\right)\Bigg\}+(2\leftrightarrow 3).
\end{align}
Then the reduction coefficients with Lorentz indices are 
    \begin{align}
        C^{\mu_1\mu_2}_{3\to 1;\widehat{2,3}}=&\frac{1}{2}\left\{\frac{1}{4}\left(X^{(2)}\right)^{\mu_1}\left[X^{(3)}\right]_2^{\mu_2}+\left(\mu_1\leftrightarrow \mu_2\right)\right\}+\left(2\leftrightarrow 3\right),\\
        C_{3 \rightarrow 1, \widehat{2,3}}^{\mu_1\mu_2\mu_3} =&\frac{1}{6}\Bigg\{\frac{D}{16(D-1)}\left(\left(X^{(2)}\right)^{\mu_1} \cdot\left[X^{(3)}\right]_2^{\mu_2} \cdot\left[\mathbf{K}\right]_2^{\mu_3}\right) +\frac{1}{16}\left(\left(X^{(2)}\right)^{\mu_1} \cdot\left[X^{(3)}\right]_2^{\mu_2}\cdot\left[\mathbf{K}\right]_{2,3}^{\mu_3}\right)\notag\\
& +\frac{D+1}{16(D-1)}\left(\left(X^{(2)}\right)^{\mu_1} \cdot\left[X^{(3)}\right]_2^{\mu_2} \cdot \mathbf{K}^{\mu_3}\right) +\frac{1}{2(D-1)}\left(\left(Y^{(2)}\right)^{\mu_1\mu_2} \cdot\left[X^{(3)}\right]_2^{\mu_3}\right)\notag\\
&+\frac{1}{4(D-1)}\left(\left(X^{(2)}\right)^{\mu_1} \cdot\left[Y^{(3)}\right]_2^{\mu_2\mu_3}\right)+\sigma(\mu_1,\mu_2,\mu_3)\Bigg\}+(2\leftrightarrow 3).
   \end{align}

\section{Comparison with PV-method and Computational Efficiency}\label{sec:compare}

\subsection{Comparison with PV-reduction}\label{subsec:PV}
In Section \ref{sec:coef with index}, we have presented the rational expressions for the reduction coefficients derived using the generating function method, and also introduced the 'translation' rules for obtaining the Lorentz-indexed forms of reduction coefficients. Moreover, we have observed that these reduction coefficients are polynomials composed of several special tensor building blocks. These tensor blocks encompass all the dynamical variables, such as $q_i \cdot q_j$ and $M_i^2$, and the coefficients $\Omega$ of these polynomials depend solely on the spacetime dimension $D$, the tensor rank $r$ and the number of propagators $n$ of the Feynman integral, and the number of propagators selected for the scalar integrals (the master integrals), $n-k$. 

Since the core of our work focuses on the analytic structure and properties of the reduction coefficients, the tensor building blocks are more important than the coefficients $\Omega$s. If we liken the process of solving for the reduction coefficients to building a LEGO structure, we know that the reduction coefficient is a complex assembly made up of the smallest building blocks: the irreducible scalar products such $q_i \cdot q_j$ and mass $M^2_i$. To improve computational efficiency, a smarter approach is to construct the entire structure from larger building blocks. This way, we can pre-build these larger blocks in parallel, and then use them to assemble the reduction coefficient. Fortunately, by employing the generating function approach, we have identified these larger building blocks.

Next, we will compare the method presented in this paper with the PV reduction in various aspects. \textbf{First, we discuss the number of tensor building blocks}. For a one-loop tensor, Feynman Integral with Lorentz indices \eqref{eq:GF fey}. In PV reduction, the tensor structure of the reduction coefficient is formed by the combination of $q_i^\mu, i=1, 2, \dots, n$ and the space-time metric $g^{\mu\nu}$, and it is symmetric with respect to the $r$ Lorentz indices $\mu_1, \mu_2, \dots, \mu_r$ permutation. We denote the number of times each external momentum $q_i$ appears in the structure as $x_i$, and the number of times the metric $g^{\mu\nu}$ appears as $g$. Therefore, the total number of structures corresponds to the number of non-negative integer solutions to the equation $x_1 + x_2 + \cdots + x_n + 2g = r$. From combinatorial mathematics, the number of non-negative integer solutions to the equation $x_1 + x_2 + \cdots + x_n = r'$ is given by $\mathbf{C}^{n-1}_{r' + n - 1}$. Clearly, the range of $g$ is $0 \to \lfloor r/2 \rfloor$, so the total number of tensor structures is:
\begin{equation}
    PV(n,r)=\sum_{g=0}^{\lfloor \frac{r}{2}\rfloor} \mathbf{C}^{n-1}_{r-2g+n-1}.
\end{equation}
Returning to the method in our paper, the tensor structures of the reduction coefficients are composed of the building blocks in equation \eqref{eq:building blocks}, and their form is given by equation \eqref{eq:red n to n-k}. Similarly, the total number is equivalent to the number of non-negative integer solutions to the equation:
\begin{equation}
    \alpha_0+\alpha_1+\cdots +\alpha_k+2(\beta_1+\cdots +\beta_k)+2r+2k-(\theta_1+\cdots +\theta_k)=r.
\end{equation}
Here, $\theta_i$ can only take the values 0 or 1. This quantity is difficult to express using binomial coefficients, so we directly provide a table for an intuitive comparison with the PV method\footnote{In this case, unlike the usual assumption of setting $q_1 = 0$, we consider the general scenario where $n$ propagators carry $n$ external momenta.}.

\begin{table}[H]
\centering
\begin{tabular}{|cc|c|l|l|l|l|l|l|l|l|l|}
\hline
\multicolumn{2}{|c|}{Tensor rank r}        &\  1 &\ 2 &\ 3 &\ 4 &\ 5 &\ 6 &\ 7 &\ 8 &\ 9 &\ 10 \\ \hline
\multicolumn{2}{|c|}{PV Method}       &\ 1 &\ 2 &\ 2 &\ 3 &\ 3 &\ 4 &\ 4 &\ 5 &\ 5 &\ 6  \\ \hline
\multicolumn{1}{|c|}{GF Method} & k=0 &\ \red{1} &\ \red{2} &\ \red{2} &\ \red{3} &\ \red{3} &\ \red{4} &\ \red{4} &\ \red{5} &\ \red{5} &\ \red{6}  \\ \hline
\end{tabular}
\caption{The number of tensor structures between the PV method and the GF method of n=1}
\end{table}

\begin{table}[H]
\centering
\begin{tabular}{|cl|c|c|c|c|c|c|c|c|c|c|}
\hline
\multicolumn{2}{|c|}{Tensor rank r}                                          & {\color[HTML]{000000} 1} & {\color[HTML]{000000} 2} & 3                        & 4                        & 5                        & 6                        & 7                        & 8                        & 9                        & 10                       \\ \hline
\multicolumn{2}{|c|}{PV Method}                                              & 2                        & 4                        & 6                        & 9                        & 12                       & 16                       & 20                       & 25                       & 30                       & 36                       \\ \hline
\multicolumn{1}{|c|}{}                            & \multicolumn{1}{c|}{k=0} & {\color[HTML]{FE0000} 1} & {\color[HTML]{FE0000} 2} & {\color[HTML]{FE0000} 2} & {\color[HTML]{FE0000} 3} & {\color[HTML]{FE0000} 3} & {\color[HTML]{FE0000} 4} & {\color[HTML]{FE0000} 4} & {\color[HTML]{FE0000} 5} & {\color[HTML]{FE0000} 5} & {\color[HTML]{FE0000} 6} \\ \cline{2-12} 
\multicolumn{1}{|c|}{\multirow{-2}{*}{GF Method}} & k=1                      & {\color[HTML]{FE0000} 1} & {\color[HTML]{FE0000} 3} & 7                        & 13                       & 22                       & 34                       & 50                       & 70                       & 95                       & 125                      \\ \hline
\end{tabular}
\caption{The number of tensor structures between the PV method and the GF method of n=2}
\end{table}

\begin{table}[H]
\centering
\begin{tabular}{|cl|c|c|c|c|c|c|c|c|c|c|}
\hline
\multicolumn{2}{|c|}{Tensor rank r}                                          & {\color[HTML]{000000} 1} & {\color[HTML]{000000} 2} & 3                        & 4                         & 5                         & 6                         & 7                         & 8                         & 9                         & 10                         \\ \hline
\multicolumn{2}{|c|}{PV Method}                                              & 3                        & 7                        & 13                       & 22                        & 34                        & 50                        & 70                        & 95                        & 125                       & 161                        \\ \hline
\multicolumn{1}{|c|}{}                            & \multicolumn{1}{c|}{k=0} & {\color[HTML]{FE0000} 1} & {\color[HTML]{FE0000} 2} & {\color[HTML]{FE0000} 2} & {\color[HTML]{FE0000} 3}  & {\color[HTML]{FE0000} 3}  & {\color[HTML]{FE0000} 4}  & {\color[HTML]{FE0000} 4}  & {\color[HTML]{FE0000} 5}  & {\color[HTML]{FE0000} 5}  & {\color[HTML]{FE0000} 6}   \\ \cline{2-12} 
\multicolumn{1}{|c|}{}                            & k=1                      & {\color[HTML]{FE0000} 1} & {\color[HTML]{FE0000} 3} &  {\color[HTML]{FE0000} 7} &  {\color[HTML]{FE0000} 13} &  {\color[HTML]{FE0000} 22} &  {\color[HTML]{FE0000} 34} & {\color[HTML]{FE0000} 50} & {\color[HTML]{FE0000} 70} & {\color[HTML]{FE0000} 95} & {\color[HTML]{FE0000} 125} \\ \cline{2-12} 
\multicolumn{1}{|c|}{\multirow{-3}{*}{GF Method}} & k=2                      & {\color[HTML]{FE0000} 0} & {\color[HTML]{FE0000} 1} & {\color[HTML]{FE0000} 5} & {\color[HTML]{FE0000} 16} & 40                        & 86                       & 166                       & 296                       & 496                       & 791                       \\ \hline
\end{tabular}
\caption{The number of tensor structures between the PV method and the GF method of n=3}
\end{table}

\begin{table}[H]
\centering
\begin{tabular}{|cc|c|c|c|c|c|c|c|c|c|c|}
\hline
\multicolumn{2}{|c|}{r}                                 & 1                        & 2                        & 3                        & 4                         & 5                         & 6                         & 7                          & 8                         & 9                         & 10                         \\ \hline
\multicolumn{2}{|c|}{PV Method}                         & 4                        & 11                       & 24                       & 46                        & 80                        & 130                       & 200                        & 295                       & 420                       & 581                        \\ \hline
\multicolumn{1}{|c|}{}                            & k=0 & {\color[HTML]{FE0000} 1} & {\color[HTML]{FE0000} 2} & {\color[HTML]{FE0000} 2} & {\color[HTML]{FE0000} 3}  & {\color[HTML]{FE0000} 3}  & {\color[HTML]{FE0000} 4}  & {\color[HTML]{FE0000} 4}   & {\color[HTML]{FE0000} 5}  & {\color[HTML]{FE0000} 5}  & {\color[HTML]{FE0000} 6}   \\ \cline{2-12} 
\multicolumn{1}{|c|}{}                            & k=1 & {\color[HTML]{FE0000} 1} & {\color[HTML]{FE0000} 3} & {\color[HTML]{FE0000} 7} & {\color[HTML]{FE0000} 13} & {\color[HTML]{FE0000} 22} & {\color[HTML]{FE0000} 34} & {\color[HTML]{FE0000} 50}  & {\color[HTML]{FE0000} 70} & {\color[HTML]{FE0000} 95} & {\color[HTML]{FE0000} 125} \\ \cline{2-12} 
\multicolumn{1}{|c|}{}                            & k=2 & {\color[HTML]{FE0000} 0} & {\color[HTML]{FE0000} 1} & {\color[HTML]{FE0000} 5} & {\color[HTML]{FE0000} 16} & {\color[HTML]{FE0000} 40} & {\color[HTML]{FE0000} 86} & {\color[HTML]{FE0000} 166} & 296                       & 496                       & 791                        \\ \cline{2-12} 
\multicolumn{1}{|c|}{\multirow{-4}{*}{GF Method}} & k=3 & {\color[HTML]{FE0000} 0} & {\color[HTML]{FE0000} 0} & {\color[HTML]{FE0000} 1} & {\color[HTML]{FE0000} 7}  & {\color[HTML]{FE0000} 29} & {\color[HTML]{FE0000} 91} & 239                        & 553                       & 1163                      & 2269                       \\ \hline
\end{tabular}
\caption{The number ot tensor structures between the PV method and the GF method of n=4}
\end{table}

\begin{table}[H]
\centering
\begin{tabular}{|cc|c|c|c|c|c|c|c|c|c|c|}
\hline
\multicolumn{2}{|c|}{r}                                 & 1                        & 2                        & 3                        & 4                         & 5                         & 6                         & 7                          & 8                          & 9                           & 10                         \\ \hline
\multicolumn{2}{|c|}{PV Method}                         & 5                        & 16                       & 40                       & 86                        & 166                       & 296                       & 496                        & 791                        & 1211                        & 1792                       \\ \hline
\multicolumn{1}{|c|}{}                            & k=0 & {\color[HTML]{FE0000} 1} & {\color[HTML]{FE0000} 2} & {\color[HTML]{FE0000} 2} & {\color[HTML]{FE0000} 3}  & {\color[HTML]{FE0000} 3}  & {\color[HTML]{FE0000} 4}  & {\color[HTML]{FE0000} 4}   & {\color[HTML]{FE0000} 5}   & {\color[HTML]{FE0000} 5}    & {\color[HTML]{FE0000} 6}   \\ \cline{2-12} 
\multicolumn{1}{|c|}{}                            & k=1 & {\color[HTML]{FE0000} 1} & {\color[HTML]{FE0000} 3} & {\color[HTML]{FE0000} 7} & {\color[HTML]{FE0000} 13} & {\color[HTML]{FE0000} 22} & {\color[HTML]{FE0000} 34} & {\color[HTML]{FE0000} 50}  & {\color[HTML]{FE0000} 70}  & {\color[HTML]{FE0000} 95}   & {\color[HTML]{FE0000} 125} \\ \cline{2-12} 
\multicolumn{1}{|c|}{}   & k=2 & {\color[HTML]{FE0000} 0} & {\color[HTML]{FE0000} 1} & {\color[HTML]{FE0000} 5} & {\color[HTML]{FE0000} 16} & {\color[HTML]{FE0000} 40} & {\color[HTML]{FE0000} 86} & {\color[HTML]{FE0000} 166} & {\color[HTML]{FE0000} 296} & {\color[HTML]{FE0000} 496}  & {\color[HTML]{FE0000} 791} \\ \cline{2-12} 
\multicolumn{1}{|c|}{}                            & k=3 & {\color[HTML]{FE0000} 0} & {\color[HTML]{FE0000} 0} & {\color[HTML]{FE0000} 1} & {\color[HTML]{FE0000} 7}  & {\color[HTML]{FE0000} 29} & {\color[HTML]{FE0000} 91} & {\color[HTML]{FE0000} 239} & {\color[HTML]{FE0000} 553} & {\color[HTML]{FE0000} 1163} & 2269                       \\ \cline{2-12} 
\multicolumn{1}{|c|}{\multirow{-5}{*}{GF Method}} &  k=4 & {\color[HTML]{FE0000} 0} & {\color[HTML]{FE0000} 0} & {\color[HTML]{FE0000} 0} & {\color[HTML]{FE0000} 1}  & {\color[HTML]{FE0000} 9}  & {\color[HTML]{FE0000} 46} & {\color[HTML]{FE0000} 147} & {\color[HTML]{FE0000} 541} & 1461                        & 3544                       \\ \hline
\end{tabular}
\caption{The number of tensor structures between the PV method and the GF method of n=5}
\end{table} 
The parts highlighted in red in the table indicate that the number of structures in the Generating Function method is not larger than that in the PV method. It is evident that the number of structures in the PV method depends only on $n$, whereas the number of structures in our method does not depend on $n$ but only on $k$. In fact, another interesting phenomenon is that when $n = 2k + 1$, the number of structures in the generating function method coincides with that in the PV method when the tensor rank $r$ differs by $k$. This can be expressed mathematically as:

\begin{equation}
    GF(k,r)=PV(2k+1,r-k).
\end{equation}
Therefore, we can conclude that the number of tensor structures in the generating function method is:
\begin{equation}
    GF(k,r)=\sum_{g=0}^{\lfloor\frac{r-k}{2}\rfloor}\mathbf{C}^{2k}_{r+k-2g}.
\end{equation}
This means that when $k < n/2$, the number of tensor structures in our method is smaller than in the PV method, regardless of how large $r$ is. Additionally, from the table, we can observe that the number of structures in the generating function method surpasses that of the PV method when $r \approx 2k\approx 2n$. We know that in QCD and QED, the highest tensor rank typically involved in one-loop calculations is around $r = n$, and even in gravity theory, the highest rank is approximately $r = 2n$. Therefore, in most general cases, the number of tensor blocks in the generating function method is significantly smaller than in the PV method. This can greatly reduce the computational cost.

Second, \textbf{let's compare the computational processes of the generating function method and the PV method}. With our generating function method, we have directly provided a general formula for the reduction coefficient with Lorentz indices, eliminating the need for recursion or solving systems of equations. In contrast, traditional PV reduction usually requires expanding the reduction coefficient into
\begin{equation}
    C^{\mu_1\mu_2\cdots \mu_r}=\sum_i A_i\cdot  t^{\mu_1\mu_2\cdots \mu_r}_i,
\end{equation}
where $t^{\mu_1\mu_2\cdots \mu_r}_i$ refers to the independent tensor structures composed of $q_i^{\mu}$ and the metric $g^{\mu\nu}$ as mentioned earlier. $A_i$ are the coefficients containing the dynamical variables. In the standard approach, we slove $A_i$ by contracting $t_{i;\mu_1\mu_2\cdots \mu_r}$ with each structures in the expansion, which can be represented in matrix form:

\begin{equation}
    \begin{pmatrix}
        t_{1;\mu_1\mu_2\cdots \mu_r}\cdot C^{\mu_1\mu_2\cdots \mu_r}\\
        t_{2;\mu_1\mu_2\cdots \mu_r}\cdot C^{\mu_1\mu_2\cdots \mu_r}\\
        \ldots\\
        t_{N;\mu_1\mu_2\cdots \mu_r}\cdot C^{\mu_1\mu_2\cdots \mu_r}
    \end{pmatrix}=M_{PV}\cdot \begin{pmatrix}
        A_1\\
        A_2\\
        \ldots\\
        A_N
    \end{pmatrix}
    \label{eq:stand pv}
\end{equation}
where $M_{PV}$ is an $N\times N$ matrix, $N$ is the corresponding number of structures from the table mentioned earlier. That is to say, even when the number of tensor blocks in our generating function method is much greater than the number of PV structures, for example, when $n=4$, $r=10$, and $k=3$, our method has 2269 blocks, while the PV method only has 581 blocks. However, the time and complexity required to solve the inverse of a $581 \times 581$ matrix far exceed the time required to directly compute the 2269 coefficients $\Omega$ using the general formula (note that our general formula in \eqref{eq:red n to n-k} is simply the sum of these 2269 terms). Moreover, the coefficients $\Omega$ in our method are simply a rational form of $D-n$ as discussed in previous section, while the left-hand side of \eqref{eq:stand pv} and the matrix $M_{PV}$ still contain a large number of dynamical variables, which gives our method a significant advantage.

At last, \textbf{let's briefly discuss a comparison with the methods provided in other papers}. 

In \cite{Feng:2021enk} and \cite{Hu:2023mgc}, we provided an improved version of the PV method, where the coefficients are solved using an iterative recurrence approach, thus avoiding the need to invert a large $N \times N$ matrix. However, in this method, each step of the iterative process still requires solving a system of $n$ equations, which is not trivial, unlike our method, which directly provides the result without the need for recurrence. 

In \cite{Goode:2024mci} and \cite{Goode:2024cfy}, the authors use the permutation symmetry of a given product of metric tensors to derive a simple ansatz for its dual/projector, expressed in terms of an orbit partition formula. Combined with the van Neerven-Vermaseren decomposition, this transforms a general tensor integrals into vacuum integrals, significantly reducing the matrix inversion computation compared to the standard PV reduction. For example, when $r=10$, instead of inverting a $945 \times 945$ matrix as in the PV reduction, this method only requires inverting a $7 \times 7$ matrix. This is a remarkable breakthrough; however, there are two main issues with their method:  (1) The van Neerven-Vermaseren decomposition, which converts the tensor integrals into a linear combination of vacuum integrals, involves considerable computational effort.  (2) Their method only reduces the tensor part that carries Lorentz indices, and the coefficients for the final master integrals still require IBP reductions, while our method directly provides the final master integrals. Of course, a significant advantage of their approach is its ability to handle higher-loop problems, while our method currently only applies to one-loop tensor reduction coefficients. 

\subsection{Computation on Mathematica}\label{subsec:mathematica}
In this subsection, we will demonstrate how to implement all the aforementioned functions using the widely utilized computational language, \textbf{Wolfram}. The Mathematica notebook file can be downloaded from 
\href{https://github.com/LinFoxEatingGrapefruit/One-Loop-Tensor-Reduction-by-Generating-Function.git}{https://github.com/LinFoxEatingGrapefruit/One-Loop-Tensor-Reduction-by-Generating-Function.git}.
It requires no other installation; simply download and use it directly.


Here are the functions we have realized:
\begin{itemize}
\item First of all, without loss of generality, we select $\mathbf{I}_k = \{1,2,\cdots,k\}$ for \eqref{eq:red n to n-k 01}. Since \eqref{eq:red n to n-k 01} is symmetric under permutations of $\mathbf{I}_k$, we consider only one specific permutation, $\{1,2,\cdots,k\}$. This expression can be represented using the command \textcolor{blue}{\texttt{ct[k,r]}}. In other words, \textcolor{blue}{\texttt{ct[k,r]}} rewrites the reduction coefficient originally expressed in terms of $x_\pm$, $X^{(b)}$, and $Y^{(b)}$ into a form using $\mathbf{K}$, $\mathbf{\Delta}$, $\mathbf{M}$, $X^{(b)}$, and $Y^{(b)}$. At this stage, the expression does not involve Lorentz indices. Our goal is to extract the coefficient $\Omega_{n\rightarrow n-k}^r(\boldsymbol{\alpha}, \boldsymbol{\beta}, \gamma, \boldsymbol{\theta})$ from it. For examples,

\texttt{\textcolor{blue}{In[$\cdot$]:=}ct[5,5]}

\texttt{\textcolor{blue}{Out[$\cdot$]:=}$\frac{1}{32} X[\{5\}] X[\{4,5\}] X[\{3,4,5\}] X[\{2,3,4,5\}] X[\{1,2,3,4,5\}]$}.

For \eqref{eq:red n to n-k 01}, the term $[X^{(b)}]_{a_1,\cdots,a_i}$ is represented in our output format as $X[\{b,a_1,\cdots,a_i\}]$, where the first entry in the brackets denotes the superscript, and the remaining entries denote the subscripts. The same convention applies to $[Y^{(b)}]_{a_1,\cdots,a_i}$. And we give another example as

\texttt{\textcolor{blue}{In[$\cdot$]:=}ct[0,4]}

\texttt{\textcolor{blue}{Out[$\cdot$]:=}$\frac{3 \Delta ^2 K[\{\}]^2}{128 (d-n+1)}+\frac{1}{256} K[\{\}]^4+\frac{3 \Delta ^4}{512 \left(\frac{1}{2} (d-n+1)+1\right) (d-n+1)}$}.

In our notation, we use the lowercase letter d to represent the spacetime dimension $d$ to avoid confusion with \textbf{Mathematica}’s differentiation operator $D$. In the output, an empty set within the brackets of $K[\{\}]$ represents $\mathbf{K}$. If the brackets contain elements, such as $K[\{a_1,\cdots,a_i\}]$, it represents $[\mathbf{K}]_{a_1,\cdots,a_i}$. The same convention applies to $\mathbf{M}$. The parameter $n$ represents the number of propagators in the integral being reduced. Since our method is universally applicable to any nn, we control the process using only the indices $k$ and $r$, treating $n$ as a parameter. Note that the result of \texttt{\textcolor{blue}{ct[k,r]}} is not always fully expanded. Readers can expand it in terms of the building blocks to verify that the odd powers of $\mathbf{\Delta}$ indeed vanishes.

    \item \textcolor{blue}{\texttt{Omega[k,r,\{alphalist,betalist,gammalist,thetalist\}]}} gives the corresponding coefficient $\Omega^r_{n\to n-k}(\boldsymbol{\alpha}, \boldsymbol{\beta}, \gamma, \boldsymbol{\theta})$, for example
    
    \texttt{\textcolor{blue}{In[$\cdot$]:=}Omega[3,8,\{\{0,0,1,0\},\{1,0,0\},\{1\},\{1,1,1\}\}]}

    \texttt{\textcolor{blue}{Out[$\cdot$]:=}$\frac{-d+n-13}{512 (d-n+4) (d-n+6) (d-n+7)}$}.

    This example illustrates the coefficient of building block 

    \begin{equation}
        \sum_{(a_1,a_2,a_3)\in \sigma(\mathbf{I}_3)}[\mathbf{K}]_{a_2a_3}\cdot \mathbf{M}\cdot \mathbf{T}\cdot [X^{(a_1)}]_{a_2a_3}[X^{(a_2)}]_{a_3}X^{(a_3)}
    \end{equation}
    
    for $k = 3$ and $r = 8$. In this case, $\boldsymbol{\alpha} = \{0,0,1,0\}$, $\boldsymbol{\beta} = \{1,0,0\}$, $\boldsymbol{\gamma} = \{1\}$, and $\boldsymbol{\theta} = \{1,1,1\}$. Note that the lengths of the four sequences $\boldsymbol{\alpha}$, $\boldsymbol{\beta}$, $\boldsymbol{\gamma}$, and $\boldsymbol{\theta}$ must be $k+1$, $k$, $1$, and $k$, respectively. These parameters must satisfy the equation $r = a_0 + 2\gamma + 2k + \sum_{i=1}^k (\alpha_i + 2\beta_i - \theta_i)$ to provide the correct number of Lorentz indices.

    \item \texttt{\textcolor{blue}{BBcoefficient[k,r]}} gives a list of coefficients $\Omega_{n\rightarrow n-k}^r(\boldsymbol{\alpha}, \boldsymbol{\beta}, \gamma, \boldsymbol{\theta})$ for all possible $\boldsymbol{\alpha}$, $\boldsymbol{\beta}$, $\gamma$ and $\boldsymbol{\theta}$ under given $k$ and $r$. For example, for $k=3$ and $r=4$, the computation is
    
    \texttt{\textcolor{blue}{In[$\cdot$]:=}BBcoefficient[3, 4]}

    \texttt{\textcolor{blue}{Out[$\cdot$]:=}$\left\{\frac{1}{8 d-8 n+24},\frac{1}{4 d-4 n+12},\frac{3}{8 (d-n+3)},\frac{d-n+6}{32 (d-n+3)},\frac{d-n+5}{32 (d-n+3)},\frac{d-n+4}{32 (d-n+3)},\frac{1}{32}\right\}$}
    
    You can also check the building block solution $\boldsymbol{\alpha}$, $\boldsymbol{\beta}$, $\gamma$ and $\boldsymbol{\theta}$ that corresponds to the coefficient in the following method.

     \texttt{\textcolor{blue}{In[$\cdot$]:=}BD[3,4][[3]]}

     \texttt{\textcolor{blue}{In[$\cdot$]:=}BBcoefficient[3,4][[3]]}

     \texttt{\textcolor{blue}{Out[$\cdot$]:=}$\{\gamma\rightarrow0,\alpha(0)\rightarrow0,\alpha(1)\rightarrow0,\alpha(2)\rightarrow0,\alpha(3)\rightarrow0,\beta(1)\rightarrow0,\beta(2)\rightarrow0,\beta(3)\rightarrow0,\theta(1)\rightarrow1,\theta(2)\rightarrow1,\theta(3)\rightarrow0\}$}

      \texttt{\textcolor{blue}{Out[$\cdot$]:=}$\frac{3}{8 (d-n+3)}$}

     Here, we can use the command \texttt{\textcolor{blue}{BD[k,r][[i]]}} to obtain the corresponding values of $\boldsymbol{\alpha}$,  $\boldsymbol{\beta}$,  $\boldsymbol{\gamma}$, and $\boldsymbol{\theta}$ for the  for the $i$-th coefficient in \texttt{\textcolor{blue}{BBcoefficient[k,r]}}.

    \item The command \texttt{\textcolor{blue}{BBWithLorentz[n,k,r,{Hatlist}]}} provides the result of \eqref{eq:red n to n-k 01} with Lorentz indices, combining the coefficient $\Omega^r_{n\to n-k}(\boldsymbol{\alpha}, \boldsymbol{\beta}, \gamma, \boldsymbol{\theta})$ with the building blocks that include Lorentz indices. For example, for $n=2$, $k=0$, $r=2$, and $\mathbf{I}_k=\{\}$, the corresponding code is:

     \texttt{\textcolor{blue}{In[$\cdot$]:=}BBWithLorentz[2, 0, 2, \{\}]}

     \texttt{\textcolor{blue}{Out[$\cdot$]:=}$\frac{1}{16} K[\{\}][u[1]] K[\{\}][u[2]] + \frac{T[\{\}][u[1], u[2]]}{16(-1 + d)}$}.

    In this result, we will use $M[\{a_1,a_2,\cdots,\a_i\}][\mu[1],\mu[2]]$ to represent the building block with and Lorentz index $[\mathbf{M}]_{a_1,a_2\cdots a_i}^{\mu_1\mu_2}$, and this notion is also applied in $\mathbf{K}$, $X^{(b)}$ and $Y^{(b)}$. It is important to emphasize that $\mathbf{T}^{\mu_i,\mu_j}$ is unique in our framework, as there is only one type of $T$-type building block. Although we omit the $\mathbf{I}_k$ in $[\mathbf{T}]_{\mathbf{I}_k}$ in the paper, our output is presented in the format $T[\{a_1,\cdots,a_k\}][\mu[i],\mu[j]]$.

    \item Any \textbf{building blocks} with Lorentz indices can be explicitl exhibited when number of propagator $n$ is fixed. One can acquire detailed form of them by "/.ReplaceRule[n]", for example,
    
    \texttt{\textcolor{blue}{In[$\cdot$]:=}K[\{2\}][$\mu$[1]]/.ReplaceRule[2]}
    
    \texttt{\textcolor{blue}{Out[$\cdot$]:=}$4 \text{q}_1^{\mu[1]}$}
    
    \texttt{\textcolor{blue}{In[$\cdot$]:=}X[\{2\}][$\mu$[1]]/.ReplaceRule[2]//Simplify}

    \texttt{\textcolor{blue}{Out[$\cdot$]:=}$\frac{\text{q}_1^{\mu[1]}-\text{q}_2^{\mu[1]}}{\text{sp}[\text{q}_1,\text{q}_1]-2\text{sp}[\text{q}_2,\text{q}_1]+\text{sp}[\text{q}_2,\text{q}_2]}$}
    
    \texttt{\textcolor{blue}{In[$\cdot$]:=}T[\{1\}][$\mu$[1],$\mu$[2]] M[\{1\}][$\mu$[3],$\mu$[4]]  /.ReplaceRule[2]}

    \texttt{\textcolor{blue}{Out[$\cdot$]:=}64m$_2^2$g$^{\mu[1]\mu[2]}$(q$_2^{\mu[3]}$q$_2^{\mu[4]}$-m$_2^2$g$^{\mu[3]\mu[4]}$)}.
    
    Here, \texttt{\textcolor{blue}{\text{sp[q$_i$, q$_j$]}}} represents $q_i \cdot q_j$. Thus, the above three examples correspond to the case of $n = 2$, which is a bubble diagram, we have
    
\begin{equation}
\begin{aligned}[\mathbf{K}]_2^{\mu_1}=&4q_1^{\mu_1},\\
[X^{(2)}]^{\mu_1}=&\frac{q_1^{\mu_1}-q_2^{\mu_1}}{q_1^2-2q_1\cdot q_2 +q_2^2},\\
[\mathbf{T}]_1^{\mu_1\mu_2}[\mathbf{M}]_1^{\mu_3\mu_4}=&64m_2^2 g^{\mu_1\mu_2}\left(-m_2^2 g^{\mu_3\mu_4}+q_2^{\mu_3}q_2^{\mu_4}\right).
\end{aligned}
\end{equation}
     \item \texttt{\textcolor{blue}{ReductionCoefficient[n,k,r,Hatlist]}} can automatically give result of the \eqref{eq:n to n with LI} or \eqref{eq:red n to n-k 01} with Lorentz indices replaced by \eqref{eq:trans K T} and \eqref{eq:trans MXY}, i.e. the final reduction coefficients with Lorentz indices. For  $C^{\mu_1\mu_2\cdots\mu_5}_{4\to 1;\widehat{1,2,3}}$, then $n=4$, $k=3$, $r=5$ and $\mathbf{I}_k=\{1,2,3\}$, the input form should be

     \texttt{\textcolor{blue}{In[$\cdot$]:=}ReductionCoefficient[4,3,5,\{1,2,3\}]}

    It is worth to note that, when $k=0$ and label list $\mathbf{I}_k$ being empty should use \texttt{\textcolor{blue}{\{\}}}, 
     
     \texttt{\textcolor{blue}{In[$\cdot$]:=}ReductionCoefficient[4,0,3,\{\}]}

    \texttt{\textcolor{blue}{In[$\cdot$]:=}ReductionCoefficient[1,0,2,\{\}]}

    \texttt{\textcolor{blue}{Out[$\cdot$]:=}$\frac{\text{m}_1^2 \text{g}^{\mu [1]\mu [2]}}{\text{d}}+\text{q}_1^{\mu [1]} \text{q}_1^{\mu [2]}$}

    \texttt{\textcolor{blue}{In[$\cdot$]:=}ReductionCoefficient[1,0,1,\{\}]}
    
    \texttt{\textcolor{blue}{Out[$\cdot$]:=}$\text{q}_1^{\mu [1]}$}

    The first two output results of the reduction coefficient commands are encouraged to be tested by the readers within the program but are not directly displayed here due to their size. The latter two examples, however, explicitly show the outputs where $C^{\mu_1\mu_2}_{1\to 1} = \frac{m_1^2 g^{\mu_1\mu_2}}{D}+q_1^{\mu_1}q_1^{\mu_2} $ and $C^{\mu_1}_{1\to 1} = q_1^{\mu_1}$.

     \item \texttt{\textcolor{blue}{SuperSimplifiedReductionCoefficient[n,k,r,Hatlist]}} produces a relatively simplified result compared with \texttt{\textcolor{blue}{ReductionCoefficient[n,k,r,Hatlist]}} when $n\leq4$, because inverse matrix $Q^{-1}$ can be analytically simplified in \eqref{eq:trans K T} and \eqref{eq:trans MXY} with a $\mathrm{Det}(Q)$, which will make the size of the result about $2/3$ of initial size. When using this command to calculate the reduction coefficients, the computation time is longer for smaller values of $r$, but it decreases as $r$ becomes larger. For example, based on our tests, when $k = 3$ and $r = 8$, the computation time using this simplified version is shorter than that of the previous command.

     \item \textbf{Numerical Evaluate}. The program supports numerical computation, mass and inner product among external momentum $q_i$ should be put in advanced. One can set the mass by \texttt{\textcolor{blue}{m$_i$=Constant}} or \texttt{\textcolor{blue}{m$_i^2$=Constant}} and inner product by \texttt{\textcolor{blue}{sp[q$_i$,q$_j$]=Constant}}. In general, we usually set the first external momentum $q_1 = 0$. If this operation is required, it can be implemented before running the program by adding an initial setup \texttt{\textcolor{blue}{Superscript[$q_1$,$\mu$\_]=0}}. For example,
     
     \texttt{\textcolor{blue}{In[$\cdot$]:=}m$_1$=1/3;m$_2$=1/5;m$_3$=1/7;m$_4$=1/11;sp[q$_1$,q$_1$]=1/13;sp[q$_1$,q$_2$]=1/17;sp[q$_1$,q$_3$]=1/19;
     sp[q$_1$,q$_4$]=1/23;sp[q$_2$,q$_2$]=1/29;sp[q$_2$,q$_3$]=1/31;sp[q$_2$,q$_4$]=1/37;sp[q$_3$,q$_3$]=1/41;
     sp[q$_3$,q$_4$]=1/43;sp[q$_4$,q$_4$]=1/47};
     
     \texttt{\textcolor{blue}{In[$\cdot$]:=}ReductionCoefficient[4,3,4,\{1,2,3\}]}
\end{itemize}

\begin{figure}[H]
    \centering
    \includegraphics[width=0.5\linewidth]{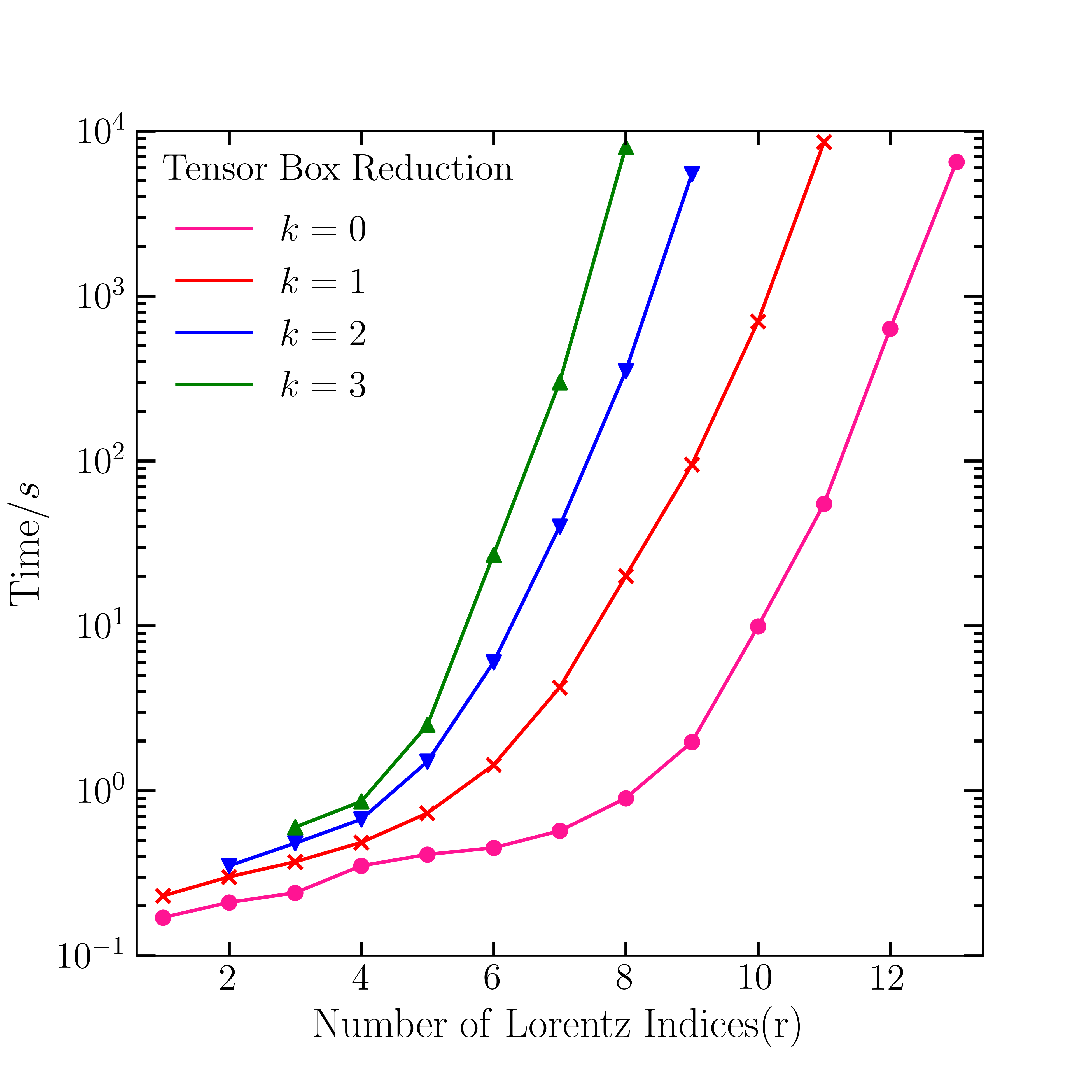}
    \caption{The runtime (in seconds) of reduction for the \textbf{full massive} tensor box. The horizontal axis represents the number of Lorentz indices, and the vertical axis, displayed on a logarithmic scale. The lines corresponding to $k=0,1,2,3$ represent the runtime for the \textbf{purely analytic} reduction of the tensor box to the master integrals of the box, triangle, bubble, and tadpole, respectively.}
    \label{fig:time testing of tensor box}
\end{figure}

In terms of computational efficiency, our method can be performed on a standard laptop\footnote{In our computational tests, we used a lightweight laptop with a processor frequency of 3 to 3.5 GHz and 32 GB of memory, meeting the typical hardware requirements of most researchers.} rather than requiring a server cluster. Figure \ref{fig:time testing of tensor box} shows the time taken to derive the analytic expressions for the reduction coefficients of the tensor box to its master integrals. Additionally, our program is entirely written in \textbf{Wolfram Mathematica}. Due to the limitations of the language, it currently performs single-core computations without parallel processing.

\section{Discussion and Outlook}\label{sec:outlook}

In this section, we present discussions and perspectives based on the work described in this paper. The primary contribution of this study is a method for reducing one-loop tensor integrals with Lorentz indices to master integrals using a generating function approach. We introduced an improved set of tensor building blocks and provided a program implementation.

In previous work, tensor integrals were typically handled by introducing an auxiliary vector $R$, which allowed for converting tensor integrals into scalar integrals. The methodology and results from \cite{Hu:2023mgc} serve as the foundation of this paper. However, that study also employed the auxiliary vector $R$. While, in principle, differentiating with respect to $R$ can yield Lorentz-indexed reductions, the presence of irrational terms in the generating function often made this operation challenging. In contrast, we derive a closed-form expression for the reduction coefficients using the generating function. By carefully analyzing the analytic structure of each term, we applied a simple variable substitution to eliminate irrational terms, making the differentiation process straightforward.

Building on this, we introduced new tensor building blocks. Compared to traditional PV tensor blocks, these new structures have fewer components in most cases, except for extreme cases with a large number of Lorentz indices. Moreover, the new tensor structures absorb all mass and external momentum information, leaving the coefficients dependent only on basic parameters such as spacetime dimensions and the number of propagators. This design enables pre-computation or parallel computation of the structures, significantly enhancing computational efficiency. Additionally, the calculation of structure coefficients is faster and yields simpler forms, providing a substantial advantage in numerical computations.

Regarding programming, existing software capable of reducing tensor integrals with Lorentz indices remains limited. Although our work is currently restricted to one-loop reductions, it serves as a promising starting point, demonstrating the potential of the generating function approach. We provide an initial implementation in \textbf{Wolfram Mathematica}, which requires no additional installation and is ready for immediate use. According to our tests, the program can handle high-order purely analytical computations effectively.

However, this work has certain limitations, primarily in program design. First, as an initial version, the program lacks optimization, meaning its efficiency can be further improved—an area for future work. Second, while our method inherently supports significant parallel computation due to its structural characteristics, the current \textbf{Mathematica}-based implementation is limited to single-core computation. Future developments could explore using other programming languages to enable parallel computing and further enhance performance.

\section*{Acknowledgments}

We would like to thank Prof. Bo Feng for useful discussions. We also extend our gratitude to Wen Chen, Tingfei Li, Xiang Li, and Yongqun Xu for their valuable advice on the assistance with programming. Yifan Hu and Jiyuan Shen acknowledge the financial support from the Hangzhou Institute for Advanced Study. Chang Hu expresses gratitude to Hebei University for providing the startup fund for young faculty research.

\begin{appendices}
    \section{Solving $\Omega$ on $n$-gon to $n$-gon}\label{sec:solve n2n}

In equation \eqref{eq:red n to n}, by replacing $x_+$ and $x_-$ with $\mathbf{\Delta}$ and $\mathbf{K}$, we obtain:
\begin{equation}
\begin{aligned}
C_{n \rightarrow n}^{(r)}&=\frac{1}{2^r}\sum_{j=0}^r \sum_{i=0}^j\left(\frac{\left(\frac{D-n}{2}\right)_i}{(D-n)_i}\left(-\mathbf{\Delta}\right)^i \left(\frac{\mathbf{\Delta}+\mathbf{K}}{2}\right)^{r-i} \mathbf{C}_{j-1}^{i-1}\right)\\ 
&=\sum_{j=0}^r \sum_{i=0}^j\sum_{l=0}^{r-i}\left(\frac{(-1)^i}{2^{2r-i}}\frac{\left(\frac{D-n}{2}\right)_i}{(D-n)_i}\mathbf{\Delta}^{l+i} \mathbf{K}^{r-i-l} \mathbf{C}_{r-i}^{l}\mathbf{C}_{j-1}^{i-1}\right). 
\end{aligned}
\end{equation}
Rewriting $l+i=k$ , we obtain:
\begin{equation}
C_{n \rightarrow n}^{(r)}=\sum_{j=0}^r \sum_{i=0}^j\sum_{k=i}^{r}\left(\frac{(-1)^i}{2^{2r-i}}\frac{\left(\frac{D-n}{2}\right)_i}{(D-n)_i}\mathbf{\Delta}^{k} \mathbf{K}^{r-k} \mathbf{C}_{r-i}^{k-i}\mathbf{C}_{j-1}^{i-1}\right). 
\end{equation}
Then, by interchanging the order of summation, we obtain:
\begin{equation}
C_{n \rightarrow n}^{(r)}=\sum_{k=0}^r \sum_{i=0}^k\left(\frac{(-1)^i}{2^{2r-i}}\frac{\left(\frac{D-n}{2}\right)_i}{(D-n)_i}\mathbf{\Delta}^{k} \mathbf{K}^{r-k} \mathbf{C}_{r-i}^{k-i}\sum_{j=i}^{r}\mathbf{C}_{j-1}^{i-1}\right). 
\end{equation}
Using the formula $\sum_{i=m}^{n}\mathbf{C}_{i}^{m}=\mathbf{C}_{n+1}^{m+1}$, we obtain:
\begin{equation}
\begin{aligned}
C_{n \rightarrow n}^{(r)}=&\sum_{k=0}^r \sum_{i=0}^k\left(\frac{(-1)^i}{2^{2r-i}}\frac{\left(\frac{D-n}{2}\right)_i}{(D-n)_i}\mathbf{\Delta}^{k} \mathbf{K}^{r-k} \mathbf{C}_{r-i}^{k-i}\mathbf{C}_{r}^{i}\right)\\
=&\frac{1}{2^{2r}}\sum_{k=0}^r\mathbf{C}_r^k\mathbf{\Delta}^k \mathbf{K}^{r-k}\left\{\sum_{i=0}^k\frac{(-2)^i}{i!}\frac{\left(\frac{D-n}{2}\right)_i}{\left(D-n\right)_i}\cdot\frac{\Gamma(k+1)}{\Gamma(k-i+1)}\right\}\\
=&\frac{1}{2^{2r}}\sum_{k=0}^r\mathbf{C}_r^k\mathbf{\Delta}^k \mathbf{K}^{r-k}\left\{\sum_{i=0}^k\frac{(2)^i}{i!}\frac{\left(\frac{D-n}{2}\right)_i(-k)_i}{\left(D-n\right)_i}\right\}\\
=&\frac{1}{2^{2r}}\sum_{k=0}^r\mathbf{C}_r^k\mathbf{\Delta}^k \mathbf{K}^{r-k}\cdot\ _2F_1\left(\begin{array}{l}
-k , \frac{D-n}{2}  \\
 D-n
\end{array} \bigg| 2\right).
\end{aligned}
\end{equation}
The last equality in the above equation is because when the integer $k< i$, we have $(-k)_i = 0$. In general, the variable $z$ of a hypergeometric function is divergent when $|z|\geq 1$. However, in the above hypergeometric function, one of the parameters in the numerator is a negative integer, so the hypergeometric function is actually a truncated finite series. For such a hypergeometric function, it can be found that:
\begin{equation}
\begin{aligned}
_2F_1\left(\begin{array}{l}
-k , \frac{D-n}{2}  \\
 D-n
\end{array} \bigg| 2\right)&=\frac{2^{-k-1}(-1)^{k} \left((-1)^k+1\right) \Gamma \left(\frac{D-n+1}{2} \right) k! }{ \Gamma \left(\frac{k+2}{2}\right) \Gamma \left(\frac{D+k-n+1}{2} \right)}\\
&=\left\{
\begin{aligned}
\frac{(k-1)!!}{2^{\frac{k}{2}}(\frac{D-n+1}{2})_{\frac{k}{2}}}, \quad\quad&\text{if } k\  \text{is even}, \\
0\quad\quad\quad, \quad\quad&\text{if } k\ \text{is odd}.
\end{aligned}
\right.
\end{aligned}
\end{equation}
After changing the viable $\mathbf{T}=\mathbf{\Delta}^2$, we can express the full reduced coefficients as:
\begin{equation}
C_{n \rightarrow n}^{(r)}=\frac{1}{2^{2r}}\sum_{k=0}^{\left\lfloor\frac{r}{2}\right\rfloor} \frac{(2 k-1)!!\mathbf{C}_r^{2 k}}{2^k\left(\frac{D-n+1}{2}\right)_k}\mathbf{T}^{ k}\mathbf{K}^{r-2 k}.
\end{equation}

\section{Result Check}\label{check}
In this section, we will present some results for the tadpole and bubble diagrams, as well as numerical coefficients of the master integrals in the triangle and box diagrams. We also compare our results with those from \texttt{AmpRed} \cite{Chen:2024xwt}.

\subsection{Tadpole}
We will start with the simplest case, tensor tadpole, the Feynman Integral is 
\begin{equation}
    I^{\mu_1\mu_2\cdots\mu_r}_1=\int d^Dl \frac{l^{\mu_1}l^{\mu_2}l^{\mu_3}\cdots l^{\mu_r}}{l^2-m_1^2}.
    \label{Tadpole}
\end{equation}
In this case, we have
\begin{equation}
    \mathbf{K}^{\mu}=0,\ \ \ \mathbf{T}^{\mu\nu}=16 m^2 g^{\mu\nu}.
\end{equation}
It is evident that when $r$ is odd, every term must necessarily contain $\mathbf{K}$, which leads to the total reduction coefficient being zero. For even $r$, however, as seen from \eqref{eq:n to n with LI}, only the term with $i = r/2$ in the summation does not vanish. Thus, when $r = 2l$, the reduction coefficient is given by:
\begin{equation}
    \begin{aligned}
        C^{\mu_1\cdots \mu_{2l}}_{1\to 1}=&\frac{(2l-1)!!}{2^{5l}\left(\frac{D}{2}\right)_l(2l)!}\sum_{\sigma\in \sigma(\mu_1,\cdots,\mu_{2l})}\mathbf{T}^{\mu_{\sigma(1)}\mu_{\sigma(2)}}\mathbf{T}^{\mu_{\sigma(3)}\mu_{\sigma(4)}}\cdots \mathbf{T}^{\mu_{\sigma(2l-1)}\mu_{\sigma(2l)}}\\
        =&\frac{m^{2l}}{D(D+2)\cdots (D+2l-2)}\frac{(2l-1)!!}{(2l)!}\sum_{\sigma\in \sigma(\mu_1,\cdots,\mu_{2l})}g^{\mu_{\sigma(1)}\mu_{\sigma(2)}}\cdots g^{\mu_{\sigma(2l-1)}\mu_{\sigma(2l)}}.
    \end{aligned}
\end{equation}

\subsection{Bubble}
The Feynman Integral is
\begin{equation}
    I^{\mu_1\mu_2\cdots\mu_r}_1=\int d^Dl \frac{l^{\mu_1}l^{\mu_2}l^{\mu_3}\cdots l^{\mu_r}}{(l^2-m_1^2)((l-q_2)^2-m_2^2)}.
    \label{Bubble}
\end{equation}
For r=1, the reduction coefficient is:
\begin{equation}
    \begin{aligned}
        C^{\mu_1}_{2\to 2}&=\frac{\left(m_1^2-m_2^2+q_2^2\right) q_2^{\mu _1}}{2 q_2^2},\quad \\
        C^{\mu_1}_{2\to 1;\hat{1}}&=\frac{q_2^{\mu _1}}{2 q_2^2},\quad \\
        C^{\mu_1}_{2\to 1;\hat{2}}&=-\frac{q_2^{\mu _1}}{2 q_2^2}.\quad  \\
    \end{aligned}
\end{equation}
For r=2, the reduction coefficient are,
    \begin{align}
    C^{\mu_1\mu_2}_{2\to 2}=&-\frac{m_{1}^{4}g^{\mu_{1}\mu_{2}}}{4\left(D-1\right)q_{2}^{2}}+\frac{m_{2}^{2}m_{1}^{2}g^{\mu_{1}\mu_{2}}}{2\left(D-1\right)q_{2}^{2}}-\frac{m_{2}^{4}g^{\mu_{1}\mu_{2}}}{4\left(D-1\right)q_{2}^{2}}+\frac{m_{1}^{2}g^{\mu_{1}\mu_{2}}}{2\left(D-1\right)}\notag\\
    &+\frac{m_2^2g^{\mu_1\mu_2}}{2(D-1)}-\frac{q_2^2g^{\mu_1\mu_2}}{4(D-1)}+\frac{Dm_1^4q_2^{\mu_1}q_2^{\mu_2}}{4(D-1)q_2^4}+\frac{Dm_1^2q_2^{\mu_1}q_2^{\mu_2}}{2(D-1)q_2^2}-\frac{m_1^2q_2^{\mu_1}q_2^{\mu_2}}{(D-1)q_2^2}\notag\\
    &-\frac{Dm_2^2m_1^2q_2^{\mu_1}q_2^{\mu_2}}{2(D-1)q_2^4}-\frac{Dm_2^2q_2^{\mu_1}q_2^{\mu_2}}{2(D-1)q_2^2}+\frac{Dm_2^4q_2^{\mu_1}q_2^{\mu_2}}{4(D-1)q_2^4}+\frac{Dq_2^{\mu_1}q_2^{\mu_2}}{4(D-1)},\\
C^{\mu_1\mu_2}_{2\to 1;\hat 1}=&\frac{m_{1}^{2}g^{\mu_{1}\mu_{2}}}{4\left(D-1\right)q_{2}^{2}}+\frac{m_{2}^{2}g^{\mu_{1}\mu_{2}}}{4\left(D-1\right)q_{2}^{2}}+\frac{g^{\mu_{1}\mu_{2}}}{4\left(D-1\right)}+\frac{Dm_{1}^{2}q_{2}{}^{\mu_{1}}q_{2}{}^{\mu_{2}}}{4\left(D-1\right)q_{2}^{4}}\notag\\
&-\frac{Dm_{2}^{2}q_{2}{}^{\mu_{1}}q_{2}{}^{\mu_{2}}}{4\left(D-1\right)q_{2}^{4}}+\frac{3Dq_{2}{}^{\mu_{1}}q_{2}{}^{\mu_{2}}}{4\left(D-1\right)q_{2}^{2}}-\frac{q_{2}{}^{\mu_{1}}q_{2}{}^{\mu_{2}}}{\left(D-1\right)q_{2}^{2}},\\
C^{\mu_1\mu_2}_{2\to 1;\hat 2}=&\frac{m_{1}^{2}g^{\mu_{1}\mu_{2}}}{4\left(D-1\right)q_{2}^{2}}-\frac{m_{2}^{2}g^{\mu_{1}\mu_{2}}}{4\left(D-1\right)q_{2}^{2}}+\frac{g^{\mu_{1}\mu_{2}}}{4\left(D-1\right)}\notag\\
&-\frac{Dm_{1}^{2}q_{2}^{\mu_{1}}q_{2}^{\mu_{2}}}{4\left(D-1\right)q_{2}^{4}}+\frac{Dm_{2}^{2}q_{2}^{\mu_{1}}q_{2}^{\mu_{2}}}{4\left(D-1\right)q_{2}^{4}}-\frac{Dq_{2}^{\mu_{1}}q_{2}^{\mu_{2}}}{4\left(D-1\right)q_{2}^{2}}.
\end{align}
For r=3, the reduction coefficient are:
    \begin{align}
    C^{\mu_1\mu_2\mu_3}_{2\to 1;\hat 1}=&-\frac{g^{\mu2\mu3} q_{2}^{\mu1} m_{1}^{4}}{8 (D-1) q_{2}^{4}}-\frac{g^{\mu1\mu3} q_{2}^{\mu2} m_{1}^{4}}{8 (D-1) q_{2}^{4}}-\frac{g^{\mu1\mu2} q_{2}^{\mu3} m_{1}^{4}}{8 (D-1) q_{2}^{4}}+\frac{D q_{2}^{\mu1} q_{2}^{\mu2} q_{2}^{\mu3} m_{1}^{4}}{8 (D-1) q_{2}^{6}}\notag\\
    &+\frac{q_{2}^{\mu1}q_{2}^{\mu2}q_{2}^{\mu3}m_{1}^{4}}{4\left(D-1\right)q_{2}^{6}}+\frac{m_{2}^{2}g^{\mu2\mu3}q_{2}^{\mu1}m_{1}^{2}}{4\left(D-1\right)q_{2}^{4}}+\frac{m_{2}^{2}g^{\mu1\mu3}q_{2}^{\mu2}m_{1}^{2}}{4\left(D-1\right)q_{2}^{4}}+\frac{m_{2}^{2}g^{\mu1\mu2}q_{2}^{\mu3}m_{1}^{2}}{4\left(D-1\right)q_{2}^{4}}\notag\\
    &+\frac{D q_2^{\mu1} q_2^{\mu2} q_2^{\mu3} m_1^2}{2 (D-1) q_2^4}-\frac{q_2^{\mu1} q_2^{\mu2} q_2^{\mu3} m_1^2}{2 (D-1) q_2^4}-\frac{D m_2^2 q_2^{\mu1} q_2^{\mu2} q_2^{\mu3} m_1^2}{4 (D-1) q_2^6}-\frac{m_2^2 q_2^{\mu1} q_2^{\mu2} q_2^{\mu3} m_1^2}{2 (D-1) q_2^6}\notag\\
    &+\frac{g^{\mu2\mu3} q_{2}^{\mu1}}{8 (D-1)}+\frac{m_{2}^{2} g^{\mu2\mu3} q_{2}^{\mu1}}{2 (D-1) q_{2}^{2}}-\frac{m_{2}^{2} g^{\mu2\mu3} q_{2}^{\mu1}}{2 (D-1) D q_{2}^{2}}-\frac{m_{2}^{4} g^{\mu2\mu3} q_{2}^{\mu1}}{8 (D-1) q_{2}^{4}}+\frac{g^{\mu1\mu3} q_{2}^{\mu2}}{8 (D-1)}\notag\\
    &+\frac{m_2^2 g^{\mu1\mu3} q_2^{\mu2}}{2 (D-1) q_2^2}-\frac{m_2^2 g^{\mu1\mu3} q_2^{\mu2}}{2 (D-1) D q_2^2}-\frac{m_2^4 g^{\mu1\mu3} q_2^{\mu2}}{8 (D-1) q_2^4}+\frac{g^{\mu1\mu2} q_2^{\mu3}}{8 (D-1)}+\frac{m_2^2 g^{\mu1\mu2} q_2^{\mu3}}{2 (D-1) q_2^2}\notag\\
    &-\frac{m_{2}^{2} g^{\mu1\mu2} q_{2}^{\mu3}}{2 (D-1) D q_{2}^{2}}-\frac{m_{2}^{4} g^{\mu1\mu2} q_{2}^{\mu3}}{8 (D-1) q_{2}^{4}}+\frac{7 D q_{2}^{\mu1} q_{2}^{\mu2} q_{2}^{\mu3}}{8 (D-1) q_{2}^{2}}-\frac{5 q_{2}^{\mu1} q_{2}^{\mu2} q_{2}^{\mu3}}{4 (D-1) q_{2}^{2}}\notag\\
    &-\frac{Dm_2^2q_2^{\mu1}q_2^{\mu2}q_2^{\mu3}}{2\left(D-1\right)q_2^4}-\frac{m_2^2q_2^{\mu1}q_2^{\mu2}q_2^{\mu3}}{2\left(D-1\right)q_2^4}+\frac{m_2^2q_2^{\mu1}q_2^{\mu2}q_2^{\mu3}}{\left(D-1\right)Dq_2^4}+\frac{Dm_2^4q_2^{\mu1}q_2^{\mu2}q_2^{\mu3}}{8\left(D-1\right)q_2^6}\notag\\
    &+\frac{m_2^4q_2^{\mu1}q_2^{\mu2}q_2^{\mu3}}{4(D-1)q_2^6},\\
    \notag \\
C^{\mu_1\mu_2\mu_3}_{2\to 1;\hat 2}=& \frac{g^{\mu_2 \mu_3} q_2^{\mu_1} m_1^4}{8(D-1) q_2^4}+\frac{g^{\mu_1 \mu_3} q_2^{\mu_2} m_1^4}{8(D-1) q_2^4}+\frac{g^{\mu_1 \mu_2} q_2^{\mu_3} m_1^4}{8(D-1) q_2^4}-\frac{D q_2^{\mu_1} q_2^{\mu_2} q_2^{\mu_3} m_1^4}{8(D-1) q_2^6}-\frac{q_2^{\mu_1} q_2^{\mu_2} q_2^{\mu_3} m_1^4}{4(D-1) q_2^6} \notag\\
& -\frac{g^{\mu_2 \mu_3} q_2^{\mu_1} m_1^2}{4(D-1) q_2^2}+\frac{g^{\mu_2 \mu_3} q_2^{\mu_1} m_1^2}{2(D-1) D q_2^2}-\frac{m_2^2 g^{\mu_2 \mu_3} q_2^{\mu_1} m_1^2}{4(D-1) q_2^4}-\frac{g^{\mu_1 \mu_3} q_2^{\mu_2} m_1^2}{4(D-1) q_2^2}+\frac{g^{\mu_1 \mu_3} q_2^{\mu_2} m_1^2}{2(D-1) D q_2^2} \notag\\
& -\frac{m_2^2 g^{\mu_1 \mu_3} q_2^{\mu_2} m_1^2}{4(D-1) q_2^4}-\frac{g^{\mu_1 \mu_2} q_2^{\mu_3} m_1^2}{4(D-1) q_2^2}+\frac{g^{\mu_1 \mu_2} q_2^{\mu_3} m_1^2}{2(D-1) D q_2^2}-\frac{m_2^2 g^{\mu_1 \mu_2} q_2^{\mu_3} m_1^2}{4(D-1) q_2^4}-\frac{D q_2^{\mu_1} q_2^{\mu_2} q_2^{\mu_3} m_1^2}{4(D-1) q_2^4} \notag\\
& +\frac{q_2^{\mu_1} q_2^{\mu_2} q_2^{\mu_3} m_1^2}{2(D-1) q_2^4}-\frac{q_2^{\mu_1} q_2^{\mu_2} q_2^{\mu_3} m_1^2}{(D-1) D q_2^4}+\frac{D m_2^2 q_2^{\mu_1} q_2^{\mu_2} q_2^{\mu_3 m_1^2}}{4(D-1) q_2^6}+\frac{m_2^2 q_2^{\mu_1} q_2^{\mu_2} q_2^{\mu_3} m_1^2}{2(D-1) q_2^6} \notag\\
& +\frac{g^{\mu_2 \mu_3} q_2^{\mu_1}}{8(D-1)}-\frac{m_2^2 g^{\mu_2 \mu_3} q_2^{\mu_1}}{4(D-1) q_2^2}+\frac{m_2^4 g^{\mu_2 \mu_3} q_2^{\mu_1}}{8(D-1) q_2^4}+\frac{g^{\mu_1 \mu_3} q_2^{\mu_2}}{8(D-1)}-\frac{m_2^2 g^{\mu_1 \mu_3} q_2^{\mu_2}}{4(D-1) q_2^2}+\frac{m_2^4 g^{\mu_1 \mu_3} q_2^{\mu_2}}{8(D-1) q_2^4} \notag\\
& +\frac{g^{\mu_1 \mu_2} q_2^{\mu_3}}{8(D-1)}-\frac{m_2^2 g^{\mu_1 \mu_2} q_2^{\mu_3}}{4(D-1) q_2^2}+\frac{m_2^4 g_1^{\mu_1 \mu_2} q_2^{\mu_3}}{8(D-1) q_2^4}-\frac{D q_2^{\mu_1} q_2^{\mu_2} q_2^{\mu_3}}{8(D-1) q_2^2}-\frac{q_2^{\mu_1} q_2^{\mu_2} q_2^{\mu_3}}{4(D-1) q_2^2} \notag\\
& +\frac{D m_2^2 q_2^{\mu_1} q_2^{\mu_2} q_2^{\mu_3}}{4(D-1) q_2^4}+\frac{m_2^2 q_2^{\mu_1} q_2^{\mu_2} q_2^{\mu_3}}{2(D-1) q_2^4}-\frac{D m_2^4 q_2^{\mu_1} q_2^{\mu_2} q_2^{\mu_3}}{8(D-1) q_2^6}-\frac{m_2^4 q_2^{\mu_1} q_2^{\mu_2} q_2^{\mu_3}}{4(D-1) q_2^6},\\
\notag \\
C^{\mu_1\mu_2\mu_3}_{2\to 2}=&-\frac{g^{\mu_{2}\mu_{3}} q_{2}^{\mu_{1}} m_{1}^{6}}{8 (D-1) q_{2}^{4}}-\frac{g^{\mu_{1}\mu_{3}} q_{2}^{\mu_{2}} m_{1}^{6}}{8 (D-1) q_{2}^{4}}-\frac{g^{\mu_{1}\mu_{2}} q_{2}^{\mu_{3}} m_{1}^{6}}{8 (D-1) q_{2}^{4}}+\frac{D q_{2}^{\mu_{1}} q_{2}^{\mu_{2}} q_{2}^{\mu_{3}} m_{1}^{6}}{8 (D-1) q_{2}^{6}}\notag\\
&+\frac{q_{2}{}^{\mu_{1}} q_{2}{}^{\mu_{2}} q_{2}{}^{\mu_{3}} m_{1}^{6}}{4 (D-1) q_{2}^{6}} + \frac{g^{\mu_{2}\mu_{3}} q_{2}{}^{\mu_{1}} m_{1}^{4}}{8 (D-1) q_{2}^{2}} + \frac{3 m_{2}^{2} g^{\mu_{2}\mu_{3}} q_{2}{}^{\mu_{1}} m_{1}^{4}}{8 (D-1) q_{2}^{4}} + \frac{g^{\mu_{1}\mu_{3}} q_{2}{}^{\mu_{2}} m_{1}^{4}}{8 (D-1) q_{2}^{2}}\notag\\
&+\frac{3 m_{2}^{2} g^{\mu1\mu3} q_{2}^{\mu2} m_{1}^{4}}{8 (D-1) q_{2}^{4}} + \frac{3 m_{2}^{2} g^{\mu1\mu2} q_{2}^{\mu3} m_{1}^{4}}{8 (D-1) q_{2}^{4}} + \frac{3 D q_{2}^{\mu1} q_{2}^{\mu2} q_{2}^{\mu3} m_{1}^{4}}{8 (D-1) q_{2}^{4}} - \frac{q_{2}^{2} g^{\mu2\mu3} q_{2}^{\mu1}}{8 (D-1)}\notag\\
&-\frac{3 q_{2}^{\mu_{1}} q_{2}^{\mu_{2}} q_{2}^{\mu_{3}} m_{1}^{4}}{4 (D-1) q_{2}^{4}}-\frac{3 D m_{2}^{2} q_{2}^{\mu_{1}} q_{2}^{\mu_{2}} q_{2}^{\mu_{3}} m_{1}^{4}}{8 (D-1) q_{2}^{6}}-\frac{3 m_{2}^{2} q_{2}^{\mu_{1}} q_{2}^{\mu_{2}} q_{2}^{\mu_{3}} m_{1}^{4}}{4 (D-1) q_{2}^{6}}\notag\\
&+\frac{g^{\mu_{2}\mu_{3}} q_{2}^{\mu_{1}} m_{1}^{2}}{8 (D-1)}+\frac{m_{2}^{2} g^{\mu_{2}\mu_{3}} q_{2}^{\mu_{1}} m_{1}^{2}}{4 (D-1) q_{2}^{2}}-\frac{3 m_{2}^{4} g^{\mu_{2}\mu_{3}} q_{2}^{\mu_{1}} m_{1}^{2}}{8 (D-1) q_{2}^{4}}+\frac{g^{\mu_{1}\mu_{3}} q_{2}^{\mu_{2}} m_{1}^{2}}{8 (D-1)}\notag\\
&+\frac{m_{2}^{2} g^{\mu_{1}\mu_{3}} q_{2}^{\mu_{2}} m_{1}^{2}}{4 (D-1) q_{2}^{2}}-\frac{3 m_{2}^{4} g^{\mu_{1}\mu_{3}} q_{2}^{\mu_{2}} m_{1}^{2}}{8 (D-1) q_{2}^{4}}+\frac{g^{\mu_{1}\mu_{2}} q_{2}^{\mu_{3}} m_{1}^{2}}{8 (D-1)}+\frac{m_{2}^{2} g^{\mu_{1}\mu_{2}} q_{2}^{\mu_{3}} m_{1}^{2}}{4 (D-1) q_{2}^{2}}\notag\\
&-\frac{3 m_{2}^{4} g^{\mu_{1}\mu_{2}} q_{2}^{\mu_{3}} m_{1}^{2}}{8 (D-1) q_{2}^{4}} + \frac{3 D q_{2}^{\mu_{1}} q_{2}^{\mu_{2}} q_{2}^{\mu_{3}} m_{1}^{2}}{8 (D-1) q_{2}^{2}} - \frac{3 q_{2}^{\mu_{1}} q_{2}^{\mu_{2}} q_{2}^{\mu_{3}} m_{1}^{2}}{4 (D-1) q_{2}^{2}} - \frac{q_{2}^{2} g^{\mu_{1}\mu_{2}} q_{2}^{\mu_{3}}}{8 (D-1) q_{2}^{2}}\notag\\
&-\frac{3 D m_{2}^{2} q_{2}^{\mu_{1}} q_{2}^{\mu_{2}} q_{2}^{\mu_{3}} m_{1}^{2}}{4 (D-1) q_{2}^{4}}+\frac{3 D m_{2}^{4} q_{2}^{\mu_{1}} q_{2}^{\mu_{2}} q_{2}^{\mu_{3}} m_{1}^{2}}{8 (D-1) q_{2}^{6}}+\frac{3 m_{2}^{4} q_{2}^{\mu_{1}} q_{2}^{\mu_{2}} q_{2}^{\mu_{3}} m_{1}^{2}}{4 (D-1) q_{2}^{6}}\notag\\
&+\frac{3 m_{2}^{2} g^{\mu_{2}\mu_{3}} q_{2}^{\mu_{1}}}{8 (D-1)} - \frac{3 m_{2}^{4} g^{\mu_{2}\mu_{3}} q_{2}^{\mu_{1}}}{8 (D-1) q_{2}^{2}} + \frac{m_{2}^{6} g^{\mu_{2}\mu_{3}} q_{2}^{\mu_{1}}}{8 (D-1) q_{2}^{4}} + \frac{3 m_{2}^{2} g^{\mu_{1}\mu_{3}} q_{2}^{\mu_{2}}}{8 (D-1)}\notag\\
&- \frac{q_{2}^{2} g^{\mu_{1}\mu_{3}} q_{2}^{\mu_{2}}}{8 (D-1)} + \frac{m_{2}^{6} g^{\mu_{1}\mu_{3}} q_{2}^{\mu_{2}}}{8 (D-1) q_{2}^{4}} + \frac{3 m_{2}^{2} g^{\mu_{1}\mu_{2}} q_{2}^{\mu_{3}}}{8 (D-1)} + \frac{q_{2}^{\mu_{1}} q_{2}^{\mu_{2}} q_{2}^{\mu_{3}}}{4 (D-1)}\notag\\
&-\frac{3 m_{2}^{4} g^{\mu_{1}\mu_{2}} q_{2}^{\mu_{3}}}{8 (D-1) q_{2}^{2}} + \frac{m_{2}^{6} g^{\mu_{1}\mu_{2}} q_{2}^{\mu_{3}}}{8 (D-1) q_{2}^{4}} + \frac{D q_{2}^{\mu_{1}} q_{2}^{\mu_{2}} q_{2}^{\mu_{3}}}{8 (D-1)} - \frac{m_{2}^{6} q_{2}^{\mu_{1}} q_{2}^{\mu_{2}} q_{2}^{\mu_{3}}}{4 (D-1) q_{2}^{6}}\notag\\
&-\frac{3 D m_{2}^{2} q_{2}^{\mu_{1}} q_{2}^{\mu_{2}} q_{2}^{\mu_{3}}}{8 (D-1) q_{2}^{2}}-\frac{3 m_{2}^{2} q_{2}^{\mu_{1}} q_{2}^{\mu_{2}} q_{2}^{\mu_{3}}}{4 (D-1) q_{2}^{2}}+\frac{3 D m_{2}^{4} q_{2}^{\mu_{1}} q_{2}^{\mu_{2}} q_{2}^{\mu_{3}}}{8 (D-1) q_{2}^{4}}\notag\\
&+\frac{3 m_{2}^{4} q_{2}^{\mu_{1}} q_{2}^{\mu_{2}} q_{2}^{\mu_{3}}}{4 (D-1) q_{2}^{4}}-\frac{D m_{2}^{6} q_{2}^{\mu_{1}} q_{2}^{\mu_{2}} q_{2}^{\mu_{3}}}{8 (D-1) q_{2}^{6}}+\frac{g^{\mu_{1}\mu_{2}} q_{2}^{\mu_{3}} m_{1}^{4}}{8 (D-1) q_{2}^{2}}-\frac{3 m_{2}^{4} g^{\mu_{1}\mu_{3}} q_{2}^{\mu_{2}}}{8 (D-1) q_{2}^{2}}.
\end{align}
Next, due to the length of the expression, we provide a semi-numerical, semi-analytical verification for the sufficiently complex case of $r = 8$. The results are presented in blocks according to the arrangement of Lorentz indices. We choose the numerical values for the kinematic variables as follows:

\begin{equation}
   m_1=1.414,\;m_2=3.14,\;D=6.23.
\end{equation}
For r=8, there are totally 5 kinds of PV tensor blocks, we view $q_2^2$ as a parameter to check it with \texttt{ApmRed}\cite{Chen:2024xwt}. For $C^{\mu_1\mu_2\mu_3\mu_4\mu_5\mu_6\mu_7\mu_8}_{2\to 1;\hat 1}$, the coefficient of $q_2^{\mu_1}q_2^{\mu_2}q_2^{\mu_3}q_2^{\mu_4}q_2^{\mu_5}q_2^{\mu_6}q_2^{\mu_7}q_2^{\mu_8}$ is:
\begin{equation}
  \begin{aligned}
& \frac{0.948681}{q_2^2}-\frac{22.8976}{q_2^4}+\frac{334.613}{q_2^6}-\frac{3064.23}{q_2^8}+ \frac{17581.5}{q_2^{10}}-\frac{61010.9}{q_2^{12}}+\frac{116431 .}{q_2^{14}}-\frac{93365.4}{q_2^{16}}.
\end{aligned}
\end{equation}
The coefficient of $q_2^{\mu_1}q_2^{\mu_2}q_2^{\mu_3}q_2^{\mu_4}q_2^{\mu_5}q_2^{\mu_6}g^{\mu_7\mu_8}$ is:
\begin{equation}
\begin{aligned}&0.00281507+\frac{1.35803}{q_2^{2}}-\frac{20.123}{q_2^{4}}+\frac{185.017}{q_2^{6}}\\&-\frac{1053.98}{q_2^{8}}+\frac{3589.56}{q_2^{10}}-\frac{6648.29}{q_2^{12}}+\frac{5121.52}{q_2^{14}}.
\end{aligned}
\end{equation}
The coefficient of $q_2^{\mu_1}q_2^{\mu_2}q_2^{\mu_3}q_2^{\mu_4}g^{\mu_5\mu_6}g^{\mu_7\mu_8}$ is:
\begin{equation}
   \begin{aligned}
&-0.000173448q_{2}^{2}+0.0145338+\frac{1.39462}{q_{2}^{2}}\\
&-\frac{12.894}{q_{2}^{4}}+\frac{72.7341}{q_{2}^{6}}-\frac{241.521}{q_{2}^{8}}+\frac{430.272}{q_{2}^{10}}-\frac{315.559}{q_{2}^{12}}.
\end{aligned}
\end{equation}
The coefficient of $q_2^{\mu_1}q_2^{\mu_2}g^{\mu_3\mu_4}g^{\mu_5\mu_6}g^{\mu_7\mu_8}$ is:
\begin{equation}
    \begin{aligned}&0.0000121889q_2^{4}-0.00108637q_2^{2}+0.0394904\\
    +&\frac{1.0692}{q_2^{2}}-\frac{5.94302}{q_2^{4}}+\frac{19.0244}{q_2^{6}}-\frac{32.162}{q_2^{8}}+\frac{22.1756}{q_2^{10}}.\end{aligned}
\end{equation}
The coefficient of $g^{\mu_1\mu_2}g^{\mu_3\mu_4}g^{\mu_5\mu_6}g^{\mu_7\mu_8}$ is:
\begin{equation}
   \begin{aligned}
-&9.96641\times10^{-7}q_2^{6}+0.0000962243q_2^{4}-0.00376731q_2^{2}\\
+& 0.076699+\frac{0.601261}{q_2^2}-\frac{1.81489}{q_2^4}+\frac{2.84872}{q_2^6}-\frac{1.81322}{q_2^8}.\end{aligned}
\end{equation}
For $C^{\mu_1\mu_2\mu_3\mu_4\mu_5\mu_6\mu_7\mu_8}_{2\to 1;\hat 2}$, the coefficient of $q_2^{\mu_1}q_2^{\mu_2}q_2^{\mu_3}q_2^{\mu_4}q_2^{\mu_5}q_2^{\mu_6}q_2^{\mu_7}q_2^{\mu_8}$ is:
\begin{equation}
   -\frac{0.0513186}{q_2^2}+\frac{3.38652}{q_2^4}-\frac{90.3351}{q_2^6}+\frac{1258.24}{q_2^8}-\frac{9863.62}{q_2^{10}}+\frac{43518.7}{q_2^{12}}-\frac{100258.}{q_2^{14}}+\frac{93365.4}{q_2^{16}}. 
\end{equation}
the coefficient of $q_2^{\mu_1}q_2^{\mu_2}q_2^{\mu_3}q_2^{\mu_4}q_2^{\mu_5}q_2^{\mu_6}g^{\mu_7\mu_8}$ is:
\begin{equation}
    \begin{aligned}
& 0.00281507-\frac{0.1946}{q_2^2}+\frac{5.36533}{q_2^4}-\frac{76.0343}{q_2^6} \\
+&\frac{596.05}{q_2^8}-\frac{2584.73}{q_2^{10}}+\frac{5761.12}{q_2^{12}}-\frac{5121.52}{q_2^{14}}.
\end{aligned}
\end{equation}
The coefficient of $q_2^{\mu_1}q_2^{\mu_2}q_2^{\mu_3}q_2^{\mu_4}g^{\mu_5\mu_6}g^{\mu_7\mu_8}$ is:
\begin{equation}
    \begin{aligned}&0.0126874-0.000173448q_2^2-\frac{0.365406}{q_2^2}\\ +&\frac{5.30475}{q_2^4}-\frac{41.5851}{q_2^6}+\frac{176.033}{q_2^8}-\frac{375.61}{q_2^{10}}+\frac{315.559}{q_2^{12}}.\end{aligned}
\end{equation}
The coefficient of $q_2^{\mu_1}q_2^{\mu_2}g^{\mu_3\mu_4}g^{\mu_5\mu_6}g^{\mu_7\mu_8}$ is:
\begin{equation}
    \begin{aligned}&0.0292453+0.0000121889q_{2}^{4}-0.000956619q_{2}^{2}\\
    -&\frac{0.440584}{q_{2}^{2}}+\frac{3.45384}{q_{2}^{4}}-\frac{14.0888}{q_{2}^{6}}+\frac{28.3207}{q_{2}^{8}}-\frac{22.1756}{q_{2}^{10}}.\end{aligned}
\end{equation}
The coefficient of $g^{\mu_1\mu_2}g^{\mu_3\mu_4}g^{\mu_5\mu_6}g^{\mu_7\mu_8}$ is:
\begin{equation}
    \begin{aligned}&0.0458058-9.96641\times10^{-7}q_{2}^{6}+0.0000856148q_{2}^{4}\\
    -&0.00285088q_{2}^{2}-\frac{0.359082}{q_{2}^{2}}+\frac{1.3734}{q_{2}^{4}}-\frac{2.53463}{q_{2}^{6}}+\frac{1.81322}{q_{2}^{8}}.\end{aligned}
\end{equation}
For $C^{\mu_1\mu_2\mu_3\mu_4\mu_5\mu_6\mu_7\mu_8}_{2\to 2}$, the coefficient of $q_2^{\mu_1}q_2^{\mu_2}q_2^{\mu_3}q_2^{\mu_4}q_2^{\mu_5}q_2^{\mu_6}q_2^{\mu_7}q_2^{\mu_8}$ is:
\begin{equation}
    \begin{aligned}
        &0.0513186-\frac{3.85539}{q_2^2}+\frac{120.644}{q_2^4}-\frac{2048.95}{q_2^6}\\
        +&\frac{20623.8}{q_2^8}-\frac{125915.}{q_2^{10}}+\frac{455615.}{q_2^{12}}-\frac{894758.}{q_2^{14}}+\frac{731912.8.}{q_2^{16}}.
    \end{aligned}
\end{equation}
The coefficient of $q_2^{\mu_1}q_2^{\mu_2}q_2^{\mu_3}q_2^{\mu_4}q_2^{\mu_5}q_2^{\mu_6}g^{\mu_7\mu_8}$ is:
\begin{equation}
    \begin{aligned}
        &0.220319-0.00281507q_2^2-\frac{7.10862}{q_2^2}+\frac{123.046}{q_2^4}\\
        -&\frac{1246.55}{q_2^6} +\frac{7561.59}{q_2^8} -\frac{26845.9}{q_2^{10}}+\frac{51131.7}{q_2^{12}}-\frac{40148.8}{q_2^{14}}.
    \end{aligned}
\end{equation}
The coefficient of $q_2^{\mu_1}q_2^{\mu_2}q_2^{\mu_3}q_2^{\mu_4}g^{\mu_5\mu_6}g^{\mu_7\mu_8}$ is:
\begin{equation}
    \begin{aligned}
    &0.479188+0.000173448q_2^4-0.0142721q_2^2-\frac{8.51093}{q_2^2}\\
    +&\frac{86.9964}{q_2^4}-\frac{523.026}{q_2^6}+\frac{1809.66}{q_2^8}-\frac{3312.26}{q_2^{10}}+\frac{2473.74}{q_2^{12}}.
    \end{aligned}
\end{equation}
The coefficient of $q_2^{\mu_1}q_2^{\mu_2}g^{\mu_3\mu_4}g^{\mu_5\mu_6}g^{\mu_7\mu_8}$ is:
\begin{equation}
    \begin{aligned}
     &0.697682-0.0000121889q_2^6+0.00106798q_2^4-0.0378354q_2^2\\
     -&\frac{7.22953}{q_2^2}+\frac{42.8749}{q_2^4}-\frac{142.886}{q_2^6}+\frac{247.857}{q_2^8}-\frac{173.84}{q_2^{10}}.
     \end{aligned}
\end{equation}
The coefficient of $g^{\mu_1\mu_2}g^{\mu_3\mu_4}g^{\mu_5\mu_6}g^{\mu_7\mu_8}$ is:
\begin{equation}
\begin{aligned}&0.755126+9.96641\times10^{-7}q_{2}^{8}-0.0000947206q_{2}^{6}+0.00362082q_{2}^{4}\\
-&0.0709358q_{2}^{2}-\frac{4.35925}{q_{2}^{2}}+\frac{13.6741}{q_{2}^{4}}-\frac{21.9827}{q_{2}^{6}}+\frac{14.2142}{q_{2}^{8}}.\end{aligned}
\end{equation}

\subsection{Triangle}

In this subsection, we present a semi-numerical verification for the triangle diagram. Due to space limitations, we only provide an example with $r = 5$. With $q_1 = 0$, the number of PV tensor blocks is 12. We assign numerical values to all scalar products of external momentum, masses, and the spacetime dimension $D$, and list the coefficients of all PV tensor blocks calculated using our method alongside the results from \texttt{AmpRed}. The Feynman integral is 

\begin{equation}
        I^{\mu_1\mu_2\cdots\mu_5}_1=\int d^Dl \frac{l^{\mu_1}l^{\mu_2}l^{\mu_3}\cdots l^{\mu_5}}{(l^2-m_1^2)((l-q_2)^2-m_2^2)((l-q_3)^2-m_3^2)}.
    \label{triangle}
\end{equation}
We choose the numeric value of the kinetic variable as follows:
\begin{equation}
    m_1=\frac14,\quad m_2=\frac54,\quad m_3=\frac32,\quad q_2^2=\frac13,\quad q_2\cdot q_3=\frac12,\quad q_3^2=1,\quad D=6.23.
\end{equation}
For $C^{\mu_1\mu_2\mu_3\mu_4\mu_5}_{3\to 1;\widehat{1,2}}$, the corresponding coefficients are shown in Table \ref{tab:3 to 1}.
\begin{table}[H]
\centering
\begin{tabular}{|c|c|c|}
\hline
Tensor Blocks                                                                & Generating Functing & AmpRed    \\ \hline
$\quad\quad q_2{}^{\mu _1} q_2{}^{\mu _2} q_2{}^{\mu _3} q_2{}^{\mu _4} q_2{}^{\mu _5}\quad\quad$ & 573.182             &$\quad\quad$ 573.182 $\quad\quad$  \\ \hline
$q_3{}^{\mu _1} q_2{}^{\mu _2} q_2{}^{\mu _3} q_2{}^{\mu _4} q_2{}^{\mu _5}$ & -227.739            & -227.739  \\ \hline
$q_3{}^{\mu _1} q_3{}^{\mu _2} q_2{}^{\mu _3} q_2{}^{\mu _4} q_2{}^{\mu _5}$ & 94.7168             & 94.7168   \\ \hline
$q_3{}^{\mu _1} q_3{}^{\mu _2} q_2{}^{\mu _3} q_3{}^{\mu _4} q_2{}^{\mu _5}$ & -39.904             & -39.904   \\ \hline
$q_3{}^{\mu _1} q_3{}^{\mu _2} q_3{}^{\mu _3} q_3{}^{\mu _4} q_2{}^{\mu _5}$ & 16.5202             & 16.5202   \\ \hline
$q_2{}^{\mu _3} q_2{}^{\mu _4} q_2{}^{\mu _5} g^{\mu _1\mu _2}$              & -4.83697            & -4.83697  \\ \hline
$q_3{}^{\mu _3} q_2{}^{\mu _4} q_2{}^{\mu _5} g^{\mu _1\mu _2}$              & 1.64732             & 1.64732   \\ \hline
$q_2{}^{\mu _5} g^{\mu _1\mu _4} g^{\mu _2\mu _3}$                           & 0.0124851           & 0.0124851 \\ \hline
$q_3{}^{\mu _1} q_3{}^{\mu _3} q_2{}^{\mu _5} g^{\mu _2\mu _4}$              & -0.509975           & -0.509975 \\ \hline
$q_3{}^{\mu _1} q_3{}^{\mu _2} q_3{}^{\mu _3} q_3{}^{\mu _4} q_3{}^{\mu _5}$ & 6.53548             & 6.53548   \\ \hline
$q_3{}^{\mu _3} q_3{}^{\mu _4} q_3{}^{\mu _5} g^{\mu _1\mu _2}$              & 0.141478            & 0.141478  \\ \hline
$q_3{}^{\mu _1} g^{\mu _2\mu _3} g^{\mu _4\mu _5}$                           & 0.0295967           & 0.0295967 \\ \hline

\end{tabular}
\caption{Numerical Check of $C^{\mu_1\mu_2\mu_3\mu_4\mu_5}_{3\to 1;\widehat{1,2}}$}
\label{tab:3 to 1}
\end{table}

For $C^{\mu_1\mu_2\mu_3\mu_4\mu_5}_{3\to 2;\hat {3}}$, the corresponding coefficients are showed in Table \ref{3 to 2}.
\begin{table}[H]
\centering
\begin{tabular}{|c|c|c|}
\hline
Tensor Blocks                                                                & Generating Functing & AmpRed     \\ \hline
$\quad\quad q_2{}^{\mu _1} q_2{}^{\mu _2} q_2{}^{\mu _3} q_2{}^{\mu _4} q_2{}^{\mu _5}\quad\quad$ & 3033.21             &$\quad\quad$ 3033.21$\quad\quad$    \\ \hline
$q_3{}^{\mu _1} q_2{}^{\mu _2} q_2{}^{\mu _3} q_2{}^{\mu _4} q_2{}^{\mu _5}$ & -1214.4             & -1214.4    \\ \hline
$q_3{}^{\mu _1} q_3{}^{\mu _2} q_2{}^{\mu _3} q_2{}^{\mu _4} q_2{}^{\mu _5}$ & 506.02              & 506.02     \\ \hline
$q_3{}^{\mu _1} q_3{}^{\mu _2} q_2{}^{\mu _3} q_3{}^{\mu _4} q_2{}^{\mu _5}$ & -219.986            & -219.986   \\ \hline
$q_3{}^{\mu _1} q_3{}^{\mu _2} q_3{}^{\mu _3} q_3{}^{\mu _4} q_2{}^{\mu _5}$ & 98.4162             & 98.4162    \\ \hline
$q_2{}^{\mu _3} q_2{}^{\mu _4} q_2{}^{\mu _5} g^{\mu _1\mu _2}$              & -24.2693            & -24.2693   \\ \hline
$q_3{}^{\mu _3} q_2{}^{\mu _4} q_2{}^{\mu _5} g^{\mu _1\mu _2}$              & 9.51377             & 9.51377    \\ \hline
$q_2{}^{\mu _5} g^{\mu _1\mu _4} g^{\mu _2\mu _3}$                           & 0.223267            & 0.223267   \\ \hline
$q_3{}^{\mu _1} q_3{}^{\mu _3} q_2{}^{\mu _5} g^{\mu _2\mu _4}$              & -3.77096            & -3.77096   \\ \hline
$q_3{}^{\mu _1} q_3{}^{\mu _2} q_3{}^{\mu _3} q_3{}^{\mu _4} q_3{}^{\mu _5}$ & 45.5059             & 45.5059    \\ \hline
$q_3{}^{\mu _3} q_3{}^{\mu _4} q_3{}^{\mu _5} g^{\mu _1\mu _2}$              & 1.63244             & 1.63244    \\ \hline
$q_3{}^{\mu _1} g^{\mu _2\mu _3} g^{\mu _4\mu _5}$                           & -0.0977989          & -0.0977989 \\ \hline
\end{tabular}
\caption{Numerical Check of $C^{\mu_1\mu_2\mu_3\mu_4\mu_5}_{3\to 2;\hat {3}}$}
\label{3 to 2}
\end{table}

For $C^{\mu_1\mu_2\mu_3\mu_4\mu_5}_{3\to 3}$, the corresponding coefficients are showed in Table \ref{3 to 3}.
\begin{table}[H]
\centering
\begin{tabular}{|c|c|c|}
\hline
Tensor Blocks                                                                & Generating Functing & AmpRed    \\ \hline
$\quad\quad q_2{}^{\mu _1} q_2{}^{\mu _2} q_2{}^{\mu _3} q_2{}^{\mu _4} q_2{}^{\mu _5}\quad\quad$ & -2406.7             &$\quad\quad$ -2406.7 $\quad\quad$  \\ \hline
$q_3{}^{\mu _1} q_2{}^{\mu _2} q_2{}^{\mu _3} q_2{}^{\mu _4} q_2{}^{\mu _5}$ & 952.399             & 952.399   \\ \hline
$q_3{}^{\mu _1} q_3{}^{\mu _2} q_2{}^{\mu _3} q_2{}^{\mu _4} q_2{}^{\mu _5}$ & -398.063              & -398.063  \\ \hline
$q_3{}^{\mu _1} q_3{}^{\mu _2} q_2{}^{\mu _3} q_3{}^{\mu _4} q_2{}^{\mu _5}$ &  172.741           & 172.741   \\ \hline
$q_3{}^{\mu _1} q_3{}^{\mu _2} q_3{}^{\mu _3} q_3{}^{\mu _4} q_2{}^{\mu _5}$ & -77.4225             & -77.4225  \\ \hline
$q_2{}^{\mu _3} q_2{}^{\mu _4} q_2{}^{\mu _5} g^{\mu _1\mu _2}$              & 19.8714            & 19.8714   \\ \hline
$q_3{}^{\mu _3} q_2{}^{\mu _4} q_2{}^{\mu _5} g^{\mu _1\mu _2}$              & -7.38205             & -7.38205  \\ \hline
$q_2{}^{\mu _5} g^{\mu _1\mu _4} g^{\mu _2\mu _3}$                           & -0.211962            & -0.211962 \\ \hline
$q_3{}^{\mu _1} q_3{}^{\mu _3} q_2{}^{\mu _5} g^{\mu _2\mu _4}$              & 2.99136            & 2.99136   \\ \hline
$q_3{}^{\mu _1} q_3{}^{\mu _2} q_3{}^{\mu _3} q_3{}^{\mu _4} q_3{}^{\mu _5}$ &35.6139             & 35.6139   \\ \hline
$q_3{}^{\mu _3} q_3{}^{\mu _4} q_3{}^{\mu _5} g^{\mu _1\mu _2}$              & -1.26151             & -1.26151  \\ \hline
$q_3{}^{\mu _1} g^{\mu _2\mu _3} g^{\mu _4\mu _5}$                           & 0.0693693          & 0.0693693 \\ \hline
\end{tabular}
\caption{Numerical Check of $C^{\mu_1\mu_2\mu_3\mu_4\mu_5}_{3\to 3}$}
\label{3 to 3}
\end{table}

\subsection{Box}
Finally, we present the most complex example: the reduction of the tensor box. Due to space constraints, we only provide two reduction coefficients with $r = 6$. With $q_1 = 0$, the number of PV tensor blocks is 50. We select 10 representative blocks for numerical verification and display their results, which we believe sufficiently demonstrate the correctness of the method and program. The corresponding Feynman integral is 
\begin{equation}
        I^{\mu_1\mu_2\cdots\mu_6}_1=\int d^Dl \frac{l^{\mu_1}l^{\mu_2}l^{\mu_3}\cdots l^{\mu_6}}{(l^2-m_1^2)((l-q_2)^2-m_2^2)((l-q_3)^2-m_3^2)((l-q_4)^2-m_4^2)}.
    \label{box}
\end{equation}
We choose the numeric value of the kinetic variable as follows:
\begin{equation}
\begin{aligned}
    &m_1=\frac14,\quad m_2=\frac54,\quad m_3=\frac32,\quad m_4=\frac34,\quad q_2^2=\frac13,\quad q_2\cdot q_3=\frac12,\quad\\& q_2\cdot q_4=\frac56,\quad q_3^2=1,\quad q_3\cdot q_4=\frac78,\quad q_4^2=\frac{10}{11},\quad D=6.23.
\end{aligned}
\end{equation}
For $C^{\mu_1\mu_2\mu_3\mu_4\mu_5\mu_6}_{4\to 4}$, the corresponding coefficients are showed in Table \ref{4 to 4}.

\begin{table}[H]
\centering
\begin{tabular}{|c|c|c|}
\hline
Tensor Blocks                                                                               & Generating Functing & AmpRed   \\ \hline
$\quad\quad q_3{}^{\mu _1} q_2{}^{\mu _2} q_2{}^{\mu _3} q_2{}^{\mu _4} q_2{}^{\mu _5} q_2{}^{\mu _6}\quad\quad$ & 49.2639             & $\quad\quad$49.2639 $\quad\quad$ \\ \hline
$q_2{}^{\mu _1} q_2{}^{\mu _2} q_2{}^{\mu _3} q_2{}^{\mu _4} q_2{}^{\mu _5} q_2{}^{\mu _6}$ & 303.803             & 303.803  \\ \hline
$q_4{}^{\mu _1} q_2{}^{\mu _2} q_2{}^{\mu _3} q_2{}^{\mu _4} q_2{}^{\mu _5} q_2{}^{\mu _6}$ & -301.998            & -301.998 \\ \hline
$q_3{}^{\mu _1} q_3{}^{\mu _2} q_2{}^{\mu _3} q_2{}^{\mu _4} q_2{}^{\mu _5} q_2{}^{\mu _6}$ & -48.0962            & -48.0962 \\ \hline
$q_3{}^{\mu _1} q_4{}^{\mu _2} q_4{}^{\mu _3} q_2{}^{\mu _4} q_2{}^{\mu _5} q_2{}^{\mu _6}$ & -31.42             & -31.42  \\ \hline
$q_4{}^{\mu _1} q_4{}^{\mu _2} q_2{}^{\mu _3} q_2{}^{\mu _4} q_2{}^{\mu _5} q_2{}^{\mu _6}$ & 236.403             & 236.403  \\ \hline
$q_2{}^{\mu _1} q_2{}^{\mu _2} q_3{}^{\mu _3} q_4{}^{\mu _4} q_2{}^{\mu _5} q_2{}^{\mu _6}$ & 3.89855             & 3.89855  \\ \hline
$q_2{}^{\mu _3} q_2{}^{\mu _4} q_2{}^{\mu _5} q_2{}^{\mu _6} g^{\mu _1\mu _2}$              & 12.7579             & 12.7579  \\ \hline
$q_2{}^{\mu _5} q_2{}^{\mu _6} g^{\mu _1\mu _2} g^{\mu _3\mu _4}$                           & 0.472526            & 0.472526 \\ \hline
$g^{\mu _1\mu _2} g^{\mu _3\mu _4} g^{\mu _5\mu _6}$                                        & 0.015153            & 0.015153 \\ \hline
\end{tabular}
\caption{Numerical Check of $C^{\mu_1\mu_2\mu_3\mu_4\mu_5\mu_6}_{4\to 4}$}
\label{4 to 4}
\end{table}

For $C^{\mu_1\mu_2\mu_3\mu_4\mu_5}_{4\to 1;\widehat {1,2,3}}$, the corresponding coefficients are showed in Table \ref{4 to 1}.
\begin{table}[H]
\centering
\begin{tabular}{|c|c|c|}
\hline
Tensor Blocks                                                                               & Generating Functing & AmpRed     \\ \hline
$\quad\quad q_3{}^{\mu _1} q_2{}^{\mu _2} q_2{}^{\mu _3} q_2{}^{\mu _4} q_2{}^{\mu _5} q_2{}^{\mu _6}\quad\quad$ & 49.2639             &$\quad\quad$ 5.23013 $\quad\quad$   \\ \hline
$q_2{}^{\mu _1} q_2{}^{\mu _2} q_2{}^{\mu _3} q_2{}^{\mu _4} q_2{}^{\mu _5} q_2{}^{\mu _6}$ & 303.803             & -1.12111   \\ \hline
$q_4{}^{\mu _1} q_2{}^{\mu _2} q_2{}^{\mu _3} q_2{}^{\mu _4} q_2{}^{\mu _5} q_2{}^{\mu _6}$ & -301.998            & 23.9192    \\ \hline
$q_3{}^{\mu _1} q_3{}^{\mu _2} q_2{}^{\mu _3} q_2{}^{\mu _4} q_2{}^{\mu _5} q_2{}^{\mu _6}$ & -48.0962            & -7.54439   \\ \hline
$q_3{}^{\mu _1} q_4{}^{\mu _2} q_4{}^{\mu _3} q_2{}^{\mu _4} q_2{}^{\mu _5} q_2{}^{\mu _6}$ & 3.89855             & 18.2918    \\ \hline
$q_4{}^{\mu _1} q_4{}^{\mu _2} q_2{}^{\mu _3} q_2{}^{\mu _4} q_2{}^{\mu _5} q_2{}^{\mu _6}$ & 236.403             & -16.3736   \\ \hline
$q_2{}^{\mu _1} q_2{}^{\mu _2} q_3{}^{\mu _3} q_4{}^{\mu _4} q_2{}^{\mu _5} q_2{}^{\mu _6}$ & 3.89855             & -10.9583   \\ \hline
$q_2{}^{\mu _3} q_2{}^{\mu _4} q_2{}^{\mu _5} q_2{}^{\mu _6} g^{\mu _1\mu _2}$              & 12.7579             & 0.874729   \\ \hline
$q_2{}^{\mu _5} q_2{}^{\mu _6} g^{\mu _1\mu _2} g^{\mu _3\mu _4}$                           & 0.472526            & 0.0052978  \\ \hline
$g^{\mu _1\mu _2} g^{\mu _3\mu _4} g^{\mu _5\mu _6}$                                        & 0.015153            & -0.0148764 \\ \hline
\end{tabular}
\caption{Numerical Check of $C^{\mu_1\mu_2\mu_3\mu_4\mu_5}_{4\to 1;\widehat {1,2,3}}$}
\label{4 to 1}
\end{table}
 \end{appendices}

	\appendix
	

	\bibliographystyle{JHEP}
    \bibliography{reference.bib}

\providecommand{\href}[2]{#2}\begingroup\raggedright\begin{thebibliography}{10}

\bibitem{Hu:2023mgc}
C.~Hu, T.~Li, J.~Shen, and Y.~Xu, {\it {An explicit expression of generating function for one-loop tensor reduction}},  {\em JHEP} {\bf 02} (2024) 158, [\href{http://arxiv.org/abs/2308.13336}{{\tt arXiv:2308.13336}}]. [Erratum: JHEP 07, 068 (2024)].

\bibitem{Ossola:2006us}
G.~Ossola, C.~G. Papadopoulos, and R.~Pittau, {\it {Reducing full one-loop amplitudes to scalar integrals at the integrand level}},  {\em Nucl. Phys. B} {\bf 763} (2007) 147--169, [\href{http://arxiv.org/abs/hep-ph/0609007}{{\tt hep-ph/0609007}}].

\bibitem{Mastrolia:2011pr}
P.~Mastrolia and G.~Ossola, {\it {On the Integrand-Reduction Method for Two-Loop Scattering Amplitudes}},  {\em JHEP} {\bf 11} (2011) 014, [\href{http://arxiv.org/abs/1107.6041}{{\tt arXiv:1107.6041}}].

\bibitem{Badger:2012dp}
S.~Badger, H.~Frellesvig, and Y.~Zhang, {\it {Hepta-Cuts of Two-Loop Scattering Amplitudes}},  {\em JHEP} {\bf 04} (2012) 055, [\href{http://arxiv.org/abs/1202.2019}{{\tt arXiv:1202.2019}}].

\bibitem{Zhang:2012ce}
Y.~Zhang, {\it {Integrand-Level Reduction of Loop Amplitudes by Computational Algebraic Geometry Methods}},  {\em JHEP} {\bf 09} (2012) 042, [\href{http://arxiv.org/abs/1205.5707}{{\tt arXiv:1205.5707}}].

\bibitem{Larsen:2015ped}
K.~J. Larsen and Y.~Zhang, {\it {Integration-by-parts reductions from unitarity cuts and algebraic geometry}},  {\em Phys. Rev. D} {\bf 93} (2016), no.~4 041701, [\href{http://arxiv.org/abs/1511.01071}{{\tt arXiv:1511.01071}}].

\bibitem{Zhang:2016kfo}
Y.~Zhang, {\it {Lecture Notes on Multi-loop Integral Reduction and Applied Algebraic Geometry}},  12, 2016.
\newblock \href{http://arxiv.org/abs/1612.02249}{{\tt arXiv:1612.02249}}.

\bibitem{Passarino:1978jh}
G.~Passarino and M.~J.~G. Veltman, {\it {One Loop Corrections for e+ e- Annihilation Into mu+ mu- in the Weinberg Model}},  {\em Nucl. Phys. B} {\bf 160} (1979) 151--207.

\bibitem{Chetyrkin:1981qh}
K.~G. Chetyrkin and F.~V. Tkachov, {\it {Integration by Parts: The Algorithm to Calculate beta Functions in 4 Loops}},  {\em Nucl. Phys. B} {\bf 192} (1981) 159--204.

\bibitem{Tkachov:1981wb}
F.~V. Tkachov, {\it {A Theorem on Analytical Calculability of Four Loop Renormalization Group Functions}},  {\em Phys. Lett. B} {\bf 100} (1981) 65--68.

\bibitem{Laporta:2000dsw}
S.~Laporta, {\it {High precision calculation of multiloop Feynman integrals by difference equations}},  {\em Int. J. Mod. Phys. A} {\bf 15} (2000) 5087--5159, [\href{http://arxiv.org/abs/hep-ph/0102033}{{\tt hep-ph/0102033}}].

\bibitem{vonManteuffel:2012np}
A.~von Manteuffel and C.~Studerus, {\it {Reduze 2 - Distributed Feynman Integral Reduction}},  \href{http://arxiv.org/abs/1201.4330}{{\tt arXiv:1201.4330}}.

\bibitem{Maierhofer:2017gsa}
P.~Maierh\"ofer, J.~Usovitsch, and P.~Uwer, {\it {Kira\textemdash{}A Feynman integral reduction program}},  {\em Comput. Phys. Commun.} {\bf 230} (2018) 99--112, [\href{http://arxiv.org/abs/1705.05610}{{\tt arXiv:1705.05610}}].

\bibitem{Smirnov:2019qkx}
A.~V. Smirnov and F.~S. Chuharev, {\it {FIRE6: Feynman Integral REduction with Modular Arithmetic}},  {\em Comput. Phys. Commun.} {\bf 247} (2020) 106877, [\href{http://arxiv.org/abs/1901.07808}{{\tt arXiv:1901.07808}}].

\bibitem{vonManteuffel:2014ixa}
A.~von Manteuffel and R.~M. Schabinger, {\it {A novel approach to integration by parts reduction}},  {\em Phys. Lett. B} {\bf 744} (2015) 101--104, [\href{http://arxiv.org/abs/1406.4513}{{\tt arXiv:1406.4513}}].

\bibitem{Bern:1994cg}
Z.~Bern, L.~J. Dixon, D.~C. Dunbar, and D.~A. Kosower, {\it {Fusing gauge theory tree amplitudes into loop amplitudes}},  {\em Nucl. Phys. B} {\bf 435} (1995) 59--101, [\href{http://arxiv.org/abs/hep-ph/9409265}{{\tt hep-ph/9409265}}].

\bibitem{Bern:1994zx}
Z.~Bern, L.~J. Dixon, D.~C. Dunbar, and D.~A. Kosower, {\it {One loop n point gauge theory amplitudes, unitarity and collinear limits}},  {\em Nucl. Phys. B} {\bf 425} (1994) 217--260, [\href{http://arxiv.org/abs/hep-ph/9403226}{{\tt hep-ph/9403226}}].

\bibitem{Britto:2004nc}
R.~Britto, F.~Cachazo, and B.~Feng, {\it {Generalized unitarity and one-loop amplitudes in N=4 super-Yang-Mills}},  {\em Nucl. Phys. B} {\bf 725} (2005) 275--305, [\href{http://arxiv.org/abs/hep-th/0412103}{{\tt hep-th/0412103}}].

\bibitem{Britto:2005ha}
R.~Britto, E.~Buchbinder, F.~Cachazo, and B.~Feng, {\it {One-loop amplitudes of gluons in SQCD}},  {\em Phys. Rev. D} {\bf 72} (2005) 065012, [\href{http://arxiv.org/abs/hep-ph/0503132}{{\tt hep-ph/0503132}}].

\bibitem{Britto:2006sj}
R.~Britto, B.~Feng, and P.~Mastrolia, {\it {The Cut-constructible part of QCD amplitudes}},  {\em Phys. Rev. D} {\bf 73} (2006) 105004, [\href{http://arxiv.org/abs/hep-ph/0602178}{{\tt hep-ph/0602178}}].

\bibitem{Anastasiou:2006gt}
C.~Anastasiou, R.~Britto, B.~Feng, Z.~Kunszt, and P.~Mastrolia, {\it {Unitarity cuts and Reduction to master integrals in d dimensions for one-loop amplitudes}},  {\em JHEP} {\bf 03} (2007) 111, [\href{http://arxiv.org/abs/hep-ph/0612277}{{\tt hep-ph/0612277}}].

\bibitem{Anastasiou:2006jv}
C.~Anastasiou, R.~Britto, B.~Feng, Z.~Kunszt, and P.~Mastrolia, {\it {D-dimensional unitarity cut method}},  {\em Phys. Lett. B} {\bf 645} (2007) 213--216, [\href{http://arxiv.org/abs/hep-ph/0609191}{{\tt hep-ph/0609191}}].

\bibitem{Britto:2006fc}
R.~Britto and B.~Feng, {\it {Unitarity cuts with massive propagators and algebraic expressions for coefficients}},  {\em Phys. Rev. D} {\bf 75} (2007) 105006, [\href{http://arxiv.org/abs/hep-ph/0612089}{{\tt hep-ph/0612089}}].

\bibitem{Britto:2007tt}
R.~Britto and B.~Feng, {\it {Integral coefficients for one-loop amplitudes}},  {\em JHEP} {\bf 02} (2008) 095, [\href{http://arxiv.org/abs/0711.4284}{{\tt arXiv:0711.4284}}].

\bibitem{Britto:2010um}
R.~Britto and E.~Mirabella, {\it {Single Cut Integration}},  {\em JHEP} {\bf 01} (2011) 135, [\href{http://arxiv.org/abs/1011.2344}{{\tt arXiv:1011.2344}}].

\bibitem{Giele:2008ve}
W.~T. Giele, Z.~Kunszt, and K.~Melnikov, {\it {Full one-loop amplitudes from tree amplitudes}},  {\em JHEP} {\bf 04} (2008) 049, [\href{http://arxiv.org/abs/0801.2237}{{\tt arXiv:0801.2237}}].

\bibitem{Mastrolia:2018uzb}
P.~Mastrolia and S.~Mizera, {\it {Feynman Integrals and Intersection Theory}},  {\em JHEP} {\bf 02} (2019) 139, [\href{http://arxiv.org/abs/1810.03818}{{\tt arXiv:1810.03818}}].

\bibitem{Frellesvig:2019uqt}
H.~Frellesvig, F.~Gasparotto, M.~K. Mandal, P.~Mastrolia, L.~Mattiazzi, and S.~Mizera, {\it {Vector Space of Feynman Integrals and Multivariate Intersection Numbers}},  {\em Phys. Rev. Lett.} {\bf 123} (2019), no.~20 201602, [\href{http://arxiv.org/abs/1907.02000}{{\tt arXiv:1907.02000}}].

\bibitem{Mizera:2019vvs}
S.~Mizera and A.~Pokraka, {\it {From Infinity to Four Dimensions: Higher Residue Pairings and Feynman Integrals}},  {\em JHEP} {\bf 02} (2020) 159, [\href{http://arxiv.org/abs/1910.11852}{{\tt arXiv:1910.11852}}].

\bibitem{Frellesvig:2020qot}
H.~Frellesvig, F.~Gasparotto, S.~Laporta, M.~K. Mandal, P.~Mastrolia, L.~Mattiazzi, and S.~Mizera, {\it {Decomposition of Feynman Integrals by Multivariate Intersection Numbers}},  {\em JHEP} {\bf 03} (2021) 027, [\href{http://arxiv.org/abs/2008.04823}{{\tt arXiv:2008.04823}}].

\bibitem{Caron-Huot:2021iev}
S.~Caron-Huot and A.~Pokraka, {\it {Duals of Feynman Integrals. Part II. Generalized unitarity}},  {\em JHEP} {\bf 04} (2022) 078, [\href{http://arxiv.org/abs/2112.00055}{{\tt arXiv:2112.00055}}].

\bibitem{Caron-Huot:2021xqj}
S.~Caron-Huot and A.~Pokraka, {\it {Duals of Feynman integrals. Part I. Differential equations}},  {\em JHEP} {\bf 12} (2021) 045, [\href{http://arxiv.org/abs/2104.06898}{{\tt arXiv:2104.06898}}].

\bibitem{Ossola:2007ax}
G.~Ossola, C.~G. Papadopoulos, and R.~Pittau, {\it {CutTools: A Program implementing the OPP reduction method to compute one-loop amplitudes}},  {\em JHEP} {\bf 03} (2008) 042, [\href{http://arxiv.org/abs/0711.3596}{{\tt arXiv:0711.3596}}].

\bibitem{Maierhofer:2018gpa}
P.~Maierh\"ofer and J.~Usovitsch, {\it {Kira 1.2 Release Notes}},  \href{http://arxiv.org/abs/1812.01491}{{\tt arXiv:1812.01491}}.

\bibitem{Klappert:2020nbg}
J.~Klappert, F.~Lange, P.~Maierh\"ofer, and J.~Usovitsch, {\it {Integral reduction with Kira 2.0 and finite field methods}},  {\em Comput. Phys. Commun.} {\bf 266} (2021) 108024, [\href{http://arxiv.org/abs/2008.06494}{{\tt arXiv:2008.06494}}].

\bibitem{Smirnov:2008iw}
A.~V. Smirnov, {\it {Algorithm FIRE -- Feynman Integral REduction}},  {\em JHEP} {\bf 10} (2008) 107, [\href{http://arxiv.org/abs/0807.3243}{{\tt arXiv:0807.3243}}].

\bibitem{Smirnov:2013dia}
A.~V. Smirnov and V.~A. Smirnov, {\it {FIRE4, LiteRed and accompanying tools to solve integration by parts relations}},  {\em Comput. Phys. Commun.} {\bf 184} (2013) 2820--2827, [\href{http://arxiv.org/abs/1302.5885}{{\tt arXiv:1302.5885}}].

\bibitem{Smirnov:2014hma}
A.~V. Smirnov, {\it {FIRE5: a C++ implementation of Feynman Integral REduction}},  {\em Comput. Phys. Commun.} {\bf 189} (2015) 182--191, [\href{http://arxiv.org/abs/1408.2372}{{\tt arXiv:1408.2372}}].

\bibitem{Smirnov:2008py}
A.~V. Smirnov and M.~N. Tentyukov, {\it {Feynman Integral Evaluation by a Sector decomposiTion Approach (FIESTA)}},  {\em Comput. Phys. Commun.} {\bf 180} (2009) 735--746, [\href{http://arxiv.org/abs/0807.4129}{{\tt arXiv:0807.4129}}].

\bibitem{Smirnov:2009pb}
A.~V. Smirnov, V.~A. Smirnov, and M.~Tentyukov, {\it {FIESTA 2: Parallelizeable multiloop numerical calculations}},  {\em Comput. Phys. Commun.} {\bf 182} (2011) 790--803, [\href{http://arxiv.org/abs/0912.0158}{{\tt arXiv:0912.0158}}].

\bibitem{Smirnov:2013eza}
A.~V. Smirnov, {\it {FIESTA 3: cluster-parallelizable multiloop numerical calculations in physical regions}},  {\em Comput. Phys. Commun.} {\bf 185} (2014) 2090--2100, [\href{http://arxiv.org/abs/1312.3186}{{\tt arXiv:1312.3186}}].

\bibitem{Smirnov:2015mct}
A.~V. Smirnov, {\it {FIESTA4: Optimized Feynman integral calculations with GPU support}},  {\em Comput. Phys. Commun.} {\bf 204} (2016) 189--199, [\href{http://arxiv.org/abs/1511.03614}{{\tt arXiv:1511.03614}}].

\bibitem{Smirnov:2021rhf}
A.~V. Smirnov, N.~D. Shapurov, and L.~I. Vysotsky, {\it {FIESTA5: Numerical high-performance Feynman integral evaluation}},  {\em Comput. Phys. Commun.} {\bf 277} (2022) 108386, [\href{http://arxiv.org/abs/2110.11660}{{\tt arXiv:2110.11660}}].

\bibitem{Klappert:2019emp}
J.~Klappert and F.~Lange, {\it {Reconstructing rational functions with FireFly}},  {\em Comput. Phys. Commun.} {\bf 247} (2020) 106951, [\href{http://arxiv.org/abs/1904.00009}{{\tt arXiv:1904.00009}}].

\bibitem{Klappert:2020aqs}
J.~Klappert, S.~Y. Klein, and F.~Lange, {\it {Interpolation of dense and sparse rational functions and other improvements in FireFly}},  {\em Comput. Phys. Commun.} {\bf 264} (2021) 107968, [\href{http://arxiv.org/abs/2004.01463}{{\tt arXiv:2004.01463}}].

\bibitem{Studerus:2009ye}
C.~Studerus, {\it {Reduze-Feynman Integral Reduction in C++}},  {\em Comput. Phys. Commun.} {\bf 181} (2010) 1293--1300, [\href{http://arxiv.org/abs/0912.2546}{{\tt arXiv:0912.2546}}].

\bibitem{Lee:2012cn}
R.~N. Lee, {\it {Presenting LiteRed: a tool for the Loop InTEgrals REDuction}},  \href{http://arxiv.org/abs/1212.2685}{{\tt arXiv:1212.2685}}.

\bibitem{Wu:2023upw}
Z.~Wu, J.~Boehm, R.~Ma, H.~Xu, and Y.~Zhang, {\it {NeatIBP 1.0, A package generating small-size integration-by-parts relations for Feynman integrals}},  \href{http://arxiv.org/abs/2305.08783}{{\tt arXiv:2305.08783}}.

\bibitem{Guan:2019bcx}
X.~Guan, X.~Liu, and Y.-Q. Ma, {\it {Complete reduction of integrals in two-loop five-light-parton scattering amplitudes}},  {\em Chin. Phys. C} {\bf 44} (2020), no.~9 093106, [\href{http://arxiv.org/abs/1912.09294}{{\tt arXiv:1912.09294}}].

\bibitem{Guan:2024byi}
X.~Guan, X.~Liu, Y.-Q. Ma, and W.-H. Wu, {\it {Blade: A package for block-triangular form improved Feynman integrals decomposition}},  \href{http://arxiv.org/abs/2405.14621}{{\tt arXiv:2405.14621}}.

\bibitem{Goode:2024mci}
J.~Goode, F.~Herzog, A.~Kennedy, S.~Teale, and J.~Vermaseren, {\it {Tensor reduction for Feynman integrals with Lorentz and spinor indices}},  {\em JHEP} {\bf 11} (2024) 123, [\href{http://arxiv.org/abs/2408.05137}{{\tt arXiv:2408.05137}}].

\bibitem{Goode:2024cfy}
J.~Goode, F.~Herzog, and S.~Teale, {\it {OPITeR: A program for tensor reduction of multi-loop Feynman Integrals}},  \href{http://arxiv.org/abs/2411.02233}{{\tt arXiv:2411.02233}}.

\bibitem{Chen:2024xwt}
W.~Chen, {\it {Semi-automatic Calculations of Multi-loop Feynman Amplitudes with AmpRed}},  \href{http://arxiv.org/abs/2408.06426}{{\tt arXiv:2408.06426}}.

\bibitem{Peraro:2019svx}
T.~Peraro, {\it {FiniteFlow: multivariate functional reconstruction using finite fields and dataflow graphs}},  {\em JHEP} {\bf 07} (2019) 031, [\href{http://arxiv.org/abs/1905.08019}{{\tt arXiv:1905.08019}}].

\bibitem{Belitsky:2023qho}
A.~V. Belitsky, A.~V. Smirnov, and R.~V. Yakovlev, {\it {Balancing act: Multivariate rational reconstruction for IBP}},  {\em Nucl. Phys. B} {\bf 993} (2023) 116253, [\href{http://arxiv.org/abs/2303.02511}{{\tt arXiv:2303.02511}}].

\bibitem{Smirnov:2020quc}
A.~V. Smirnov and V.~A. Smirnov, {\it {How to choose master integrals}},  {\em Nucl. Phys. B} {\bf 960} (2020) 115213, [\href{http://arxiv.org/abs/2002.08042}{{\tt arXiv:2002.08042}}].

\bibitem{Usovitsch:2020jrk}
J.~Usovitsch, {\it {Factorization of denominators in integration-by-parts reductions}},  \href{http://arxiv.org/abs/2002.08173}{{\tt arXiv:2002.08173}}.

\bibitem{Anastasiou:2004vj}
C.~Anastasiou and A.~Lazopoulos, {\it {Automatic integral reduction for higher order perturbative calculations}},  {\em JHEP} {\bf 07} (2004) 046, [\href{http://arxiv.org/abs/hep-ph/0404258}{{\tt hep-ph/0404258}}].

\bibitem{Abreu:2018zmy}
S.~Abreu, J.~Dormans, F.~Febres~Cordero, H.~Ita, and B.~Page, {\it {Analytic Form of Planar Two-Loop Five-Gluon Scattering Amplitudes in QCD}},  {\em Phys. Rev. Lett.} {\bf 122} (2019), no.~8 082002, [\href{http://arxiv.org/abs/1812.04586}{{\tt arXiv:1812.04586}}].

\bibitem{Ellis:2007qk}
R.~K. Ellis and G.~Zanderighi, {\it {Scalar one-loop integrals for QCD}},  {\em JHEP} {\bf 02} (2008) 002, [\href{http://arxiv.org/abs/0712.1851}{{\tt arXiv:0712.1851}}].

\bibitem{Carrazza:2016gav}
S.~Carrazza, R.~K. Ellis, and G.~Zanderighi, {\it {QCDLoop: a comprehensive framework for one-loop scalar integrals}},  {\em Comput. Phys. Commun.} {\bf 209} (2016) 134--143, [\href{http://arxiv.org/abs/1605.03181}{{\tt arXiv:1605.03181}}].

\bibitem{Denner:2002ii}
A.~Denner and S.~Dittmaier, {\it {Reduction of one loop tensor five point integrals}},  {\em Nucl. Phys. B} {\bf 658} (2003) 175--202, [\href{http://arxiv.org/abs/hep-ph/0212259}{{\tt hep-ph/0212259}}].

\bibitem{Denner:2016kdg}
A.~Denner, S.~Dittmaier, and L.~Hofer, {\it {Collier: a fortran-based Complex One-Loop LIbrary in Extended Regularizations}},  {\em Comput. Phys. Commun.} {\bf 212} (2017) 220--238, [\href{http://arxiv.org/abs/1604.06792}{{\tt arXiv:1604.06792}}].

\bibitem{Mastrolia:2016dhn}
P.~Mastrolia, T.~Peraro, and A.~Primo, {\it {Adaptive Integrand Decomposition in parallel and orthogonal space}},  {\em JHEP} {\bf 08} (2016) 164, [\href{http://arxiv.org/abs/1605.03157}{{\tt arXiv:1605.03157}}].

\bibitem{Feng:2021enk}
B.~Feng, T.~Li, and X.~Li, {\it {Analytic tadpole coefficients of one-loop integrals}},  {\em JHEP} {\bf 09} (2021) 081, [\href{http://arxiv.org/abs/2107.03744}{{\tt arXiv:2107.03744}}].

\bibitem{Hu:2021nia}
C.~Hu, T.~Li, and X.~Li, {\it {One-loop Feynman integral reduction by differential operators}},  {\em Phys. Rev. D} {\bf 104} (2021), no.~11 116014, [\href{http://arxiv.org/abs/2108.00772}{{\tt arXiv:2108.00772}}].

\bibitem{Feng:2022uqp}
B.~Feng, T.~Li, H.~Wang, and Y.~Zhang, {\it {Reduction of general one-loop integrals using auxiliary vector}},  {\em JHEP} {\bf 05} (2022) 065, [\href{http://arxiv.org/abs/2203.14449}{{\tt arXiv:2203.14449}}].

\bibitem{Feng:2022iuc}
B.~Feng and T.~Li, {\it {PV-reduction of sunset topology with auxiliary vector}},  {\em Commun. Theor. Phys.} {\bf 74} (2022), no.~9 095201, [\href{http://arxiv.org/abs/2203.16881}{{\tt arXiv:2203.16881}}].

\bibitem{Feng:2022rwj}
B.~Feng, J.~Gong, and T.~Li, {\it {Universal treatment of the reduction for one-loop integrals in a projective space}},  {\em Phys. Rev. D} {\bf 106} (2022), no.~5 056025, [\href{http://arxiv.org/abs/2204.03190}{{\tt arXiv:2204.03190}}].

\bibitem{Feng:2022rfz}
B.~Feng, C.~Hu, T.~Li, and Y.~Song, {\it {Reduction with degenerate Gram matrix for one-loop integrals}},  {\em JHEP} {\bf 08} (2022) 110, [\href{http://arxiv.org/abs/2205.03000}{{\tt arXiv:2205.03000}}].

\bibitem{Li:2022cbx}
T.~Li, {\it {Nontrivial one-loop recursive reduction relation}},  {\em JHEP} {\bf 07} (2023) 051, [\href{http://arxiv.org/abs/2209.11428}{{\tt arXiv:2209.11428}}].

\bibitem{Zhang:2023jzv}
L.~Zhang, {\it {Tensor loop reduction via the Baikov representation and an auxiliary vector}},  {\em Phys. Rev. D} {\bf 110} (2024), no.~1 016013, [\href{http://arxiv.org/abs/2309.00930}{{\tt arXiv:2309.00930}}].

\bibitem{Feng:2022hyg}
B.~Feng, {\it {Generation function for one-loop tensor reduction}},  {\em Commun. Theor. Phys.} {\bf 75} (2023), no.~2 025203, [\href{http://arxiv.org/abs/2209.09517}{{\tt arXiv:2209.09517}}].

\bibitem{Ablinger:2014yaa}
J.~Ablinger, J.~Bl\"umlein, C.~Raab, C.~Schneider, and F.~Wi\ss{}brock, {\it {Calculating Massive 3-loop Graphs for Operator Matrix Elements by the Method of Hyperlogarithms}},  {\em Nucl. Phys. B} {\bf 885} (2014) 409--447, [\href{http://arxiv.org/abs/1403.1137}{{\tt arXiv:1403.1137}}].

\bibitem{Kosower:2018obg}
D.~A. Kosower, {\it {Direct Solution of Integration-by-Parts Systems}},  {\em Phys. Rev. D} {\bf 98} (2018), no.~2 025008, [\href{http://arxiv.org/abs/1804.00131}{{\tt arXiv:1804.00131}}].

\bibitem{Guan:2023avw}
X.~Guan, X.~Li, and Y.-Q. Ma, {\it {Exploring the linear space of Feynman integrals via generating functions}},  \href{http://arxiv.org/abs/2306.02927}{{\tt arXiv:2306.02927}}.

\bibitem{Feng:2024qsa}
B.~Feng, C.~Hu, J.~Shen, and Y.~Zhang, {\it {General One-loop Generating Function by IBP relations}},  \href{http://arxiv.org/abs/2403.16040}{{\tt arXiv:2403.16040}}.

\bibitem{Li:2024rvo}
T.~Li, Y.~Song, and L.~Zhang, {\it {Solving arbitrary one-loop reduction via generating function}},  \href{http://arxiv.org/abs/2404.04644}{{\tt arXiv:2404.04644}}.

\bibitem{Chen:2024bpf}
G.~Chen, J.-W. Kim, and T.~Wang, {\it {Systematic integral evaluation for spin-resummed binary dynamics}},  {\em Phys. Rev. D} {\bf 111} (2025), no.~2 L021701, [\href{http://arxiv.org/abs/2406.17658}{{\tt arXiv:2406.17658}}].

\bibitem{Brandhuber:2024qdn}
A.~Brandhuber, G.~R. Brown, G.~Chen, G.~Travaglini, and P.~Vives~Matasan, {\it {Spinning waveforms in cubic effective field theories of gravity}},  {\em JHEP} {\bf 12} (2024) 039, [\href{http://arxiv.org/abs/2408.00587}{{\tt arXiv:2408.00587}}].

\end{thebibliography}\endgroup

\end{document}